\documentclass[11pt]{article}

\usepackage{epsfig}

\usepackage{float}

\usepackage{tikz}
\usetikzlibrary{shapes,arrows} 
\usepackage{epstopdf, epsfig}

\usepackage[label font=it]{subfig}
\usepackage{caption}
\usepackage{multirow,tabularx}

\newsavebox{\twofigures}
\usepackage[export]{adjustbox}
\usepackage{calc}
\usepackage[version=4]{mhchem}
\usepackage{siunitx}
\usepackage{longtable,tabularx}
\usepackage{tabularray}
\usepackage{lmodern,pdftexcmds,amsmath}

\usepackage[title]{appendix}
\usepackage{titlesec}
\titleformat{\paragraph}
{\normalfont\normalsize}{\theparagraph}{1em}{}
\titlespacing*{\paragraph}
{0pt}{3.25ex plus 1ex minus .2ex}{1.5ex plus .2ex}
\counterwithin*{equation}{section}
\DeclareMathAlphabet{\mathsfbi}{OT1}{\sfdefault}{bx}{sl}
\DeclareMathVersion{sfletters}
\SetSymbolFont{letters}{sfletters}{OML}{ntxsfmi}{b}{it}

\makeatletter
\newcommand{\mathbfsbilow}[1]{%
  \text{\mathversion{sfletters}$\m@th#1$}%
}

\DeclareRobustCommand{\tensor}[1]{%
  \begingroup
  \ifcat\noexpand #1\relax
    \edef\greek@test{\detokenize{#1}}%
    \edef\greek@test{\expandafter\@cdr\greek@test\@nil}%
    \edef\greek@test{\expandafter\@car\greek@test\@nil}%
    \edef\x{\the\lccode\expandafter`\greek@test}%
    \edef\y{\number\expandafter`\greek@test}%
    \ifnum\x=\y\relax
      \mathbfsbilow{#1}%
    \else
      \mathsfbi{#1}%
    \fi
  \else
    \mathsfbi{#1}%
  \fi
  \endgroup
}
\makeatother

\usepackage{orcidlink}

\usepackage{tabularx, caption}
 \makeatletter
 \newcommand*{\compress}{\@minipagetrue}
 \makeatother

\newsavebox{\bigimage}

\usepackage{graphicx}
\usepackage{newtxtext}
\usepackage{newtxmath}
\usepackage{natbib}
\usepackage{hyperref}

\usepackage{algorithm}
\usepackage[noend]{algpseudocode}

\usepackage{etoolbox} 
\makeatletter
\patchcmd{\NAT@citex}
  {\@citea\NAT@hyper@{%
     \NAT@nmfmt{\NAT@nm}%
     \hyper@natlinkbreak{\NAT@aysep\NAT@spacechar}{\@citeb\@extra@b@citeb}%
     \NAT@date}}
  {\@citea\NAT@nmfmt{\NAT@nm}%
   \NAT@aysep\NAT@spacechar\NAT@hyper@{\NAT@date}}{}{}

\patchcmd{\NAT@citex}
  {\@citea\NAT@hyper@{%
     \NAT@nmfmt{\NAT@nm}%
     \hyper@natlinkbreak{\NAT@spacechar\NAT@@open\if*#1*\else#1\NAT@spacechar\fi}%
       {\@citeb\@extra@b@citeb}%
     \NAT@date}}
  {\@citea\NAT@nmfmt{\NAT@nm}%
   \NAT@spacechar\NAT@@open\if*#1*\else#1\NAT@spacechar\fi\NAT@hyper@{\NAT@date}}
  {}{}
\makeatother

\hypersetup{
    colorlinks = true,
     citecolor  = blue,
    allcolors=blue,
}

\newcommand{\RomanNumeralCaps}[1]

\usepackage{multirow,tabularx}

\usepackage[export]{adjustbox}
\usepackage{calc}
\usepackage[version=4]{mhchem}
\usepackage{siunitx}
\usepackage{longtable,tabularx}
\usepackage{tabularray}
 \UseTblrLibrary{amsmath}
 \usepackage{lmodern,pdftexcmds,amsmath}

\usepackage[noblocks]{authblk}
\usepackage[margin=1in]{geometry}
\usepackage[title]{appendix}

\newcommand\affiliation[1]{\gdef\@affiliation{\let\aff\aff@inst#1}}
\gdef\@affiliation{}
\def\email#1{Email address for correspondence: #1}
\def\aff#1{\ignorespaces\textsuperscript{#1}}
\def\corresp#1{\unskip\thanks{#1}}
\numberwithin{equation}{section}

\renewenvironment{abstract}
{\begin{quote}
\noindent \rule{\linewidth}{.5pt}\par{\bfseries \abstractname.}}
{\medskip\noindent \rule{\linewidth}{.5pt}
\end{quote}
}

\hypersetup{
   colorlinks=true,
   linkcolor={blue!100!black},
   citecolor={blue!100!black},
   plainpages=false,
}

\title{\bf Resolvent-based estimation and control of a laminar airfoil wake}

\author[1]{\bf Junoh Jung\corresp{\email{junohj@umich.edu}}}
\author[1, 2]{\bf Rutvij Bhagwat}
\author[1]{\bf Aaron Towne}

\affil[1]{\normalsize Department of Mechanical Engineering, University of Michigan, Ann Arbor, MI, USA  }
\affil[2]{\normalsize Department of Mechanical Engineering, Florida State University, Tallahassee, FL 32310, USA \vspace{-1cm}}

\date{}
\begin{document}
\maketitle

\begin{abstract}
We develop an optimal resolvent-based estimator and controller to predict and attenuate unsteady vortex shedding fluctuations in the laminar wake of a NACA 0012 airfoil at an angle of attack of 6.5 degrees, chord-based Reynolds number of 5000, and Mach number of 0.3. The resolvent-based estimation and control framework offers several advantages over standard methods. Under equivalent assumptions, the resolvent-based estimator and controller reproduce the Kalman filter and LQG controller, respectively, but at substantially lower computational cost using either an operator-based or data-driven implementation. Unlike these methods, the resolvent-based approach can naturally accommodate forcing terms (nonlinear terms from Navier-Stokes) with colored-in-time statistics, significantly improving estimation accuracy and control efficacy. Causality is optimally enforced using a Wiener-Hopf formalism. We integrate these tools into a high-performance-computing-ready compressible flow solver and demonstrate their effectiveness for estimating and controlling velocity fluctuations in the wake of the airfoil immersed in clean and noisy freestreams, the latter of which prevents the flow from falling into a periodic limit cycle. Using four shear-stress sensors on the surface of the airfoil, the resolvent-based estimator predicts a series of downstream targets with approximately $3\%$ and $30\%$ error for the clean and noisy freestream conditions, respectively. For the latter case, using four actuators on the airfoil surface, the resolvent-based controller reduces the turbulent kinetic energy in the wake by $98\%$. \\  
\end{abstract}


\section{Introduction}\label{sec:intro}

The laminar flow over an airfoil is a canonical problem in fluid mechanics due to its role in aerodynamics and status as a prototypical problem for studying wakes.  Unsteady perturbations in the wake are of particular interest for several reasons.  First, these perturbations are intimately tied to the separation bubble that forms over the suction side of the airfoil, which in turn increases drag \citep{alam_ultra-low_2010, chang_shedding_2022}.  Second, unsteady fluctuations can degrade aerodynamic performance in many aircraft flight control scenarios, such as maneuvers at high angles of attack, landing, takeoff, or encountering atmospheric turbulence.  Third, wake perturbations significantly contribute to aerodynamic noise, which is a concern for wind turbines \citep{wagner_wind_1996, agrawal_towards_2015} and rotorcrafts, including drones.  In all these scenarios, accurate estimation and effective closed-loop control of wake perturbations are critical to improve engineering performance, i.e., to reduce drag, enhance flight control, and mitigate aerodynamic noise.\\
\indent A dominant feature of the laminar flow over an airfoil is vortex shedding in the wake. Vortices form on each side of the airfoil and shed periodically, creating downstream flow patterns such as the Karman vortex street. Factors including the shape of the object, the angle of attack, and the Reynolds number influence this vortex shedding. In NACA0012 airfoils at moderate angles of attack (6-10 degrees), an anticlockwise vortex is generated at the trailing edge, and a clockwise vortex at the front suction side separates and is entrained downstream \citep{chang_shedding_2022} -- accordingly, both the front section of the suction side and the trailing edge impact vortex-shedding. 
Many studies have been conducted to suppress vortex shedding in the wakes behind cylinders \citep{jin_feedback_2020,deda_neural_2023,lin_feedback_2024} and airfoils \citep{colonius_control_2011,broglia_output_nodate}.\\
\indent An accurate estimator is essential for successful closed-loop control \citep{RobertFStengel1994, brunton_closed-loop_2015}.  Standard estimation and control methods such as the Kalman filter \citep{Kalman_1960} and the linear-quadratic-Gaussian (LQG) controller have been applied to fluid mechanics problems over the past two decades, e.g., Kalman filter \citep{rafiee_kalman_2009, colburn_state_2011, An_liftcoeff_estimation_2021} and LQG control \citep{bagheri_inputoutput_2009,fabbiane_adaptive_2014,fabbiane_role_2015,fabbiane_energy_2017,sasaki_wave-cancelling_2018}.  However, these standard methods have two significant limitations when applied to flow-control problems.  First, solving the Riccati equations required to obtain the Kalman and LQG gains scales poorly with problem size and becomes computationally expensive, or in some cases prohibitive, for the large systems typically obtained when discretizing the Navier-Stokes equations.  This issue can be partially mitigated by reducing the size of the system a priori via some model-reduction method \citep{pasquale_robust_2017, gomez_unsteady_2019}, but this potentially degrades the performance of the controller.  Second, standard methods model the nonlinear terms of the Navier-Stokes equations as a white noise forcing on the linear dynamics.  These terms are not white \citep{zare_colour_2017, Towne2017statistical, Morra2021colour}, and treating them as such deteriorates estimation and control performance \citep{martini_resolvent-based_2020, amaral_resolvent-based_2021}.\\
\indent In this paper, we utilize a recently developed class of estimation and control methods based on resolvent analysis \citep{towne_resolvent-based_2020, martini_resolvent-based_2020, martini_resolvent-based_2022}.  Resolvent analysis, or input-output analysis, is a powerful methodology based on a linear mapping between forcing (input) and response (output) modes and the gain between them.  These modes and gains are obtained from a singular value decomposition of the resolvent operator obtained from the linearized Navier-Stokes equations.  Early studies viewed the input as an external forcing on the Navier-Stokes equations linearized about some laminar fixed point \citep{jovanovic_componentwise_2005, sharma_stabilising_2006, sipp_dynamics_2010}.  \cite{mckeon_critical-layer_2010} extended resolvent analysis to turbulent flows by interpreting the nonlinear terms from the Navier-Stokes equations as a forcing on the linear dynamics.  \cite{towne_spectral_2018} showed that resolvent modes and spectral proper orthogonal decomposition (SPOD) modes are identical when the forcing is white noise, directly linking resolvent modes to coherent structures observed in flow data.  However, resolvent analysis poses computational challenges for high-dimensional problems.  To address these difficulties, \cite{ribeiro_randomized_2020} applied randomized SVD.  \cite{martini_resolvent-based_2020} used a time-stepping approach to obtain the action of the resolvent operator on a specified forcing, eliminating the need to compute the inverse of large matrices.  \cite{farghadan_scalable_2023,farghadan_efficient_2024} further extended the range of problems amenable to resolvent analysis by combining randomized SVD with an approach to minimize the cost of the time-stepping method and successfully applied the method to several three-dimensional problems.\\
\indent Resolvent analysis has been extensively used to study the flow over airfoils at different Reynolds numbers and angles of attack. \cite{thomareis_resolvent_2018} investigated the physics of separated and attached flows around a NACA 0012 airfoil at $\emph{Re}_{L_{c}}=50,000$ and angle of attack $5^{\circ}$. \cite{symon_tale_2019} investigated two angles of attack, $0^{\circ}$ and $10^{\circ}$, for a NACA 0018 airfoil and showed that these two cases behave as an oscillator and amplifier \citep{huerre_local_nodate}, respectively. \cite{kojima_resolvent_2020} identified the origin of the two-dimensional transonic buffet over a NACA 0012 airfoil at $\emph{Re}_{L_{c}}=2,000$, $Ma_{\infty}=0.85$ and $\alpha=3^{\circ}$. \cite{marquet_hysteresis_2022} studied the flow over a NACA 0012 at $\emph{Re}_{L_{c}}=5,000$ for angles of attack between $\alpha=6.5^{\circ}$ and $9^{\circ}$ using an incompressible Navier-Stokes linear operator with the mean flow obtained from a numerical simulation, which was validated against experimental data.\\
\indent Motivated by the ability of resolvent modes to efficiently represent coherent structures, \cite{towne_resolvent-based_2020} introduced a resolvent-based method to estimate space-time flow statistics from limited sensor measurements.  \cite{martini_resolvent-based_2020} extended this method to reconstruct the time series of the flow state using limited measurements.  \cite{amaral_resolvent-based_2021} used this method to estimate the velocity field in a channel flow using pressure and shear-stress measurements at the channel wall.  The aforementioned resolvent-based estimators are non-causal, i.e., they use both past and future measurements to determine the current flow state.  \cite{martini_resolvent-based_2022} derived a causal resolvent-based estimator and controller by enforcing causality using a Wiener-Hopf formalism \citep{martinelli_feedback_nodate}, making them applicable for closed-loop flow control.\\
\indent The causal resolvent-based estimator and controller reproduce the Kalman filter and LQG controller, respectively, under equivalent assumptions, namely white-noise forcing.  However, the resolvent-based estimator and controller have two crucial advantages over these standard methods.  First, they can be efficiently computed even for large systems using a time-stepping method similar to that described earlier for computing resolvent modes or via a data-driven approach.  This makes the method applicable to the large problems typical of fluid mechanics without the need for detrimental a priori model reduction.  Second, the resolvent-based methods can accommodate colored-in-time forcing to statistically account for the nonlinear terms from the Navier-Stokes equations, improving estimation accuracy and control efficacy.\\
\begin{figure}[]
  \begin{center}
      \begin{tikzpicture}[baseline]
        \tikzstyle{every node}=[font=\small]
        \tikzset{>=latex}
        \node[anchor=south west,inner sep=0] (image) at (0,0) {
          \includegraphics[scale=1,width=1\textwidth]{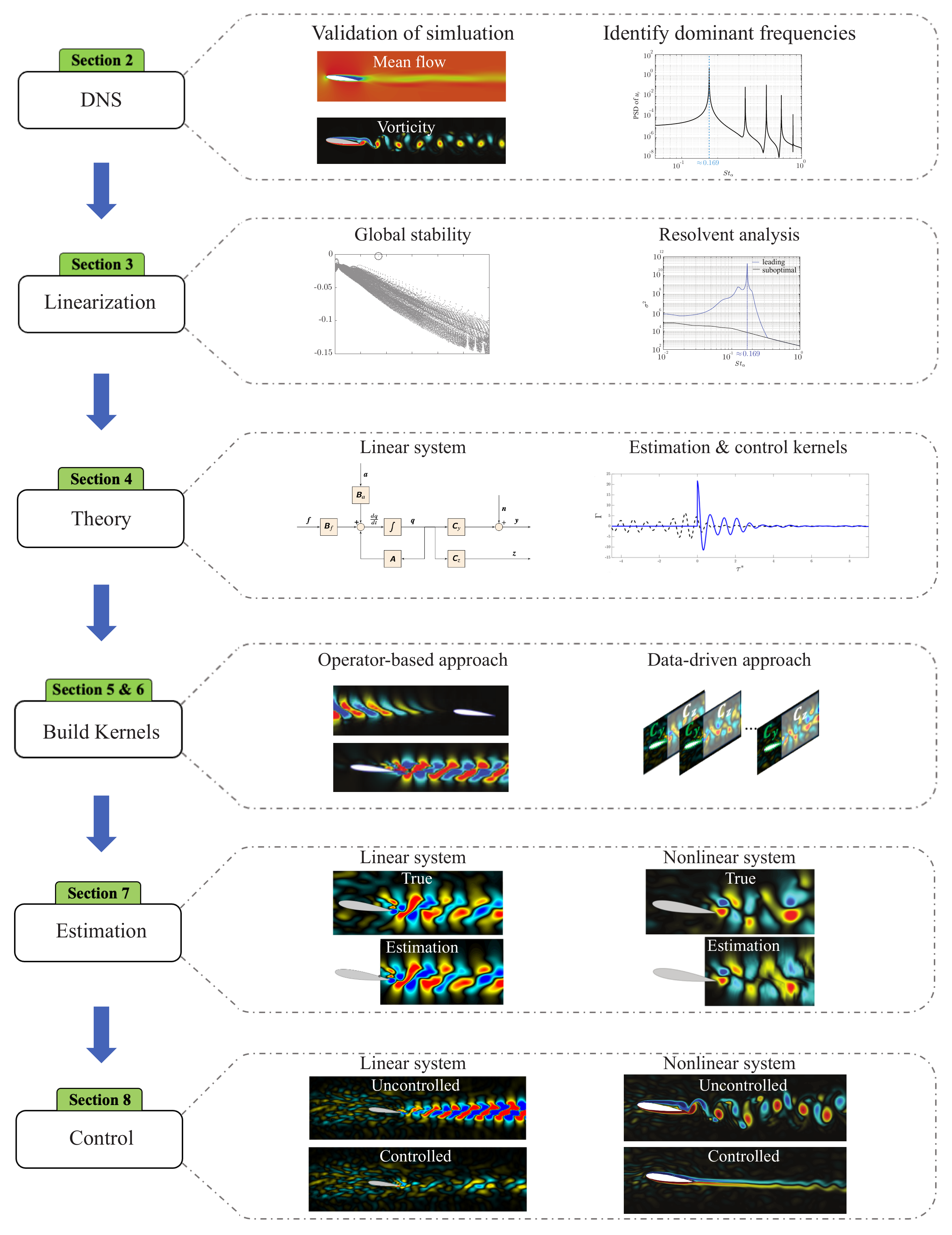}
        };
      \end{tikzpicture}
    \caption[]{\label{Fig_Roadmap} Roadmap for resolvent-based estimation and control of the laminar flow over an airfoil.}
  \end{center}
\end{figure}
\indent In this paper, we aim to estimate and control unsteady fluctuations in the wake of a two-dimensional NACA 0012 airfoil at $Ma_{\infty} = 0.3$, $Re_{L_{c}} = 5000$, and $\alpha = 6.5^\circ$ using the resolvent-based methods. The overall roadmap of the paper is depicted schematically in figure \ref{Fig_Roadmap}.  We begin in $\S$~\ref{sec_Problem} by computing the flow using direct numerical simulation (DNS) and validate the simulation against results in the literature.  In $\S$~\ref{sec_GlobalStabilityAnalysis_RSV}, we analyze the global eigenmodes and resolvent modes of the flow to validate the linearized compressible Navier-Stokes operator and highlight the key flow physics.  We then introduce the resolvent-based estimation and control methods in $\S$~\ref{sec_Methodology}, discuss how the kernels are computed in $\S$~\ref{sec_Computing_est_con_kernels}, and detail our implementation of these tools into a computational fluids dynamics (CFD) code in $\S$~\ref{sec_Implmentation_Resolvent-based control tools}.  The resolvent-based estimator is applied to both the linearized and nonlinear problems for clean and noisy freestream conditions, the latter of which is designed to disrupt the periodic vortex-shedding and induce chaotic fluctuations in the wake, in $\S$~\ref{sec_Resolvent-based estimation}.  Following this, we demonstrate the effectiveness of resolvent-based control for the same systems and flow conditions in $\S$~\ref{sec_Resolvent-based control}.  Finally, $\S$~\ref{sec_Conclusion} summarizes the paper and discusses future research directions.


\section{Problem setup and simulation}\label{sec_Problem}


We aim to use a resolvent-based approach to estimate and mitigate chaotic velocity fluctuations in a laminar airfoil wake.  Following \cite{marquet_hysteresis_2022}, we consider the flow around a NACA0012 airfoil at a low chord-based Reynolds number of $Re_{L_{c}} = 5000$, Mach number of $Ma_{\infty} = 0.3$, and an angle of attack of $\alpha = 6.5^{\circ}$.  We consider two different freestream conditions: (i) a clean freestream with no ambient fluctuations and (ii) a noisy freestream with substantial ambient fluctuations generated by random forcing upstream of the airfoil.  The flow falls into a periodic limit cycle due to vortex shedding for the clean freestream; the noisy freestream kicks the flow out of this limit cycle, leading to chaotic fluctuations in the wake -- a far more challenging problem for estimation and control.  

\begin{figure}[t]
  \begin{center}
      \begin{tikzpicture}[baseline]
        \tikzstyle{every node}=[font=\small]
        \tikzset{>=latex}
        \node[anchor=south west,inner sep=0] (image) at (0,0) {
          \includegraphics[scale=0.9,width=1\textwidth]{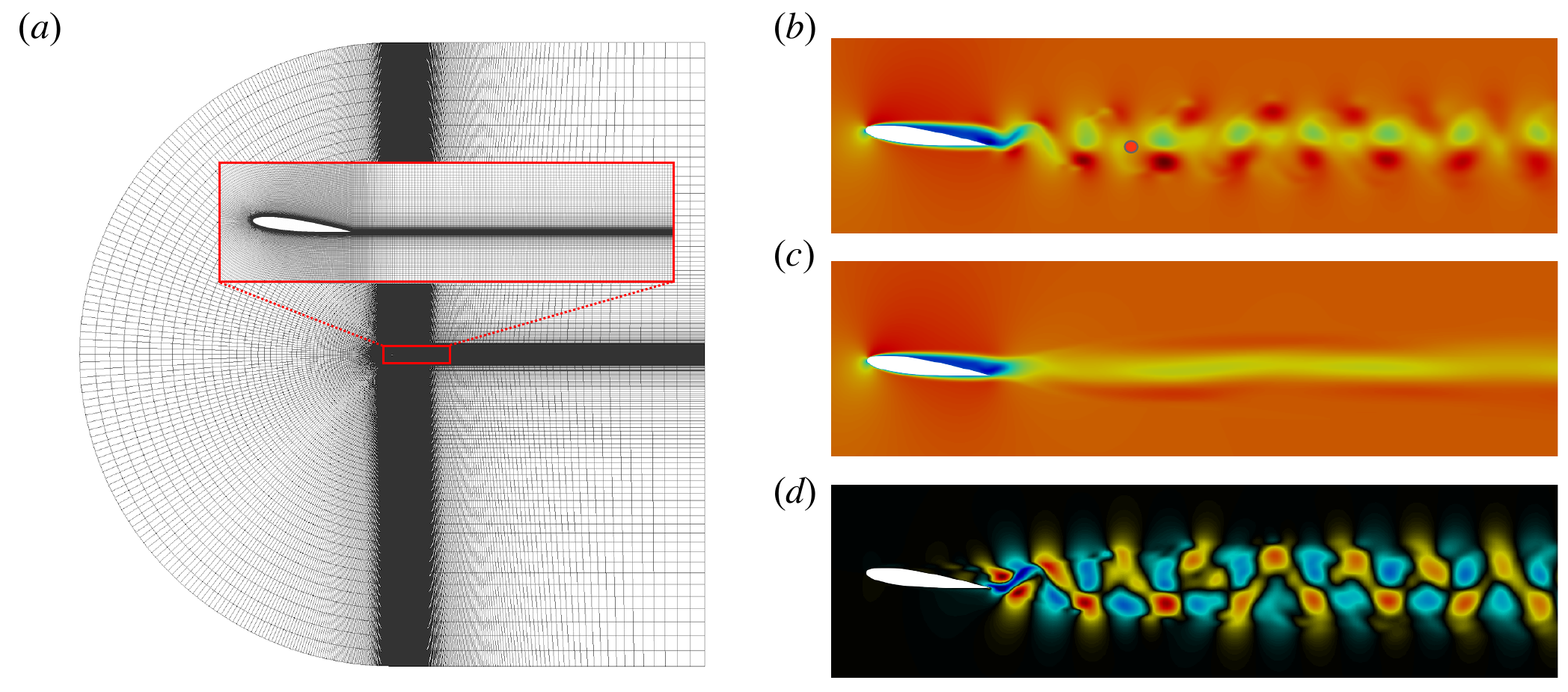}
        };
      \end{tikzpicture}
    \caption[]{\label{Fig_DNS} Direct numerical simulation: (\textit{a}) the full computational C-shaped grid with a close-up view for the wall and wake regions, (\textit{b}) a snapshot of the instantaneous streamwise velocity $u_x$ with the red dot indicating the probe location at $(x, y) / L_c = (2.1, -0.11)$ for the power spectral density in figure \ref{Fig_PSD}, (\textit{c}) the mean streamwise velocity $\bar{u}_x$, and (\textit{d}) a snapshot of the instantaneous streamwise velocity fluctuation $u_x'$.}
  \end{center}
\end{figure}

We simulate the flow via a direct numerical simulation (DNS) using the compressible flow solver CharLES \citep{bres_unstructured_2017}. A C-shape mesh is created using Pointwise, as shown in figure \ref{Fig_DNS}(\textit{a}), where the computational grid near the airfoil is also shown in the red box. The leading edge of the airfoil is located at the origin, $x/L_{c}=y/L_{c}=0$. The size of the domain in the streamwise and normal direction is $x/L_{c} \in [-49,50] $ and $y/L_{c} \in [-50,50] $, respectively. To create a two-dimensional simulation, the spanwise direction is one cell thick ($z/L_c \in [0, 0.1]$) with symmetry boundary conditions. The grid consists of approximately 148,000 cells, with a finer grid resolution applied in the wake region, as studied in appendix $\ref{appB}$. Characteristic far-field boundary conditions are applied along the outer boundary of the domain, and a sponge layer is used in the region $x/L_{c}\in[30,50]$ to prevent reflections and ensure an effective outflow boundary condition as the wake exits the domain.  
Time integration is performed using a third-order total-variation-diminishing Runge-Kutta scheme \citep{gottlieb_total_1998}. A constant timestep is maintained by setting the Courant–Friedrichs–Lewy (CFL) number to approximately 1. After passing initial transients, data are collected for the duration $t U_{\infty} / L_c \in [0, 350]$.  This extended time window ensures convergence of the mean and the second-order space-time statistics of vortex shedding in the wake.

Figure \ref{Fig_DNS}(\textit{b}) shows an instantaneous snapshot of the streamwise velocity field around the airfoil. A separation bubble on the suction side of the airfoil and vortex shedding in the wake are clearly observed. The mean streamwise velocity is shown in Figure \ref{Fig_DNS}(\textit{c}).  This mean flow (along with the mean of the other state variables) is used to define the linearization used to construct the estimation and control kernels.  Figure \ref{Fig_DNS}(\textit{d}) shows an instantaneous snapshot of the streamwise velocity fluctuation, i.e., the mean-subtracted velocity.  We aim to estimate and control (suppress) such fluctuations.

\begin{figure}[]
  \begin{center}
      \begin{tikzpicture}[baseline]
        \tikzstyle{every node}=[font=\small]
        \tikzset{>=latex}
        \node[anchor=south west,inner sep=0] (image) at (0,0) {
          \includegraphics[scale=0.6,width=0.6\textwidth]{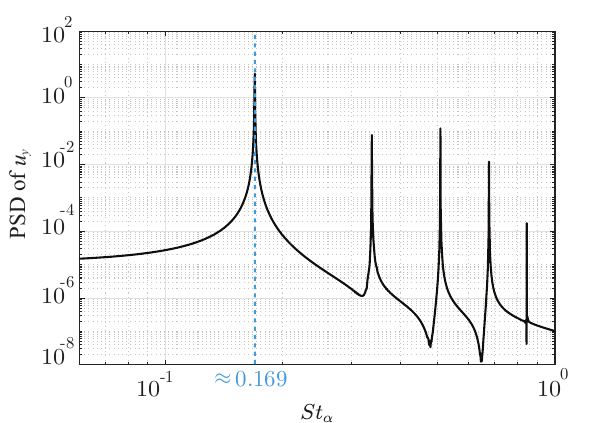}
        };
      \end{tikzpicture}
    \caption[]{\label{Fig_PSD} Power spectral density (PSD) of the wall-normal velocity at $(x,y)/L_{c}=(2.1,-0.11)$.}
  \end{center}
\end{figure}

We validate the DNS via comparisons of the aerodynamic forces and vortex-shedding frequency against the results of \cite{marquet_hysteresis_2022}, who considered incompressible flow over the same airfoil at the same Reynolds number and angle of attack. The time-averaged drag and lift coefficients, 
\begin{equation}\label{eq2.1}
C_{D}=\frac{F_{D}}{\frac{1}{2} \rho_{\infty} U_{\infty}^{2} A}, C_{L}=\frac{F_{L}}{\frac{1}{2} \rho_{\infty} U_{\infty}^{2} A},
\end{equation}
are reported in Table \ref{tab1}.  Our results match those of \cite{marquet_hysteresis_2022} within a 2\% error.  The power spectral density (PSD) of the transverse velocity at $x/L_{c}=2.1,y/L_{c}=-0.11$ is shown in figure \ref{Fig_PSD}.  The vortex-shedding frequency $St_{\alpha} \equiv \omega_{r}(L_{c} sin\alpha)/(2 \pi Ma_{\infty})$ is approximately 0.169 in the present study, close to the value of 0.18 found by \cite{marquet_hysteresis_2022}.  This slight difference in the aerodynamic forces and vortex-shedding frequency could result from minor differences between the incompressible and compressible flows or differences in the grid refinement (the present grid is more finely resolved). 

\begin{table}[t]
\centering
 \begin{tabular}{ c c c c}
\hline
& \quad Present study  & \quad \cite{marquet_hysteresis_2022} & Error   \\
\hline
  $\bar{C}_{D}$ & \quad 0.0862 & \quad 0.088 & 2.05\%   	\\
  $\bar{C}_{L}$ & \quad 0.2941 & \quad 0.289 & 1.76\% \\
 \hline
\end{tabular}
\caption{The comparison of the time-averaged drag and lift coefficients at $\alpha = 6.5{^\circ}$ with the results from incompressible periodic solution for a NACA 0012 airfoil at $\emph{Re}_{L_{c}} = 5,000$ and $Ma_{\infty} =0.3$.}
\label{tab1}
\end{table}

\section{Global stability and resolvent analyses}\label{sec_GlobalStabilityAnalysis_RSV}
 
\subsection{Linearization and global stability analysis}\label{subsec_Linearization_and_global_stability_analysis}
 We start with the compressible Navier-Stokes equations written as
\begin{equation}\label{eq3.1}
\frac{\partial \boldsymbol{q}}{\partial t}= \mathcal{F} (\boldsymbol{q}),
\end{equation}
where $\boldsymbol{q}$ is a state vector of flow variables $[\rho,\rho u_{x},\rho u_{y},\rho u_{z},\rho E]^{T}$ and $\mathcal{F}$ is the nonlinear Navier-Stokes operator. The equations are linearized using a Reynolds decomposition, giving

\begin{equation}\label{eq3.2}
\frac{\partial \boldsymbol{q}'}{\partial t}-\mathsfbi{A}\boldsymbol{q}'=\mathsfbi{B}\boldsymbol{f}(\boldsymbol{\bar{q}},\boldsymbol{q}'),
\end{equation}
where $\bar{\boldsymbol{q}}$ and $\boldsymbol{q}'$ represent the mean and perturbation state vectors, respectively, $\mathsfbi{A} = \frac{\partial \mathcal{F} (\boldsymbol{\bar{q}})}{\partial \boldsymbol{q}}$ is the linearized Navier-Stokes operator, $\boldsymbol{f}$ comprises the remaining nonlinear terms and any exogenous forcing, and $\mathsfbi{B}$ is an input matrix that can be used to restrict the form of $\boldsymbol{f}$. For convenience, we omit $(\cdot )^{'}$ for perturbation from this point on. Our approach for constructing $\mathsfbi{A}$ is detailed in $\S$\ref{sec_Implmentation_Resolvent-based control tools}.

 \begin{figure}[t]
  \begin{center}
      \begin{tikzpicture}[baseline]
        \tikzstyle{every node}=[font=\small]
        \tikzset{>=latex}
        \node[anchor=south west,inner sep=0] (image) at (0,0) {
          \includegraphics[scale=0.9,width=1\textwidth]{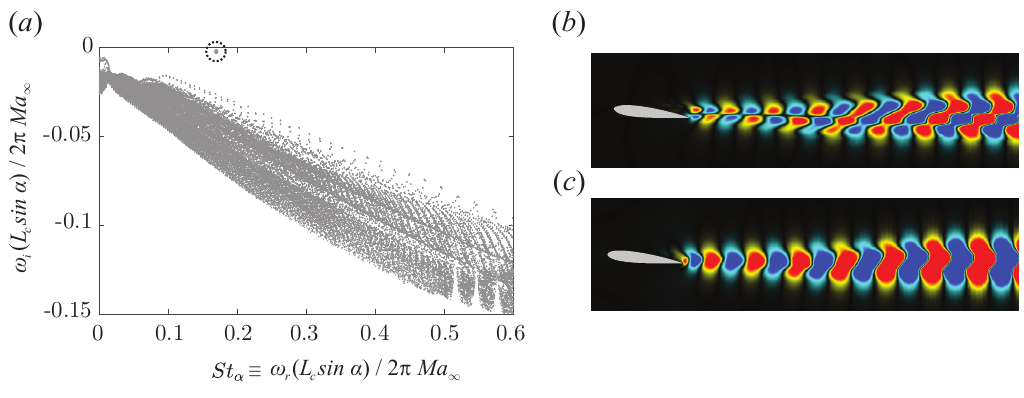}
        };
      \end{tikzpicture}

    \caption[]{\label{Fig_Eigenspectrum}Eigenspectrum: (\textit{a}) Eigenspectrum. The dotted circle shows the dominant wake eigenmode at the vortex-shedding frequency $St_{\alpha}\approx0.17$, (\textit{b}) the corresponding streamwise velocity eigenmode, and (\textit{c}) cross-streamwise velocity eigenmode.}
    
  \end{center}
\end{figure}  

Resolvent-based estimation and control are nominally applicable and robust only for globally stable systems \citep{schmid_linear_2016, martini_resolvent-based_2020, martini_resolvent-based_2022}. Thus, we first conduct a global stability analysis to ensure that our linear operator is stable. Figure \ref{Fig_Eigenspectrum}(\textit{a}) shows the spectrum of the linearized Navier-Stokes operator $\mathsfbi{A}$, represented in terms of the complex frequency $\omega = i \lambda$, where $\lambda$ is the eigenvalue of $\mathsfbi{A}$.  The imaginary part $\omega_i$ is negative for all eigenvalues, indicating that the flow around the airfoil is globally stable.  The least-damped eigenvalue appears at the frequency $St_{\alpha} \approx 0.169$, which matches the vortex-shedding frequency observed in the PSD in figure \ref{Fig_PSD}.  The corresponding eigenmode, shown in \ref{Fig_Eigenspectrum}(\textit{b}) and (\textit{c}), exhibits the expected characteristics of vortex shedding \citep{Noack2003hierarchy}.

\subsection{Resolvent analysis}\label{subsec_Resolvent_analysis}

Resolvent analysis is central to our estimation and control methods. Therefore, we compute the leading resolvent modes for the airfoil flow as a preliminary step to validate our implementation and to gain insights into the flow physics and the appropriate sensor and actuator placements. After transforming the linear system~(\ref{eq3.2}) into the frequency domain and solving for the state, we obtain
\begin{equation}\label{eq_q_Rf}
\boldsymbol{\hat{q}} = \mathsfbi{R} \mathsfbi{B} \boldsymbol{\hat{f}},
\end{equation}
where the resolvent operator is defined as $\mathsfbi{R} = (-i \omega \mathsfbi{I} - \mathsfbi{A})^{-1}$. The notation ($\hat{\cdot }$) indicates a quantity in the frequency domain throughout this paper.

Resolvent analysis seeks input and output modes that maximize the resolvent gain 
\begin{equation}
\label{eq:resolvent_gain_def}
\sigma^{2} = \frac{\| \hat{\boldsymbol{y}} \|^2}{\| \hat{\boldsymbol{f}} \|^2},
\end{equation}
where $\hat{\boldsymbol{y}} = \mathsfbi{C}\hat{\boldsymbol{q}}$ is an output of interest extracted from the state by the output matrix $\mathsfbi{C}$.  The norm is induced by the inner product $\langle \boldsymbol{q}_{1},\boldsymbol{q}_{2}\rangle = \boldsymbol{q}_{1}^{*} \mathsfbi{W} \boldsymbol{q}_{2}$, where $\mathsfbi{W}$ is a weight matrix used to set a desired norm and $(\cdot)^{*}$ indicates the conjugate transpose.  We use the Chu compressible energy norm \citep{chu_energy_1965} and set $\mathsfbi{B}$ and $\mathsfbi{C}$ to the identity matrix for the preliminary resolvent analysis in this section (non-identity values will be used later for estimation and control).        


After defining the weighted resolvent operator
\begin{equation}\label{eq3.4}
\mathsfbi{\tilde{R}}= \mathsfbi{W}^{\frac{1}{2}} \mathsfbi{C}\mathsfbi{R} \mathsfbi{B} \mathsfbi{W}^{-\frac{1}{2}},
\end{equation}
the resolvent gains and modes are obtained from the SVD
\begin{equation}\label{eq3.6}
\mathsfbi{\tilde{R}}=\tilde{\boldsymbol{U}}  \Sigma \tilde{\boldsymbol{V}}^{*}.
\end{equation}
The resolvent gains are contained within the diagonal matrix $\Sigma = diag[\sigma_{1},\sigma_{2},.....,\sigma_{n}]$. The forcing and response modes that maximize~(\ref{eq:resolvent_gain_def}) are recovered as $\boldsymbol{V}=\mathsfbi{W}^{-\frac{1}{2}}\tilde{\boldsymbol{V}}$ and $\boldsymbol{U}=\mathsfbi{W}^{-\frac{1}{2}}\tilde{\boldsymbol{U}}$, respectively \cite{towne_spectral_2018}.  We compute the resolvent modes by transforming the SVD into an equivalent eigenvalue problem, which is solved using an Arnoldi iteration in which The action of $\tilde{\mathsfbi{R}}$ is obtained by computing its LU decomposition \citep{jeun_input-output_2016,schmidt_spectral_2018}.

\begin{figure}[t]
  \begin{center}
      \begin{tikzpicture}[baseline]
        \tikzstyle{every node}=[font=\small]
        \tikzset{>=latex}
        \node[anchor=south west,inner sep=0] (image) at (0,0) {
          \includegraphics[scale=0.9,width=1\textwidth]{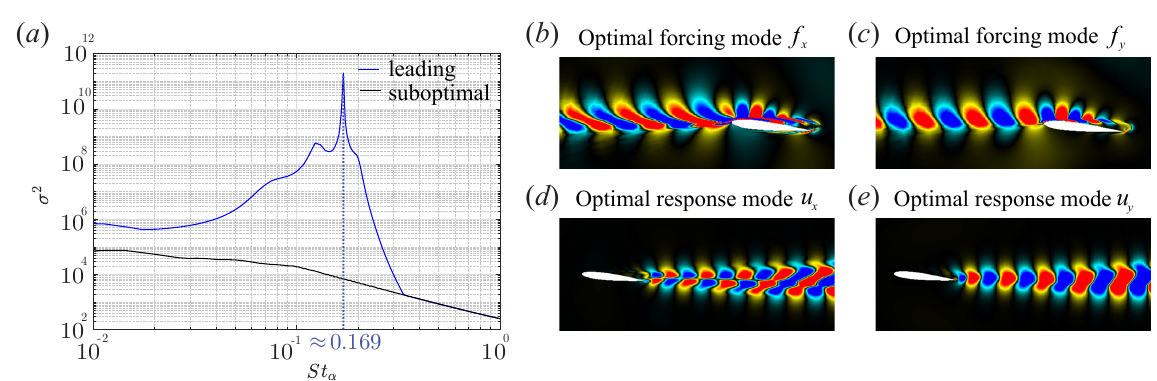}
        };
      \end{tikzpicture}
    \caption[]{\label{Fig_RSV_gains}Resolvent gains, optimal forcing and response modes: (a) leading and second optimal gains, (\textit{b}) optimal forcing mode of $u_{x}$, (\textit{c}) optimal forcing mode of $ u_{y}$, (\textit{d}) optimal response mode of $u_{x}$, and (\textit{e}) optimal response mode of $u_{y}$.}
  \end{center}
\end{figure}

In figure \ref{Fig_RSV_gains}, the peak of the leading resolvent gain is observed at the vortex shedding frequency $St_{\alpha} \approx 0.169$. The streamwise and transverse velocity components of the optimal forcing and response modes are shown in figure \ref{Fig_RSV_gains}(\textit{b})-(\text{e}). The optimal forcing mode is primarily located upstream and above the airfoil, while the optimal response mode is clearly observed downstream in the wake, as expected for a convective flow. The optimal response mode is similar to the dominant eigenmode.  

\section{Resolvent-based estimation and control framework}\label{sec_Methodology}

In this section, we provide a brief overview of the resolvent-based estimation and control framework developed by \cite{towne_resolvent-based_2020} and \cite{martini_resolvent-based_2020,martini_resolvent-based_2022}.

\subsection{System set-up}\label{subsec_Methodology}

As a generalization of~(\ref{eq3.2}), we consider the linear time-invariant system
\begin{subequations}\label{eq_LTIsystem}
\begin{alignat}{3}
\frac{d\boldsymbol{q}}{dt}(t)&=
\mathsfbi{A}\boldsymbol{q}(t)+\mathsfbi{{B}}_{f}\boldsymbol{f}(t)+\mathsfbi{{B}}_{a}\boldsymbol{a}(t),\\
\boldsymbol{y}(t)&=
\mathsfbi{{C}}_{y}\boldsymbol{q}(t)+\boldsymbol{n}(t),\\
\boldsymbol{z}(t)&=
\mathsfbi{{C}}_{z}\boldsymbol{q}(t).
\end{alignat}
\end{subequations}
The system matrix $\mathsfbi{A} \in \mathbb{C}^{n \times n} $ is the linearized compressible Navier-Stokes operator, $\boldsymbol{q} \in \mathbb{C}^{n}$ is the full state of flow, and $n$ is the total size of the state.  The forcing $\boldsymbol{f} \in \mathbb{C}^{n_{f}}$ represents the nonlinear terms from the Navier-Stokes equations and any exogenous forcing. The forcing matrix $\mathsfbi{{B}}_{f} \in \mathbb{C}^{n \times n_f}$ restricts the form of the forcing $\boldsymbol{f}$, e.g., localizes it to a certain spatial region.  The actuation signal $\boldsymbol{a} \in \mathbb{C}^{n_{a}}$, which will be the output of the controller, is mapped onto the system by the actuation matrix $\mathsfbi{{B}}_{a} \in \mathbb{C}^{n \times n_a}$, and $n_a$ is the total number of actuators.  The sensor measurement $\boldsymbol{y} \in \mathbb{C}^{n_y}$ is extracted from the state by the measurement matrix $\mathsfbi{{C}}_{y} \in \mathbb{C}^{n \times n_y}$, which defines the sensor locations and types, and $n_y$ is the total number of sensors.  The sensor measurements are corrupted by the sensor noise $\boldsymbol{n} \in \mathbb{C}^{n_{y}}$.  The target $\boldsymbol{z} \in \mathbb{C}^{n_z}$, i.e., the quantity that we wish to estimate and control, is extracted from the state by the target matrix $\boldsymbol{{C}}_{z} \in \mathbb{C}^{n \times n_z}$, and $n_z$ is the total number of targets.

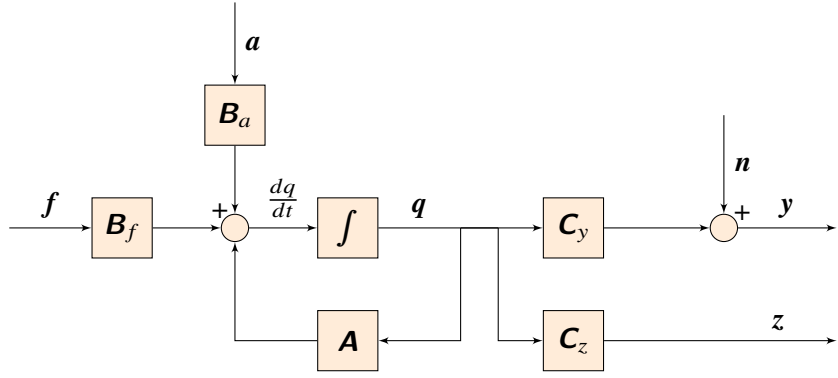
\begin{figure}[t]
\centering
 \tikzstyle{block} = [draw, rectangle, 
    minimum height=8mm, minimum width=8mm, fill=orange!15]
\tikzstyle{sum} = [draw, circle, node distance=1cm, fill=orange!15]
\tikzstyle{input} = [coordinate]
\tikzstyle{output} = [coordinate]
\tikzstyle{pinstyle} = [pin edge={to-,thin,black}]

\begin{tikzpicture}[auto, node distance=1.5cm,>=latex']
    \node [input, name=input] {};
    \node [block, right of=input] (matrix) {$\mathsfbi{{B}}_f$};
    \node [sum={+}{-}{+}{+}, right of=matrix,node distance=1.5cm] (sum) {};
    \node [block, right of=sum] (integral) {$\boldsymbol{\int}$};
    \node [output, right of=integral] (output_tmp) {};
    \node [output, right of=output_tmp,node distance=0.5cm] (output_tmp2) {};
    \node [output, right of=output_tmp2,node distance=0.5cm] (output_tmp3) {};
    \node [block, right of=output_tmp3,
            node distance=0.5cm] (Cy) {$\mathsfbi{{C}}_{y}$};
    \node [block, above of=sum,
            node distance=1.5cm] (Ba) {$\mathsfbi{{B}}_{a}$};
    \node [input, above of=Ba] (act_signal) {};
    \node [sum, right of=Cy, node distance=2cm] (sum2) {};
    \node [output, right of=sum2] (output) {a};
    \node [block, below of=integral] (system_mat) {$\mathsfbi{A}$};
    \node [block, below of=Cy] (target_mat) {$\mathsfbi{{C}}_z$};
    \node [input, above of=sum2] (noise) {};
    \node [output, right of=target_mat,node distance=3.5cm] (output2) {};
    \draw [draw,->] (input) -- node {$\boldsymbol{f}$} (matrix);
    \draw [->] (matrix) -- node[pos=0.99] {$+$} (sum);
    \draw [->] (sum) -- node {$\frac{dq}{dt}$} (integral);
    \draw [-] (integral) -- node[name=q] {$\boldsymbol{q}$} (output_tmp);
    \draw [-] (output_tmp) -- (output_tmp2);
    \draw [-] (output_tmp) -- (output_tmp3);
    \draw [->] (output_tmp3) -- (Cy);
    \draw [->] (Cy) -- node {} (sum2);
    \draw [->] (sum2) -- node [name=y] {$\boldsymbol{y}$}(output);
     \draw [->] (noise) -- node[pos=0.99] {$+$} node {$\boldsymbol{n}$} (sum2);
    \draw [->] (output_tmp) |- (system_mat);
    \draw [->] (output_tmp2) |- (target_mat);
    \draw [->] (system_mat) -|  (sum);
    \draw [->] (target_mat) -- node[near end] {$\boldsymbol{z}$} (output2);
    \draw [draw,->] (act_signal) -- node {$\boldsymbol{a}$} (Ba);
     \draw [->] (Ba) -- (sum);

\end{tikzpicture}
\caption[]{\label{figure_Plant}Block-diagram representation of the linear system.}
\end{figure}

Following \cite{martini_resolvent-based_2022}, we split the system (\ref{eq_LTIsystem}) into two parts: the so-called forcing system that is driven by the forcing $\boldsymbol{f}(t)$ and corrupted by the sensor noise $\boldsymbol{n}(t)$,
\begin{subequations}\label{eq_ForcingLTI}
\begin{alignat}{3}
\frac{d\boldsymbol{q}_{f}}{dt}(t)&=
\mathsfbi{A}\boldsymbol{q}_{f}(t)+\mathsfbi{B}_{f}\boldsymbol{f}(t),\\
\boldsymbol{y}_{f}(t)&=
\mathsfbi{C}_{y}\boldsymbol{q}_{f}(t)+\boldsymbol{n}(t),\\
\boldsymbol{z}_{f}(t)&=
\mathsfbi{C}_{z}\boldsymbol{q}_{f}(t),
\end{alignat}
\end{subequations}
and the actuation system that is driven only by the actuation signal $\boldsymbol{a}(t)$,
\begin{subequations}\label{eq_ACTLTI}
\begin{alignat}{3}
\frac{d\boldsymbol{q}_{a}}{dt}(t)&=\mathsfbi{A}\boldsymbol{q}_{a}(t)+\mathsfbi{B}_{a}\boldsymbol{a}(t),\\
\boldsymbol{y}_{a}(t)&=\mathsfbi{C}_{y}\boldsymbol{q}_{a}(t),\\
\boldsymbol{z}_{a}(t)&=\mathsfbi{C}_{z}\boldsymbol{q}_{a}(t).
\end{alignat}
\end{subequations}
The state, sensor measurements, and targets for the full system are recovered as
\begin{equation}\label{eq_split_flowstates}
\boldsymbol{q}=\boldsymbol{q}_{f}+\boldsymbol{q}_{a}, \qquad \boldsymbol{y}=\boldsymbol{y}_{f}+\boldsymbol{y}_{a}, \quad
\textrm{and} \quad \boldsymbol{z}=\boldsymbol{z}_{f}+\boldsymbol{z}_{a}.
\end{equation}

 By applying the Fourier transform to (\ref{eq_ForcingLTI}) and (\ref{eq_ACTLTI}) and using the resolvent operator defined in (\ref{eq_q_Rf}), we obtain the frequency-domain representations of sensor measurements and targets of the forcing and actuation systems,
\begin{subequations}\label{eq4.5}
\begin{alignat}{4}
\boldsymbol{\hat{y}}_{f}&= \mathsfbi{{R}}_{yf}\boldsymbol{\hat{f}}+\boldsymbol{\hat{n}},\\
\boldsymbol{\hat{z}}_{f}&= \mathsfbi{{R}}_{zf}\boldsymbol{\hat{f}},\\
\boldsymbol{\hat{y}}_{a}&= \mathsfbi{{R}}_{ya}\boldsymbol{\hat{a}},\\
\boldsymbol{\hat{z}}_{a}&= \mathsfbi{{R}}_{za}\boldsymbol{\hat{a}}.
\end{alignat}
\end{subequations}
Here, $\mathsfbi{R}_{yf}=\mathsfbi{C}_{y}\mathsfbi{R}\mathsfbi{{B}}_{f}$,
$\mathsfbi{R}_{zf}=\mathsfbi{C}_{z}\mathsfbi{R}\mathsfbi{{B}}_{f}$,
$\mathsfbi{R}_{ya}=\mathsfbi{C}_{y}\mathsfbi{R}\mathsfbi{{B}}_{a}$, and $\mathsfbi{R}_{za}=\mathsfbi{C}_{z}\mathsfbi{R}\mathsfbi{{B}}_{a}$ are modified resolvent operators (sometimes called input-output operators) that will appear in the resolvent-based estimation and control kernels.

\subsection{Resolvent-based estimation}\label{subsec_Resolvent-based estimation}

The resolvent-based estimates of the target $\tilde{\boldsymbol{z}}$ are obtained via a convolution between the sensor measurements and a kernel.  Here, we define three distinct resolvent-based estimation kernels: non-causal, truncated non-causal, and causal kernels. 

First, we define a non-causal estimator
\begin{equation}\label{eq_z_nc_est}
\Tilde{\boldsymbol{z}}_{nc}(t)=\int_{-\infty}^{\infty} \mathsfbi{T}_{nc}(t-\tau) \ \boldsymbol{y}(\tau)d\tau,
\end{equation}
where $\mathsfbi{T}_{nc} \in \mathbb{C}^{n_z \times n_y}$ is a non-causal estimation kernel. The optimal estimation kernel is obtained by minimizing a cost function defined as the time-integrated expected value of the estimation error,
\begin{equation}\label{eq_J_nc}
{J}_{nc}=\int_{-\infty}^{\infty} \mathbb{E} \{ \boldsymbol{e}(t)^{\dagger}  \boldsymbol{e}(t) \} \ dt,
\end{equation}
where the estimation error \( \boldsymbol{e}(t) = \tilde{\boldsymbol{z}}(t) - \boldsymbol{z}(t) \) is the difference between the estimated and true target, \((\cdot)^{\dagger}\) denotes the adjoint operator using a suitable inner product, and $\mathbb{E} \{ \cdot \}$ is the expectation operator. The cost function \( J_{nc} \) is minimized by setting its derivative with respect to \( \mathsfbi{T}_{nc} \) to zero, yielding the optimal non-causal estimation kernel \citep{martini_resolvent-based_2020}
\begin{equation}\label{eq_T_nc}
\mathsfbi{\hat{T}}_{nc}(\omega)= \mathsfbi{{R}}_{zf} \mathsfbi{\hat{F}} \mathsfbi{{R}}_{yf}^{\dagger} (\mathsfbi{{R}}_{yf} \mathsfbi{\hat{F}} \mathsfbi{{R}}_{yf}^{\dagger} + \mathsfbi{\hat{N}})^{-1},
\end{equation}
where \(\mathsfbi{\hat{F}} = \mathbb{E} \{ \hat{\boldsymbol{f}} \hat{\boldsymbol{f}}^{\dagger} \}\) and \(\mathsfbi{\hat{N}} = \mathbb{E} \{ \hat{\boldsymbol{n}}  \hat{\boldsymbol{n}}^{\dagger} \}\) are the CSDs of the forcing and sensor noise, respectively.  The time-domain estimation kernel $\mathsfbi{T}_{nc}$ is obtained by inverse Fourier transforming the frequency domain kernel $\mathsfbi{\hat{T}}_{nc}$.  In general, this kernel will be non-causal, i.e., $\mathsfbi{T}_{nc}(\tau)$ will not be strictly zero for $\tau<0$, therefore requiring future sensor data, $\boldsymbol{y}(\tau>0)$, which is unavailable for real-time applications, to evaluate~(\ref{eq_z_nc_est}).

Second, we define a truncated non-causal estimation kernel as a simple baseline approach to address the aforementioned lack of causality.  This is done by simply truncating the integral in (\ref{eq_z_nc_est}), 
\begin{equation}\label{eq_z_tnc_est}
\Tilde{\boldsymbol{z}}_{tnc}(t)=\int_{-\infty}^{0} \mathsfbi{T}_{tnc}(t-\tau) \ \boldsymbol{y}(\tau)d\tau,
\end{equation}
which is equivalent to truncating the non-causal part of the kernel,
\begin{eqnarray}\label{eq_T_tnc}
    \mathsfbi{T}_{tnc}(\tau)= 
\begin{cases}
    \mathsfbi{T}_{nc}(\tau),& \tau\geq 0,\\
    0,              & \tau<0.
\end{cases}
\end{eqnarray}
The downside of this approach is that the ex post facto truncation of the optimal non-causal kernel ruins its optimality.

Third, we define an optimal causal resolvent-based estimation kernel by enforcing causality using the Wiener-Hopf formalism \citep{martinelli_feedback_nodate,martini_resolvent-based_2022}. The causal estimator 
\begin{equation}\label{eq_z_c_est}
\Tilde{\boldsymbol{z}}_{c}(t)=\int_{-\infty}^{0} \mathsfbi{T}_{c}(t-\tau) \ \boldsymbol{y}(\tau)d\tau,
\end{equation}
is defined in terms of the causal estimation kernel $\mathsfbi{T}_{c} \in \mathbb{C}^{n_{z} \times n_{y}}$. To find the optimal kernel under the constraint of causality, we modify the cost function (\ref{eq_J_nc}) to read
\begin{equation}\label{eq_J_c}
{J}_{c}=\int_{-\infty}^{\infty} \mathbb{E}\{ e(t)^{\dagger}  e(t)\} + (\mathsfbi{\Lambda}_{-}(t) \mathsfbi{T}_{c} (t) + \mathsfbi{\Lambda}_{-}^{\dagger}(t) \mathsfbi{T}_{c}^{\dagger} (t)) \ dt,
\end{equation}
where $\mathsfbi{\Lambda}$ is a Lagrange multiplier that is used to force the causal kernel to be zero for the non-causal part ($\tau<0$). The ($+$) and ($-$) subscripts indicate that the non-causal ($\tau<0$) and causal ($\tau>0$) parts, respectively, of a matrix or function are set to zero using a Wiener-Hopf factorization.  An introduction to the Wiener-Hopf method is described in appendix $\ref{appB}$.  Similar to the derivation of (\ref{eq_T_nc}), we minimize the causal cost function (\ref{eq_J_c}) by setting its derivative with respect to \(\mathsfbi{T}_{c}\) to zero. In doing so, we encounter the Wiener-Hopf problem (\ref{eq_WHproblems1}) with \(\mathsfbi{\hat{G}}=\mathsfbi{R}_{zf}\mathsfbi{\hat{F}}\mathsfbi{R}_{yf}^{\dagger}\) and \(\mathsfbi{\hat{H}}=\mathsfbi{R}_{yf}\mathsfbi{\hat{F}}\mathsfbi{R}_{yf}^{\dagger} + \mathsfbi{\hat{N}}\).  Using the solution of this Wiener-Hopf problem given by (\ref{eq_causal1}), the causal estimation kernel \citep{martini_resolvent-based_2022} is 
\begin{eqnarray}\label{eq_T_c}
\mathsfbi{\hat{T}}_{c}(\omega) &=&  \left( \mathsfbi{R}_{zf} \mathsfbi{\hat{F}} \mathsfbi{R}_{yf}^{\dagger} (\mathsfbi{R}_{yf} \mathsfbi{\hat{F}} \mathsfbi{R}_{yf}^{\dagger} + \mathsfbi{\hat{N}})^{-1}_{-} \right)_{+} \left(\mathsfbi{R}_{yf} \mathsfbi{\hat{F}} \mathsfbi{R}_{yf}^{\dagger} + \mathsfbi{\hat{N}}\right)^{-1}_{+}.
\end{eqnarray}

\subsection{Resolvent-based control}\label{subsec_Resolvent-based control}

Analogous to the estimation problem, the actuation signal $\boldsymbol{a}(t)$ that will be used to control the flow is obtained via a convolution between the sensor measurements and a control kernel (also referred to as control law in some studies).  Again, we consider non-causal, truncated non-causal, and causal variations of the resolvent-based control kernels \citep{martini_resolvent-based_2022}. 

The non-causal controller takes the form
\begin{equation}\label{eq_a_nc}
\boldsymbol{a}_{nc}(t)=\int_{-\infty}^{\infty} \mathsfbi{\Gamma}_{nc}(t-\tau) \ \boldsymbol{y}_{f}(\tau)d\tau,
\end{equation}
where $\mathsfbi{\Gamma}_{nc} \in \mathbb{C}^{n_a \times n_y}$ is the noncausal control kernel. The optimal kernel is obtained by minimizing the cost function 
\begin{equation}\label{eq_J_nc_con}
{J}_{nc}=\int_{-\infty}^{\infty} \mathbb{E}\{\boldsymbol{z}(t)^{\dagger} \boldsymbol{z}(t) + \boldsymbol{a}(t)^{\dagger} \mathsfbi{P} \boldsymbol{a}(t)\}\ dt,
\end{equation}
where $\mathsfbi{P}$ is a weight matrix that penalizes the actuation effort. Minimizing this cost function yields the optimal non-causal control kernel
\begin{eqnarray}\label{eq_Gamma_nc}
\mathsfbi{\hat{\Gamma}}_{nc}(\omega) &=& (\mathsfbi{{R}}_{za}^{\dagger}\mathsfbi{{R}}_{za} + \mathsfbi{\hat{P}})^{-1} (-\mathsfbi{{R}}_{za}^{\dagger})  \mathsfbi{{R}}_{zf} \mathsfbi{\hat{F}} \mathsfbi{{R}}_{yf}^{\dagger} (\mathsfbi{{R}}_{yf} \mathsfbi{\hat{F}} \mathsfbi{{R}}_{yf}^{\dagger} + \mathsfbi{{\hat{N}}})^{-1}.
\end{eqnarray}

The truncated non-causal control kernel is defined as
\begin{eqnarray}\label{eq_Gamma_tnc}
    \mathsfbi{\Gamma}_{tnc}(\tau)= 
\begin{cases}
    \mathsfbi{\Gamma}_{nc}(\tau),& \tau\geq 0,\\
    0,              & \tau<0,
\end{cases}
\end{eqnarray}
from which the actuation signal is computed as
\begin{equation}\label{eq_a_tnc}
\boldsymbol{a}_{tnc}(t)=\int_{-\infty}^{0} \mathsfbi{\Gamma}_{tnc}(t-\tau) \ \boldsymbol{y}_{f}(\tau)d\tau.
\end{equation}
As before, truncating the optimal non-causal kernel yields a non-optimal causal kernel.

Finally, the optimal causal controller is
\begin{equation}\label{eq_a_c}
\boldsymbol{a}_{c}(t)=\int_{-\infty}^{0} \mathsfbi{\Gamma}_{c}(t-\tau) \ \boldsymbol{y}_{f}(\tau)d\tau,
\end{equation}
and the optimal causal control kernel is obtained by using Lagrange multipliers to enforce causality, expressed by the cost function
\begin{equation}\label{eq_J_c_con}
{J}_{c}=\int_{-\infty}^{\infty} \mathbb{E}\{\boldsymbol{z}(t)^{\dagger} \boldsymbol{z}(t) + \boldsymbol{a}(t)^{\dagger} \mathsfbi{P} \boldsymbol{a}(t) + (\mathsfbi{\Lambda}_{-}(t) \mathsfbi{\Gamma}_{c} (t) + \mathsfbi{\Lambda}_{-}^{\dagger}(t) \mathsfbi{\Gamma}_{c}^{\dagger} (t)) \} dt.
\end{equation}
Minimizing (\ref{eq_J_c_con}) leads to the Wiener-Hopf problem (\ref{eq_WHproblems2}) with \(\mathsfbi{\hat{K}}=\mathsfbi{R}_{za}^{\dagger}\mathsfbi{R}_{za} + \mathsfbi{\hat{P}}\) and $\mathsfbi{\hat{L}}=-\mathsfbi{R}_{za}^{\dagger}$.  Using the solution of this Wiener-Hopf problem given in (\ref{eq_causal2}), the optimal causal resolvent-based control kernel is 
\begin{eqnarray}\label{eq_Gamma_c}
\mathsfbi{\hat{\Gamma}}_{c}(\omega)&=&
\big(\mathsfbi{R}_{za}^{\dagger}\mathsfbi{R}_{za}+\mathsfbi{\hat{P}})_{+}^{-1}\Big((\mathsfbi{R}_{za}^{\dagger}\mathsfbi{R}_{za}+\mathsfbi{\hat{P}})_{-}^{-1}\nonumber (-\mathsfbi{R}_{za}^{\dagger})\mathsfbi{R}_{zf}\mathsfbi{\hat{F}}\mathsfbi{R}_{yf}^{\dagger}\\
&&(\mathsfbi{R}_{yf}\mathsfbi{\hat{F}}\mathsfbi{R}_{yf}^{\dagger}+ \mathsfbi{\hat{N}})_{-}^{-1}\Big)_{+}(\mathsfbi{R}_{yf}\mathsfbi{\hat{F}}\mathsfbi{R}_{yf}^{\dagger}+ \mathsfbi{\hat{N}})_{+}^{-1}.
\end{eqnarray}

In (\ref{eq_a_nc}), (\ref{eq_a_tnc}), and (\ref{eq_a_c}), the actuation signal is obtained as a convolution between the control kernel and the sensor measurement $\boldsymbol{y}_{f}$ for the forcing system~(\ref{eq_ForcingLTI}), i.e., the control kernels were derived using a measurement excluding the response of the system (\ref{eq_ACTLTI}) to actuation.  In practice, only the complete measurement $\boldsymbol{y}=\boldsymbol{y}_f + \boldsymbol{y}_a$ is available.  Considering the full (combined) linear system (\ref{eq_LTIsystem}), the final closed-loop controller takes the form
\begin{equation}\label{eq_act_signal_fullSys}
\boldsymbol{a}(t)=\int_{-\infty}^{0} \mathsfbi{\Gamma}_{cl}(t-\tau) \ \boldsymbol{y}(\tau)d\tau,
\end{equation}
where the final closed-loop kernel is
\begin{eqnarray}\label{eq_Gamma_cl}
\mathsfbi{\hat{\Gamma}}_{cl} &=&
(\mathsfbi{I} + \mathsfbi{\hat{\Gamma}} \mathsfbi{R}_{ya})^{-1} \mathsfbi{\hat{\Gamma}},
\end{eqnarray}
with \(\mathsfbi{\hat{\Gamma}}\) replaced by \(\mathsfbi{\hat{\Gamma}}_{tnc}\) or \(\mathsfbi{\hat{\Gamma}}_{c}\) for truncated non-causal and optimal causal resolvent-based control, respectively. 

\section{Computing estimation and control kernels}\label{sec_Computing_est_con_kernels}

In this section, we present two approaches to compute resolvent-based estimation and control kernels \citep{martini_resolvent-based_2022}. First, we describe an operator-based approach that allows for efficient implementation of linear simulations without the need for inverting the operator or performing prior model reduction, making it particularly efficient for large-scale problems. Second, we explain a data-driven approach that does not require the construction of the linearized Navier-Stokes operator. Instead, this method utilizes training data from simulations or experiments to build cross-spectral densities, which are then used to compute the components of the estimation and control kernels. 

\subsection{Operator-based approach}\label{subsec_Operator-based approach}

The resolvent operator \(\mathsfbi{R}\) is defined in terms of an inverse, the cost of which scales poorly with the problem dimension $n$ and becomes computationally expensive for large systems. To circumvent this, we employ a time-stepping approach that avoids the inverse operation and instead constructs the necessary modified resolvent operators that appear in the estimation and control kernels by solving linear equations in the time domain \citep{martini_resolvent-based_2020, martini_resolvent-based_2022, farghadan_scalable_2023}. For both cases, the cost of this approach scales linearly with the problem dimension, avoiding the need to reduce the system via a priori model reduction and the associated loss of accuracy.  Individual modified resolvent operators such as \(\mathsfbi{R}_{za}\) and \(\mathsfbi{R}_{ya}\) can be obtained using a single-stage run of the direct linear equations.  Products of modified resolvent operators (and the forcing CSD) such as \(\mathsfbi{R}_{zf}\mathsfbi{\hat{F}}\mathsfbi{R}_{yf}^{\dagger}\) and \(\mathsfbi{R}_{yf} \mathsfbi{\hat{F}}\mathsfbi{R}_{yf}^{\dagger}\) can be obtained with greater efficiently using two-stage runs of both the adjoint and direct linear equations.

\subsubsection{Single-stage run}\label{subsubsec_Single-stage run}

The operators $\mathsfbi{R}_{za}$ and $\mathsfbi{R}_{ya}$ appearing in~(\ref{eq_Gamma_c}) and~(\ref{eq_Gamma_cl}), respectively, can be constructed by computing a series of impulse responses of the actuation system (\ref{eq_ACTLTI}),
\begin{subequations}\label{eq_singlestagerun}
\begin{alignat}{3}\label{eq_singlestagerun_a}
\frac{d\boldsymbol{q}_{a,k}}{dt}(t) &= \mathsfbi{A}\boldsymbol{q}_{a,k}(t)+\mathsfbi{B}_{a,k}\delta(t),\\\label{eq_singlestagerun_b}
\boldsymbol{y}_{a,k}(t)&= \mathsfbi{C}_{y}\boldsymbol{q}_{a,k}(t),\\\label{eq_singlestagerun_c}
\boldsymbol{z}_{a,k}(t)&= \mathsfbi{C}_{z}\boldsymbol{q}_{a,k}(t).
\end{alignat}
\end{subequations}
Here, $\boldsymbol{y}_{a,k} \in\mathbb{C}^{n_{y}}$ and $\boldsymbol{z}_{a,k} \in \mathbb{C}^{n_{z}}$  are the sensor and target measurement of the direct system forced by an impulse $\delta(t)$ located at $k$-th actuator, as encoded by the $k$-th column of the actuation matrix, $\mathsfbi{B}_{a,k}$. By collecting these data for each actuator $k=1,\dots,n_a$ and taking a Fourier transform, we obtain
\begin{subequations}\label{eq_Ya_Za}
\begin{alignat}{2}\label{eq_Ya_Za_a}
\mathsfbi{\hat{Y}}_{a} &=& 
\begin{bmatrix}
\hat{\boldsymbol{y}}_{a,1}&\hat{\boldsymbol{y}}_{a,2} & \dots & \hat{\boldsymbol{y}}_{a,n_{a}}
\end{bmatrix}
= 
\mathsfbi{R}_{ya},\\\label{eq_Ya_Za_b}
\mathsfbi{\hat{Z}}_{a} &=& 
\begin{bmatrix}
\hat{\boldsymbol{z}}_{a,1}&\hat{\boldsymbol{z}}_{a,2}& \dots & \hat{\boldsymbol{z}}_{a,n_{a}}
\end{bmatrix}
= 
\mathsfbi{R}_{za},
\end{alignat}
\end{subequations}
with $\mathsfbi{\hat{Y}}_{a} \in \mathbb{C}^{n_{y} \times n_{a}}$ and $\mathsfbi{\hat{Z}}_{a} \in  \mathbb{C}^{n_{z} \times n_{a}}$. That is, the Fourier  transform of each measurement $\boldsymbol{y}_{a,k}$ and $\boldsymbol{z}_{a,k}$ yields a column of the modified resolvent operators $\mathsfbi{R}_{ya}$ and $\mathsfbi{R}_{za}$, respectively.

\subsubsection{Two-stage run}\label{subsubsec_Two-stage run}

While single-stage runs could be used to construct all of the modified resolvent operators in the estimation and control kernels, certain products thereof can be constructed more efficiently using pairs of adjoint and direct runs.  The procedure begins with solving the adjoint system
\begin{subequations}\label{eq_Two_stageruns}
\begin{alignat}{2}\label{eq_Two_stageruns_a}
-\frac{d\boldsymbol{q}_{f,i}}{dt}(t) &= \mathsfbi{A}^{\dagger}\boldsymbol{q}_{f,i}(t)+\mathsfbi{C}_{y,i}^{\dagger}\delta(t),\\\label{eq_Two_stageruns_b}
\boldsymbol{s}_{i}(t)&= \mathsfbi{B}_{f}^{\dagger}\boldsymbol{q}_{f,i}(t),
\end{alignat}
\end{subequations}
where \(\mathsfbi{A}^{\dagger}\) is the adjoint linearized Navier-Stokes operator and the subscript \(i\) indicates the sensor defined by the \(i\)-th row of the sensor measurement matrix \(\mathsfbi{C}_y\). The output $\boldsymbol{s}_{i}$ of the adjoint run is used as a forcing in a corresponding direct run of the forcing system (\ref{eq_ForcingLTI}),  
\begin{subequations}\label{eq_dirrun_inTwostageRun}
\begin{alignat}{3}\label{eq_dirrun_inTwostageRun_a}
\frac{d\boldsymbol{q}_{f,i}}{dt}(t) &= \mathsfbi{A}\boldsymbol{q}_{f,i}(t)+\mathsfbi{B}_{f}\boldsymbol{s}_{i}(t),\\\label{eq_dirrun_inTwostageRun_b}
\boldsymbol{y}_{f,i}(t) &= \mathsfbi{C}_{y}\boldsymbol{q}_{f,i}(t) +\boldsymbol{n}_{i}(t),\\\label{eq_dirrun_inTwostageRun_c}
\boldsymbol{z}_{f,i}(t) &= \mathsfbi{C}_{z}\boldsymbol{q}_{f,i}(t),
\end{alignat}
\end{subequations}
As in the single-stage run, by collecting each of the final sensor and target measurements and taking a Fourier-transforming, we obtain 
\begin{subequations}\label{eq_Yf_Zf}
\begin{alignat}{2}
\mathsfbi{\hat{Y}}_{f} &=& 
\begin{bmatrix}
\hat{\boldsymbol{y}}_{1}&\hat{\boldsymbol{y}}_{2} & \dots & \hat{\boldsymbol{y}}_{n_{y}}\label{eq_Yf_Zfa}
\end{bmatrix}
= 
\mathsfbi{R}_{yf}\mathsfbi{R}_{yf}^{\dagger},\\
\mathsfbi{\hat{Z}}_{f} &=& 
\begin{bmatrix}
\hat{\boldsymbol{z}}_{1}&\hat{\boldsymbol{z}}_{2}& \dots & \hat{\boldsymbol{z}}_{n_{y}}\label{eq_Yf_Zfb}
\end{bmatrix}
= 
\mathsfbi{R}_{zf} \mathsfbi{R}_{yf}^{\dagger},
\end{alignat}
\end{subequations}
with $\mathsfbi{\hat{Y}}_{f} \in \mathbb{C}^{n_{y} \times n_{y}}$ and $\mathsfbi{\hat{Z}}_{f} \in  \mathbb{C}^{n_{z} \times n_{y}}$. To account for the nonlinearity of the flow, the colored forcing statistics, $\mathsfbi{\hat{F}}$, can be incorporated into (\ref{eq_Yf_Zf}) during adjoint and direct simulations, resulting in $\mathsfbi{R}_{yf} \mathsfbi{\hat{F}} \mathsfbi{R}_{yf}^{\dagger}$ and $\mathsfbi{R}_{zf} \mathsfbi{\hat{F}} \mathsfbi{R}_{yf}^{\dagger}$   \citep{Jung_RSV}.

Finally, (\ref{eq_Ya_Za}) and (\ref{eq_Yf_Zf}) are used to write the estimation kernels in (\ref{eq_T_nc}) and (\ref{eq_T_c}) and control kernels in (\ref{eq_Gamma_nc}) and (\ref{eq_Gamma_c}) in terms of $\mathsfbi{\hat{Y}}_{a}$, $\mathsfbi{\hat{Z}}_{a}$, $\mathsfbi{\hat{Y}}_{f}$, and $\mathsfbi{\hat{Z}}_{f}$.  The final operator-based estimation and control kernels are 
\begin{subequations}\label{eq_Tnc_Tc_O}
\begin{alignat}{2}
\mathsfbi{\hat{T}}_{nc,O} &= \mathsfbi{\hat{Z}}_{f} ( \mathsfbi{\hat{Y}}_{f} + \mathsfbi{\hat{N}})^{-1}, \\
\mathsfbi{\hat{T}}_{c,O} &= (\mathsfbi{\hat{Z}}_{f} ( \mathsfbi{\hat{Y}}_{f} + \mathsfbi{\hat{N}} )^{-1}_{-})_{+} (\mathsfbi{\hat{Y}}_{f}+ \mathsfbi{\hat{N}})_{+}^{-1},
\end{alignat}
\end{subequations}
and
\begin{subequations}\label{eq5.7}
\begin{alignat}{2}
\mathsfbi{\hat{\Gamma}}_{nc,O} &= (\hat{\boldsymbol{Z}}_{a}^{\dagger} \hat{\boldsymbol{Z}}_{a} +\hat{\boldsymbol{P}})^{-1} (-\hat{\boldsymbol{Z}}_{a}^{\dagger}) \hat{\boldsymbol{Z}}_{f} (\hat{\boldsymbol{Y}}_{f}+ \mathsfbi{\hat{N}})^{-1}, \\\label{eq5.7a}
\mathsfbi{\hat{\Gamma}}_{c,O} &= \big(\mathsfbi{\hat{Z}}_{a}^{\dagger} \mathsfbi{\hat{Z}}_{a}+\mathsfbi{\hat{P}})_{+}^{-1}\Big((\mathsfbi{\hat{Z}}_{a}^{\dagger} \mathsfbi{\hat{Z}}_{a}+\mathsfbi{\hat{P}})_{-}^{-1} (-\mathsfbi{\hat{Z}}_{a}^{\dagger})\mathsfbi{\hat{Z}}_{f}(\mathsfbi{\hat{Y}}_{f}+ \mathsfbi{\hat{N}})_{-}^{-1}\Big)_{+}(\mathsfbi{\hat{Y}}_{f}+ \mathsfbi{\hat{N}})_{+}^{-1}.
\end{alignat}
\end{subequations}

\subsection{Data-driven approach}\label{subsec_Data-driven approach}

When the linearized Navier-Stokes operator is unavailable, a data-driven approach \citep{martini_resolvent-based_2022} can be employed to build the required modified resolvent operators using CSDs computed from data \citep{towne_spectral_2018, towne_resolvent-based_2020}. This approach extends the applicability of resolvent-based estimation and control to experimental settings and circumvents the need for adjoint solvers. Additionally, when the CSD is computed from the dataset of a nonlinear system, the resulting kernels automatically include the forcing CSD $\mathsfbi{\hat{F}}$, which statistically accounts for the nonlinearity of the flow, thereby improving the estimation and control performance for the nonlinear system.

The nonlinear terms in the Navier-Stokes equations act as a forcing of the resolvent operator \citep{mckeon_critical-layer_2010}, and their influence is crucial in complex dynamic systems \citep{amaral_resolvent-based_2021}. To explicitly address the nonlinear terms, we split the forcing vector $\boldsymbol{f}$ into two components: the external forcing $\boldsymbol{f}_{ext}$ and the nonlinear terms $\boldsymbol{f}_{nl}$. This distinction helps us better understand their impact when building the CSDs from the data. The overall forcing term in~(\ref{eq_LTIsystem}) can be split as
\begin{equation}\label{eq_Bf_Bext_f_ext_f_nl}
\mathsfbi{B}_{f}\boldsymbol{f} \\
=
\begin{bmatrix}
\mathsfbi{B}_{ext} & \mathsfbi{B}_{nl} 
\end{bmatrix}
\begin{bmatrix}
\boldsymbol{f}_{ext} \\
\boldsymbol{f}_{nl} 
\end{bmatrix}
\end{equation}
where $\mathsfbi{B}_{ext} \in \mathbb{C}^{n \times n_{ext}}$ and $\mathsfbi{B}_{nl} \in \mathbb{C}^{n \times n}$. Typically, the region subject to external forcing is smaller than the overall domain, such that $n_{ext}<n$. In a linear system, the nonlinear terms $\boldsymbol{f}_{nl}$ are not included, allowing us to analyze the linear dynamics of the flow and to build linear estimators and controllers. However, our ultimate goal is to manipulate the unsteady fluctuations inherent in the actual flow, which necessitates considering nonlinearity. For the nonlinear system, we collect data from the systems without and with external forcing to better capture the behavior of the systems influenced by external forcing. The forcing system (\ref{eq_ForcingLTI}) without and with the external forcing can be expressed as
\begin{equation}\label{eq_forcing_wo_EF}
\frac{d\boldsymbol{q}}{dt}(t) = \mathsfbi{A}\boldsymbol{q}(t)+\mathsfbi{B}_{nl}\boldsymbol{f}_{nl}(t),
\end{equation}
\begin{equation}\label{eq_forcing_w_EF}
\frac{d\boldsymbol{q}_{e}}{dt}(t) = \mathsfbi{A}\boldsymbol{q}_{e}(t)+\mathsfbi{B}_{ext}\boldsymbol{f}_{ext}+\mathsfbi{B}_{nl}\boldsymbol{f}_{nl,e}(t).
\end{equation}
The subscript $e$ indicates the flow quantity that contains the development of nonlinearity, which was impacted by the external forcing. The $\boldsymbol{f}_{nl,e}$ term is evolved by the external forcing in time and space, so in the nonlinear system, the nonlinear effect can not be neglected. Equation (\ref{eq_forcing_wo_EF}) is the DNS or LES system without any source term. We assume external and nonlinear forcings are uncorrelated. Then we obtain

\begin{equation}\label{eq_y_fnl_zfnl}
\begin{bmatrix}
\hat{\boldsymbol{y}}_{f,nl}\\
\hat{\boldsymbol{z}}_{f,nl}
\end{bmatrix}
=
\begin{bmatrix}
\mathsfbi{R}_{yf,nl}  \\
\mathsfbi{R}_{zf,nl} 
\end{bmatrix}
\hat{\boldsymbol{f}}_{nl} 
+
\begin{bmatrix}
\hat{\boldsymbol{n}} \\
 0
\end{bmatrix},
\end{equation}
\begin{equation}\label{eq_y_f_ext_nl_z_f_ext_nl}
\begin{bmatrix}
\hat{\boldsymbol{y}}_{f,ext,nl}\\
\hat{\boldsymbol{z}}_{f,ext,nl}
\end{bmatrix}
=
\begin{bmatrix}
\mathsfbi{R}_{yf,ext} & \mathsfbi{R}_{yf,nl}  \\
\mathsfbi{R}_{zf,ext} & \mathsfbi{R}_{zf,nl} 
\end{bmatrix}
\begin{bmatrix}
\hat{\boldsymbol{f}}_{ext} \\
\hat{\boldsymbol{f}}_{nl,e} 
\end{bmatrix}
+
\begin{bmatrix}
\hat{\boldsymbol{n}} \\
 0
\end{bmatrix}
.
\end{equation}

Computing the cross-spectral density of $\left[ \hat{\boldsymbol{y}} \quad \hat{\boldsymbol{z}} \right]^{T}$ from (\ref{eq_y_fnl_zfnl}) and (\ref{eq_y_f_ext_nl_z_f_ext_nl}) gives
\begin{equation}\label{eq_Syy1_Szy1}
\begin{bmatrix}
\mathsfbi{S}_{yy} \\
\mathsfbi{S}_{zy} 
\end{bmatrix}
\triangleq
\begin{bmatrix}
\mathsfbi{S}_{yy,f,nl}  \\
\mathsfbi{S}_{zy,f,nl} 
\end{bmatrix}
=
\begin{bmatrix}
 \mathsfbi{R}_{yf,nl} \mathsfbi{\hat{F}}_{nl} \mathsfbi{R}_{yf,nl}^{\dagger} + \mathsfbi{\hat{N}} \\
\mathsfbi{R}_{zf,nl} \mathsfbi{\hat{F}}_{nl} \mathsfbi{R}_{yf,nl}^{\dagger}  
\end{bmatrix},
\end{equation}
\begin{equation}\label{eq_Syy2_Szy2}
\begin{bmatrix}
\mathsfbi{S}_{yy,e} \\
\mathsfbi{S}_{zy,e} 
\end{bmatrix}
\triangleq
\begin{bmatrix}
\mathsfbi{S}_{yy,f,ext,nl} \\
\mathsfbi{S}_{zy,f,ext,nl} 
\end{bmatrix}
=
\begin{bmatrix}
\mathsfbi{R}_{yf,nl} \mathsfbi{\hat{F}}_{nl,r} \mathsfbi{R}_{yf,nl}^{\dagger} + \mathsfbi{R}_{yf,ext} \mathsfbi{\hat{F}}_{ext}\mathsfbi{R}_{yf,ext}^{\dagger} + \mathsfbi{\hat{N}} \\
\mathsfbi{R}_{zf,nl} \mathsfbi{\hat{F}}_{nl,r} \mathsfbi{R}_{yf,nl}^{\dagger} + \mathsfbi{R}_{zf,ext} \mathsfbi{\hat{F}}_{ext}\mathsfbi{R}_{yf,ext}^{\dagger}
\end{bmatrix}.
\end{equation}
with $\mathsfbi{S}_{yy}=\mathbb{E} \{ \hat{\boldsymbol{y}} \hat{\boldsymbol{y}}^{\dagger} \}$ and $\mathsfbi{S}_{zy}=\mathbb{E} \{ \hat{\boldsymbol{z}} \hat{\boldsymbol{y}}^{\dagger} \}$. Since the right-hand sides of (\ref{eq_Syy1_Szy1}) and (\ref{eq_Syy2_Szy2}) contain the terms needed to build the estimation and control kernels (\ref{eq_T_nc}), (\ref{eq_T_c}), (\ref{eq_Gamma_nc}), and (\ref{eq_Gamma_c}), this shows that the CSDs on the left-hand side can be used in their place. Note that the CSDs inherently contain statistical information about the nonlinearity of the flow within the forcing CSD matrix.

The data-driven non-causal and causal estimation kernels in (\ref{eq_T_nc}) and (\ref{eq_T_c}) are computed using the CSDs from (\ref{eq_Syy1_Szy1}), yielding
\begin{subequations}\label{eq_T_nc_c_D}
\begin{alignat}{2}\label{eq_T_nc_D}
\mathsfbi{\hat{T}}_{nc,D} &= \mathsfbi{S}_{zy} ( \mathsfbi{S}_{yy} + \mathsfbi{\hat{N}})^{-1},\\\label{eq_c_D}
\mathsfbi{\hat{T}}_{c,D} &= (\mathsfbi{S}_{zy} ( \mathsfbi{S}_{yy} + \mathsfbi{\hat{N}})^{-1}_{-})_{+} (\mathsfbi{S}_{yy} + \mathsfbi{\hat{N}})_{+}^{-1}.
\end{alignat}
\end{subequations}

The control kernels require $\mathsfbi{R}_{ya}$ and $\mathsfbi{R}_{za}$, which do not include the forcing CSD matrix and can be obtained by imposing an impulse forcing at the actuator location in the nonlinear system,
\begin{eqnarray}\label{eq5.24}
\frac{d\boldsymbol{q}_{a,k}}{dt}(t) &=& \mathsfbi{A}\boldsymbol{q}_{a,k}(t)+\mathsfbi{B}_{a,k}\delta (t)+\mathsfbi{B}_{nl}\boldsymbol{f}_{nl,k}(t).
\end{eqnarray}
The Fourier-transformed sensor and target reading from the nonlinear system forced by impulses at the actuator locations are subtracted by the same quantities, $\hat{\boldsymbol{y}}_{f,nl}$ and $\hat{\boldsymbol{z}}_{f,nl}$, from the nonlinear system without the actuators. Then we obtain
\begin{equation}\label{eq5.16}
\begin{bmatrix}
\mathsfbi{\hat{Y}}_{s}\\
\mathsfbi{\hat{Z}}_{s}
\end{bmatrix}
=
\begin{bmatrix}
\hat{\boldsymbol{y}}_{a}-\hat{\boldsymbol{y}}_{f,nl}\\
\hat{\boldsymbol{z}}_{a}-\hat{\boldsymbol{z}}_{f,nl}
\end{bmatrix}
\approx
\begin{bmatrix}
\mathsfbi{R}_{ya}   \\
\mathsfbi{R}_{za} 
\end{bmatrix}
,
\end{equation} 
where the subscript $s$ denotes the Fourier-transformed readings computed through subtraction using the data-driven approach.
The resulting CSDs are
\begin{equation}\label{eq_S_yy_zz_3}
\begin{bmatrix}
\mathsfbi{S}_{yy,s} \\
\mathsfbi{S}_{zz,s} 
\end{bmatrix}
=
\begin{bmatrix}
 \mathsfbi{R}_{ya} \mathsfbi{R}_{ya}^{\dagger} \\
\mathsfbi{R}_{za} \mathsfbi{R}_{za}^{\dagger}  
\end{bmatrix}
\approx
\begin{bmatrix}
 \mathsfbi{Y}_{a} \mathsfbi{Y}_{a}^{\dagger}  \\
\mathsfbi{Z}_{a} \mathsfbi{Z}_{a}^{\dagger}  
\end{bmatrix}.
\end{equation}

Using this result, we can finally obtain the data-driven control kernels,
\begin{subequations}\label{eq_Gamma_nc_c_D}
\begin{alignat}{2}\label{eq_Gamma_nc_D}
\mathsfbi{\hat{\Gamma}}_{nc,D} &= (\mathsfbi{S}_{zz,s}^{\dagger} +\mathsfbi{\hat{P}})^{-1}  (-\mathsfbi{\hat{Z}}_{s}^{\dagger})  \mathsfbi{S}_{zy}(\mathsfbi{S}_{yy}+\mathsfbi{\hat{N}})^{-1}, 
\\\label{eq_Gamma_c_D}
\mathsfbi{\hat{\Gamma}}_{c,D}&=
\big(\mathsfbi{S}_{zz,s}^{\dagger} +\mathsfbi{\hat{P}})_{+}^{-1}\Big((\mathsfbi{S}_{zz,s}^{\dagger} +\mathsfbi{\hat{P}})_{-}^{-1} (-\mathsfbi{\hat{Z}}_{s}^{\dagger})\mathsfbi{S}_{zy}(\mathsfbi{S}_{yy}+\mathsfbi{\hat{N}})_{-}^{-1}\Big)_{+}(\mathsfbi{S}_{yy}+\mathsfbi{\hat{N}})_{+}^{-1}.
\end{alignat}
\end{subequations}
To apply these kernels to the system that is excited by the external forcing, $\mathsfbi{S}_{yy}$ and $\mathsfbi{S}_{zy}$ can be replaced with $\mathsfbi{S}_{yy,r}$ and $\mathsfbi{S}_{zy,r}$.

\section{Implementation of a resolvent-based estimation and control tool}\label{sec_Implmentation_Resolvent-based control tools}

\begin{figure}
  \begin{center}
      \begin{tikzpicture}[baseline]
        \tikzstyle{every node}=[font=\small]
        \tikzset{>=latex}
        \node[anchor=south west,inner sep=0] (image) at (0,0) {
          \includegraphics[scale=1,width=1\textwidth]{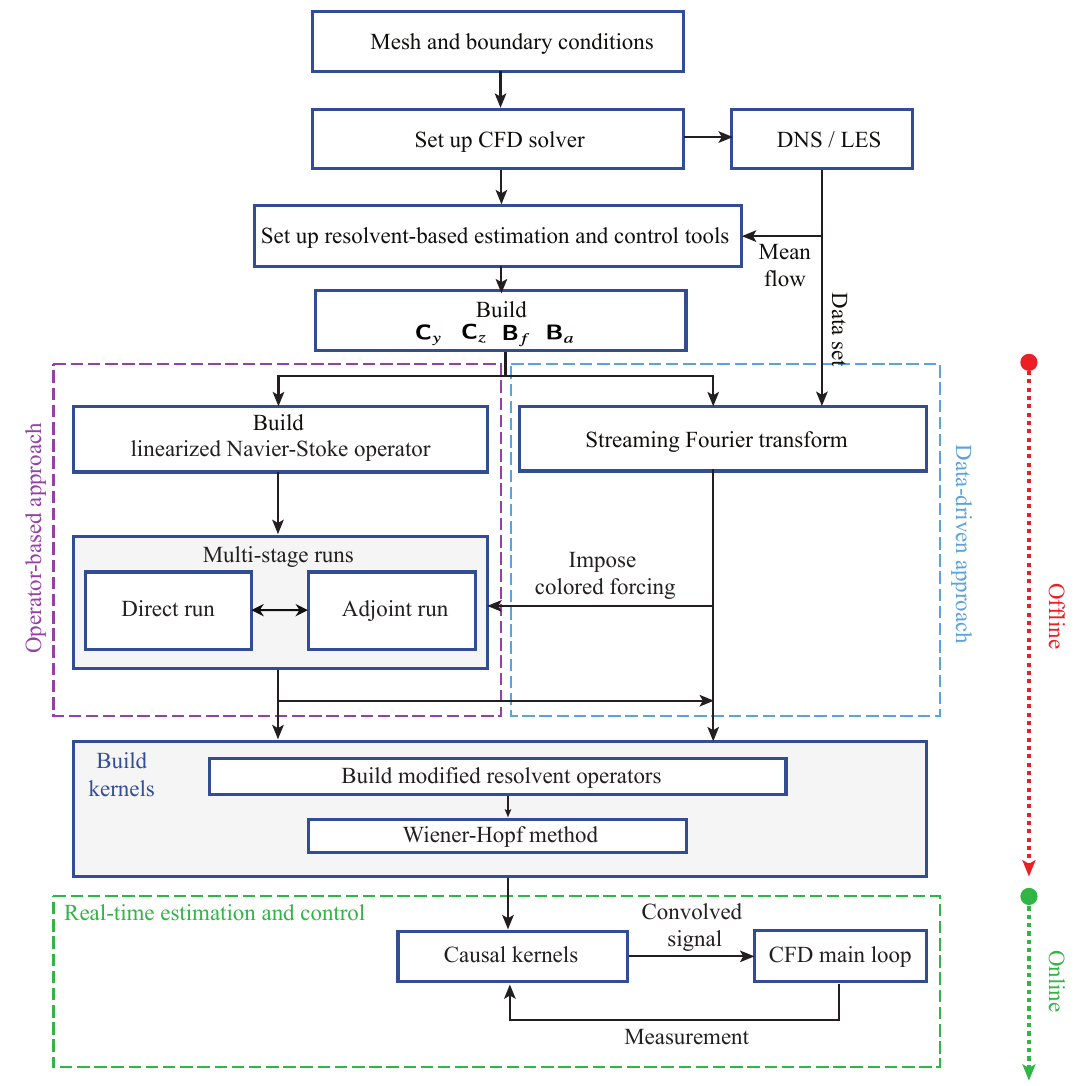}
        };
      \end{tikzpicture}
    \caption[]{\label{Fig_FlowChart_RSVTOOLS} Flow chart for the new implementation of resolvent-based estimation and control tools within the compressible flow solver CharLES.}
  \end{center}
\end{figure}

\begin{figure}[t]
  \begin{center}
      \begin{tikzpicture}[baseline]
        \tikzstyle{every node}=[font=\small]
        \tikzset{>=latex}
        \node[anchor=south west,inner sep=0] (image) at (0,0) {
          \includegraphics[scale=0.9,width=1\textwidth]{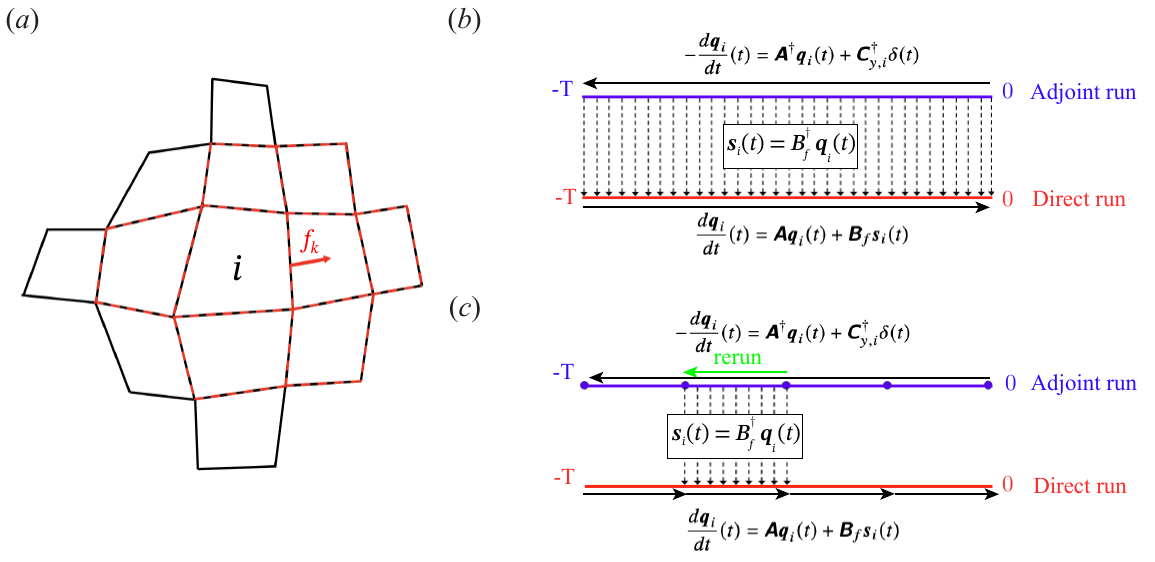}
        };
      \end{tikzpicture}
    \caption[]{\label{Fig_stensil_Chcekpointing} Linearization within the CharLES solver: (\textit{a}) The computational stencil for a control volume ($\text{CV}_i$) labeled $i$. The computational stencil needed to compute the flux at a specific face $k$ of $\text{CV}_i$ is shown with red dashed lines. The CV-based stencil is the union of all face-based stencils for a given CV. A schematic of the two-stage adjoint-direct run: (\textit{b}) without checkpointing and (\textit{c}) with checkpointing. Checkpoints are marked with solid blue circles.}
  \end{center}
\end{figure}

In this section, we describe our implementation of the resolvent-based estimation and control tools described in $\S$\ref{sec_Methodology} and $\S$\ref{sec_Computing_est_con_kernels} within CharLES, an unstructured compressible flow solver for large-scale problems within a high-performance computing environment. The conceptional flow chart for the software is depicted in figure \ref{Fig_FlowChart_RSVTOOLS}. These tools can be used for any flows simulated within the CharLES solver \citep{bres_unstructured_2017} and can be linked with any external packages written in C/C++. To parallelize linear algebra computation and streaming Fourier transform, we integrated the solver with PETSc \citep{PETSC2019} and FFTW \citep{frigo_design_2005}.

\subsection{Linearization}

The operator-based approach in $\S$\ref{sec_Computing_est_con_kernels} requires access to the linear operator $\mathsfbi{A}$ or the actions of both the linear and adjoint operators. We use a matrix-forming approach in which the linear operator is directly computed and stored within the nonlinear solver. This approach accounts for the numerical schemes and boundary conditions employed in the nonlinear simulations and ensures that the linear operator is readily available at run-time.  Extracting the linear operator from large-scale, compressible CFD solvers is not a trivial task, and several approaches have been suggested in the literature  \citep{nielsen_efficient_2006, fosas_de_pando_efficient_2012,cook_2018,bhagwat_2020}.  The approach we adopt most closely mirrors that of \citet{nielsen_efficient_2006} and \citet{cook_2018}.

The CharLES solver uses a control-volume-based finite volume method. A straightforward way to extract the linear operator numerically is to use a finite-difference approximation of the Jacobian, computing the linear operator one column at a time. For example, the \( j^{\text{th}} \) column of the linear operator can be extracted using a second-order approximation as
\begin{equation}\label{eq_Naive_perturb}
\mathsfbi{A}(:,j) = \frac{\boldsymbol{\mathcal{F}}(\boldsymbol{\bar{q}}+\epsilon \boldsymbol{e}_j) - \boldsymbol{\mathcal{F}}(\boldsymbol{\bar{q}}-\epsilon {\boldsymbol{e}}_j)}{2\epsilon} \,,
\end{equation}
where \( j \) refers to the $j$-th degree of freedom and \( \boldsymbol{{\mathcal{F}}} \) represents the right-hand-side of CFD code (the discretized nonlinear compressible Navier-Stokes operator). However, this approach is computationally expensive, as the number of the global right-hand-side evaluations ($\boldsymbol{{\mathcal{F}}}$) required to form the operator equals the problem dimension \( n \).

Instead, we use an approach adopted by \citet{nielsen_efficient_2006} and more recently applied by \citet{cook_2018} that relies on perturbing multiple degrees of freedom (DOF) simultaneously. The key insight is that perturbing an element of the state vector \( \boldsymbol{q} \) affects only a small number of nearby control volumes, known as its computational stencil. This stencil is determined by the numerics used in the Charles solver \citep{bres_unstructured_2017}. Thus, it is feasible to perturb multiple elements of the state vector at once without the perturbations interfering with each other. This approach allows us to compute multiple columns of \( \mathsfbi{A} \) simultaneously, significantly reducing the computational cost.  This is accomplished by replacing unit vector $\boldsymbol{e}_j$ in~(\ref{eq_Naive_perturb}) with \(\Tilde{\boldsymbol{e}}_{k}\), which represents the multiple degrees of freedom perturbed simultaneously.  Using this approach, the number of right-hand-side evaluations $\boldsymbol{\mathcal{F}}(\boldsymbol{q})$ scales with the extent of the computational stencil, not with the size of the problem. In the spatial discretization in CharLES, the flux at a face depends on the adjoining control volumes and their immediate face neighbors, as illustrated in figure \ref{Fig_stensil_Chcekpointing}(\textit{a}).  

To optimize the process of building the \(\Tilde{\boldsymbol{e}}_{k}\) vectors, we sort the computational grid into lists of non-overlapping degrees of freedom on a single processor and then broadcast this information to all other processors. The linearized compressible Navier-Stokes operator \(\mathsfbi{A}\) is extracted and saved at the initial step, and the routine is performed only once before the time-stepping or nonlinear runs (DNS or LES). Typically, the number of right-hand-side evaluations required for standard hexahedral/quadrilateral grids is approximately 125 and 300 for 2D and 3D problems, respectively. The small parameter \( \epsilon \) in~(\ref{eq_Naive_perturb}) is empirically chosen to minimize the error in the numerical derivatives. We have found \( \epsilon = \epsilon_0 ||\boldsymbol{q}|| \) with \( \epsilon_0 = 10^{-6} \) to be an effective and robust choice. The \(\|\boldsymbol{q}\|\) is computed separately for each quantity (density, velocities, and energy). 

\subsection{Efficient implementation of the estimation and control tools}

To effectively store and utilize the linear operator for large-scale numerical linear algebra computations such as matrix-vector products, we parallelize our implementation using the open-source linear algebra package PETSc \citep{PETSC2019}. PETSc leverages the underlying domain decomposition used by the CFD solver to partition the computational grid. Once the matrices such as $\mathsfbi{A}$, $\mathsfbi{C}_{y}$, $\mathsfbi{C}_{z}$, $\mathsfbi{B}_{a}$, and $\mathsfbi{B}_{f}$ are constructed, PETSc is used to advance the linear dynamics for the single or two-stage runs described in $\S$\ref{subsec_Operator-based approach}. To advance these equations in time, we use the TVD-RK3 scheme \citep{gottlieb_total_1998}, the same scheme used by the nonlinear solver.

In the two-stage run described in $\S$\ref{subsubsec_Two-stage run}, the time-series data of \(\boldsymbol{s}_{i}(t)\) from (\ref{eq_Two_stageruns_b}) must be stored to serve as the forcing term \(\mathsfbi{B}_f \boldsymbol{s}_{i}(t)\) in (\ref{eq_dirrun_inTwostageRun_a}), as shown schematiclly in figure \ref{Fig_stensil_Chcekpointing}(\textit{b}).  However, storing all snapshots from the initial step of the adjoint run to the time $T$ where the direct run begins becomes prohibitively expensive over long time horizons.  We use checkpointing, as shown in figures \ref{Fig_stensil_Chcekpointing}(\textit{c}), to address this issue by only storing snapshots at particular intervals during the adjoint run in (\ref{eq_Two_stageruns_a}).  After completing the first full adjoint run, the direct run is advanced in chunks. The adjoint run is then rerun between the last two checkpoints, using only the stored snapshots within that interval, before conducting the direct run through the same interval. This approach reduces memory usage, which is particularly beneficial for large-scale problems. For example, if advancing \(N_t\) timesteps in the adjoint run, the storage requirement can be reduced from \(O(N_t)\) to \(O(W_t + N_t/W_t)\), where \(W_t\) is the length of the interval between two checkpoints, and \(N_t/W_t\) roughly indicates the number of checkpoints. The minimum memory requirement is achieved when \(W_t \approx \sqrt{N_t}\). Thus, checkpointing reduces the memory required from \(O(N_t)\) to \(O(2\sqrt{N_t})\).


Constructing the data-driven estimation and control kernels (\ref{eq_T_nc_c_D}) and (\ref{eq_Gamma_nc_c_D}) requires the computation of CSDs, as described in $\S$\ref{subsec_Data-driven approach}.  Typically, CSDs are computed using FFTs, which require simultaneous access to many snapshots of the state (to be precise, the number of desired frequencies $n_{freq}$).  However, this approach quickly becomes infeasible when each snapshot is large, i.e., when the state dimension $n$ is large.   To reduce data size and memory usage, we employ streaming discrete Fourier transforms (DFTs), as proposed by \cite{schmidt_efficient_2019} within the context of a streaming algorithm for spectral proper orthogonal decomposition (SPOD) and further utilized by \cite{farghadan_scalable_2023} within a scalable time-stepping algorithm for resolvent analysis.

The streaming algorithm requires access to only one instantaneous snapshot of the state at a time, avoiding the need to store the entire time-series. This is achieved by using the definition of the discrete Fourier transform (DFT), which yields results equivalent to the FFT. Each snapshot contributes to the summation of the Fourier modes as
\begin{equation}\label{streaming_eq}
\hat{\boldsymbol{f}}_{k}^{l} = \sum_{j=1}^{n_{freq}} \boldsymbol{f}_{j}^{l} p_{jk},
\end{equation}
where \( p_{jk} = e^{(k-1)(j-1)(-i2\pi/n_{freq})} \), with \(k\) representing the \(k\)-th frequency, \(j\) the \(j\)-th snapshot, and \(l\) the \(l\)-th block of data. The full time series data is divided into multiple blocks, each windowed with a 50\% overlap. Each snapshot is multiplied by the complex scalar $p_{jk}$ and added to the summation of the Fourier modes.  Our implementation is integrated with FFTW \citep{frigo_design_2005} and stores the DFT matrix $\mathbb{C}^{n_{freq} \times n_{freq}}$ during the initialization step of the CFD solver. 

\subsection{Extracting nonlinear terms}\label{subsubsec_Extract_nlterms}

Extracting the nonlinear terms ($\boldsymbol{f}_{nl}$ in (\ref{eq_Bf_Bext_f_ext_f_nl})) of Navier-Stokes equations is useful to investigate the nonlinear interactions. The nonlinear terms are extracted within the application developed for the resolvent-based estimation and control tool. The principle is described here.

The nonlinear Navier-Stokes operator \(\mathcal{F}\) can be expressed as
\begin{equation}\label{eq6.9}
\mathcal{F}(\boldsymbol{q}) =  \mathcal{F}(\boldsymbol{\bar{q}}) +   \frac{\partial \mathcal{F} (\boldsymbol{\bar{q}})}{\partial \boldsymbol{q}}\boldsymbol{q'} + nl (\boldsymbol{q'}),
\end{equation}
where \(nl(\boldsymbol{q'})\) represents all remaining nonlinear terms after linearization. The forcing vector that accounts for the nonlinear terms can be derived as follows
\begin{equation}\label{eq6.10}
\boldsymbol{f}_{nl}(\boldsymbol{q'}) = \mathcal{F}(\bar{\boldsymbol{q}}) +   nl (\boldsymbol{q'}) = \underbrace{\mathcal{F}(\boldsymbol{q})}_{\text{from DNS}} - \underbrace{\mathsfbi{A}\boldsymbol{q'}}_{\text{from linear run}}.
\end{equation}

We run the nonlinear simulation (DNS or LES) in time and, within the same loop, compute the term \(\mathsfbi{A}\boldsymbol{q}'\) to subtract from \(\mathcal{F}(\boldsymbol{q})\). The resulting nonlinear term \(\boldsymbol{f}_{nl}\) is saved in the solver. Computing the CSDs of the nonlinear term requires significant memory due to the large \(n_{nl}\). Therefore, to efficiently compute \(\mathsfbi{\hat{F}}\), we utilize a streaming Fourier transform in (\ref{streaming_eq}), which doesn't require to save all the snapshot $\boldsymbol{f}_{nl}$.

\section{Resolvent-based estimation results}\label{sec_Resolvent-based estimation}

In this section, we use the resolvent-based estimation framework to estimate velocity fluctuations in the wake of the airfoil.  
We expect this framework to be well-suited for this task since the flow is globally stable and resolvent modes capture the vortex shedding, as demonstrated in figures \ref{Fig_Eigenspectrum} and \ref{Fig_RSV_gains}, respectively.  

Before considering the actual nonlinear flow of interest, we first evaluate the performance of the resolvent-based estimator applied to the linear system (excited by upstream disturbances) on which the estimator is based.  We then turn our attention to the true, nonlinear system with clean and noisy freestream conditions.  As discussed previously, the noisy freestream prevents the flow from falling into a periodic limit cycle, resulting in chaotic fluctuations in the wake.

The measurements $\boldsymbol{y}$ used by the estimator correspond to one or more shear stress sensors on the surface of the airfoil, extracted from the state $\boldsymbol{q}$ by an appropriately defined measurement matrix $\mathsfbi{C}_{y}$.  To mimic the effect of sensors of finite size and to facilitate convergence on a finite grid, the sensors and targets have Gaussian spatial support of the form
\begin{equation}\label{eq6.2}
\alpha e^{-(x-x_{c})^{2}/ 2 \sigma_{x} ^{2} -(y-y_{c})^{2}/ 2 \sigma_{y} ^{2} },
\end{equation}
where $\sigma_{x}$ and $\sigma_{y}$ set the width of the Gaussian function and  the constant $\alpha$ is set so that the kernel integrates to one.

\subsection{Linear system}\label{subsec_Linear system}
\begin{figure}[t]
  \begin{center}
      \begin{tikzpicture}[baseline]
        \tikzstyle{every node}=[font=\small]
        \tikzset{>=latex}
        \node[anchor=south west,inner sep=0] (image) at (0,0) {
          \includegraphics[scale=0.9,width=1\textwidth]{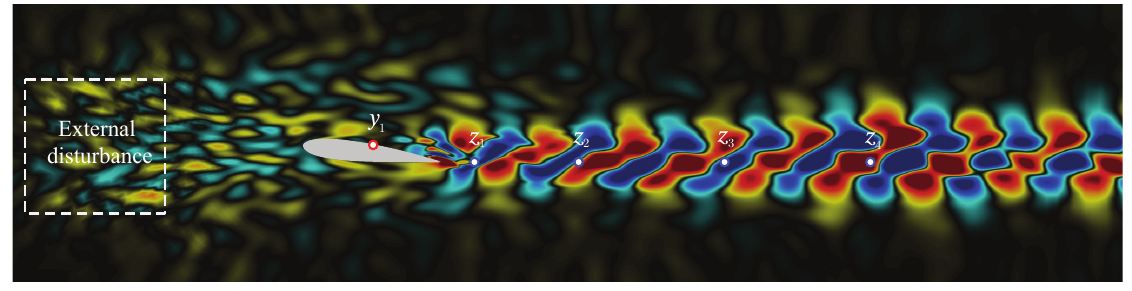}
        };
      \end{tikzpicture}
    \caption[]{\label{Fig_Est_setup} Estimation set up for the linear system, showing the locations of the upstream forcing (white box), sensors (red circle), and targets (blue circle).  The contours show an instantaneous snapshot of the streamwise velocity fluctuation.}
  \end{center}
\end{figure}

Figure \ref{Fig_Est_setup} shows an instantaneous snapshot of the streamwise velocity fluctuation for the linear system driven by an external disturbance in the upstream region. The dominant wake mode, closely resembling the least stable eigenmode and the optimal resolvent response mode, is clearly visible. 

\subsubsection{Building estimation kernels with white noise forcing}\label{subsub_Est_kernels_whitenoise}

\begin{figure}[t]
  \begin{center}
      \begin{tikzpicture}[baseline]
        \tikzstyle{every node}=[font=\small]
        \tikzset{>=latex}
        \node[anchor=south west,inner sep=0] (image) at (0,0) {
          \includegraphics[scale=1,width=1\textwidth]{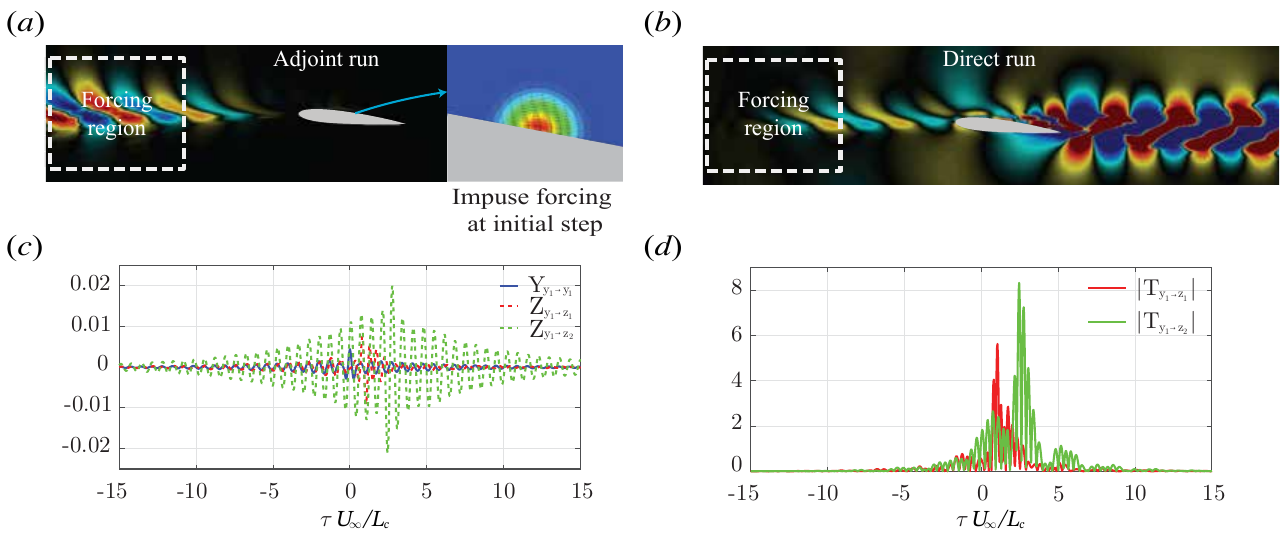}
        };
      \end{tikzpicture}
    \caption[]{\label{Fig_adjdir}Operator-based estimation approach: (\textit{a}) Snapshot of the adjoint run at a specific time instant, including a zoom-in of the impulsive forcing applied at time zero; (\textit{b}) Snapshot of the direct run forced by the \(\mathsfbi{B}_{f}\) readings from the adjoint run; (\textit{c}) sensor and target readings \(y_1\) \([\text{x}/L_c = 0.5]\), \(z_1\) \([\text{x}/L_c = 1.2]\), and \(z_2\) \([\text{x}/L_c = 2.0]\) from the direct run in (\textit{b}); (\textit{d}) Non-causal estimation kernels constructed using the readings from (\textit{c}).}
  \end{center}
\end{figure}

We begin by demonstrating how to construct the estimation kernel using the operator-based approach described in $\S$\ref{subsec_Operator-based approach}. A sensor (\(y_1\)) is placed on the suction surface of the airfoil at \(x/L_c = 0.5\), and four targets (\(z_1, z_2, z_3, z_4\)) are positioned in the wake aligned with the trailing edge at \(x/L_c = [1.2, 2.0, 3.0, 4.0]\) and \(y/L_c = -0.11\), as illustrated in figure \ref{Fig_Est_setup}.  In figure~\ref{Fig_adjdir}, we show an example of the two-stage run used to build the operator-based kernel between sensor \( y_1 \) and targets \( z_1 \) or \( z_2 \). An impulsive forcing is applied to the adjoint system at the sensor location at time zero, as shown in figure \ref{Fig_adjdir}(\textit{a}). This forcing has Gaussian temporal and spatial support, with $\sigma_{t} = 12.5$ and $\sigma_{x} = \sigma_{y} = 0.02$, and the sensors and targets have the same spatial support. The input matrix $\mathsfbi{B}_{f}$ is defined such that the external forcing $\boldsymbol{f}_{ext}$ is defined only in the prescribed upstream region. The forcing CSD matrix is assumed to be white noise, i.e., \( \mathsfbi{{F}}_{ext} = \mathsfbi{{I}} \), implying that the forcing is uncorrelated in space and time. The sensor noise CSD is $ \mathsfbi{\hat{N}} = \epsilon I$, with $\epsilon$ equal to $0.1$ of the maximum value of \( \mathsfbi{\hat{Y}} \). While sensor noise amplitude can affect the smoothness of the estimated data, it does not significantly impact its amplitude or phase. 

The readings $\boldsymbol{s}_{i}$ from the adjoint run are then used to force the direct run, leading to the response shown in figure \ref{Fig_adjdir}(\textit{b}). The corresponding sensor and target measurements are recorded in $\mathsfbi{Y}$ and $\mathsfbi{Z}$, as defined in (\ref{eq_Yf_Zf}) and shown in \ref{Fig_adjdir}(\textit{c}).  $\mathsfbi{{Y}}$ is symmetric about $\tau U_{\infty} / L_{c}=0$ as it is autocorrelation. The perturbation grows as it travels downstream, so $Z_{y_{1} \to z_{2}}$ is generally greater than $Z_{y_{1} \to z_{1}}$. To achieve convergence in the adjoint and direct runs, we simulate the two-stage run over a time span \( t U_{\infty}/L_{c} \in [-36, 36] \) with a time step twice that of the DNS.

The Fourier-transformed data,  $\mathsfbi{\hat{Y}}$ and $\mathsfbi{\hat{Z}}$, are equivalent to $\mathsfbi{R}_{yf}\mathsfbi{R}_{yf}^{\dagger}$ and $\mathsfbi{R}_{zf}\mathsfbi{R}_{yf}^{\dagger}$, as shown in~(\ref{eq_Yf_Zf}). Finally, the Forier-transformed data are used to form the resolvent-based kernels using~(\ref{eq_Tnc_Tc_O}). Figure \ref{Fig_adjdir}(\textit{d}) illustrates the estimation kernel $\mathsfbi{T}_{nc}$ in (\ref{eq_T_nc}) computed from $\mathsfbi{Y}$ and $\mathsfbi{Z}$. The peaks of $T_{y_{1} \to z_1}$ and $T_{y_1 \to z_2}$ can be explained as the travel time from the sensor location (where the impulse forcing is imposed) to the target. A second dominant peak is observed for $T_{y_1 \to z_1}$ and $T_{y_1 \to z_2}$ on the left side of the main peaks, which may be the result of acoustic waves, which are faster than the hydrodynamic waves. For the estimation kernels, $\tau U_{\infty} / L_{c}>0$ and $\tau U_{\infty} / L_{c}<0$  represent past and future times, respectively. Both kernels shown in figure \ref{Fig_adjdir}(\textit{d}) have non-zero amplitude mainly for $\tau U_{\infty}/ L_{c}>0$, i.e., they are nearly causal.   If a significant non-causal part is present, truncating it will degrade the performance of the estimator; optimality, under the constraint of causality, can be restored using the Wiener-Hopf decomposition. This impact is more significant for the nonlinear system, so we will discuss it in greater detail in that section. 

\subsubsection{Estimation results for the linear system}\label{subsub_Est_results_Linsys}

We present the causal resolvent-based estimation result only using a single sensor, comparing the true streamwise velocity fluctuation \(u_{x}'\) with the estimated value over time. Additionally, we estimate the fluctuations of both the streamwise \(u_{x}'\) and cross-streamwise \(u_{y}'\) velocity components in an extended region of the targets using a small number of sensors. To quantify the accuracy of the estimates, we calculate the estimation error 
\begin{equation}\label{eq6.3}
\textit{E} = \frac{\Sigma_{i} \int(\boldsymbol{\tilde{z}_{i}}(t)-\boldsymbol{z}_{i}(t))^{2} dt}{\Sigma_{i} \int(\boldsymbol{z}_{i}(t))^{2} dt},
\end{equation}
where \(\tilde{\boldsymbol{z}}_{i}\) and \(\boldsymbol{z}_{i}\) represent the estimated and true values for the \(i\)-th target, respectively. In computing the estimation error, we assume the system is ergodic, allowing the ensemble average to be replaced by the time average. 


\begin{figure}[t]
  \begin{center}
      \begin{tikzpicture}[baseline]
        \tikzstyle{every node}=[font=\small]
        \tikzset{>=latex}
        \node[anchor=south west,inner sep=0] (image) at (0,0) {
          \includegraphics[scale=1,width=0.9\textwidth]{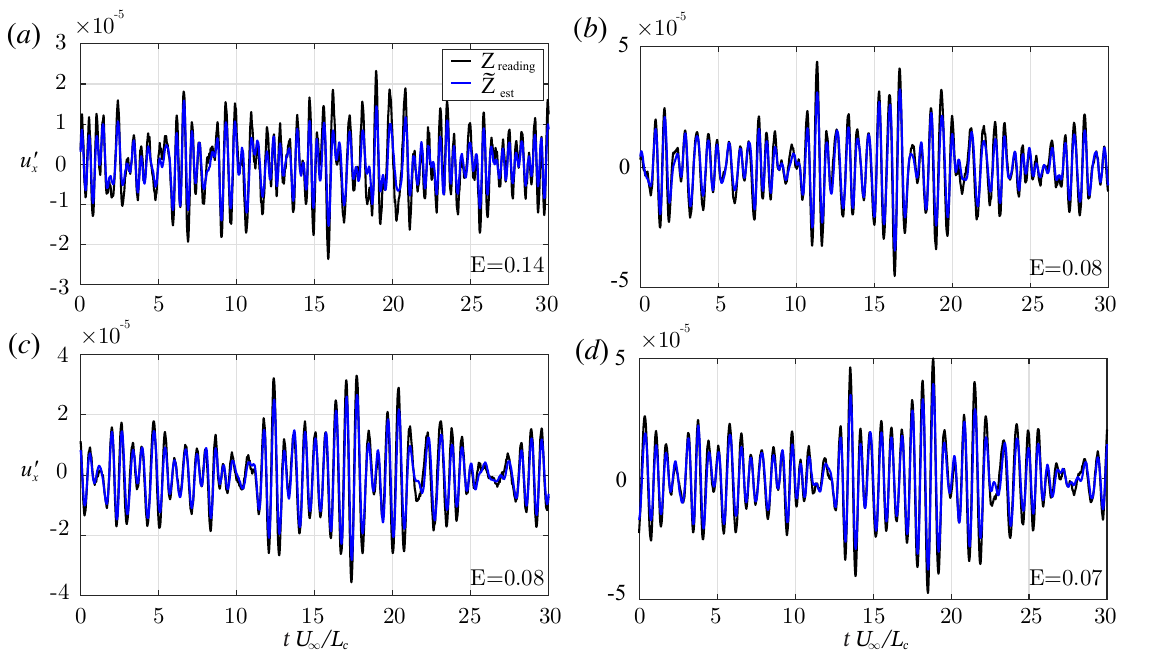}
        };
      \end{tikzpicture}
    \caption[]{\label{Fig_estResult}  Causal estimation using an operator-based approach for the linear system at the targets: (\textit{a}) \(z_{1}\), (\textit{b}) \(z_{2}\), (\textit{c}) \(z_{3}\), and (\textit{d}) \(z_{4}\) at positions [\(x/L_{c} = 1.2, 2.0, 3.0, 4.0\)], as shown in Figure \ref{Fig_Est_setup}. The estimation error (\ref{eq6.3}) is reported for each case.}
  \end{center}
\end{figure}
\begin{figure}[!t]
  \begin{center}
      \begin{tikzpicture}[baseline]
        \tikzstyle{every node}=[font=\small]
        \tikzset{>=latex}
        \node[anchor=south west,inner sep=0] (image) at (0,0) {
          \includegraphics[scale=0.7,width=0.65 \textwidth]{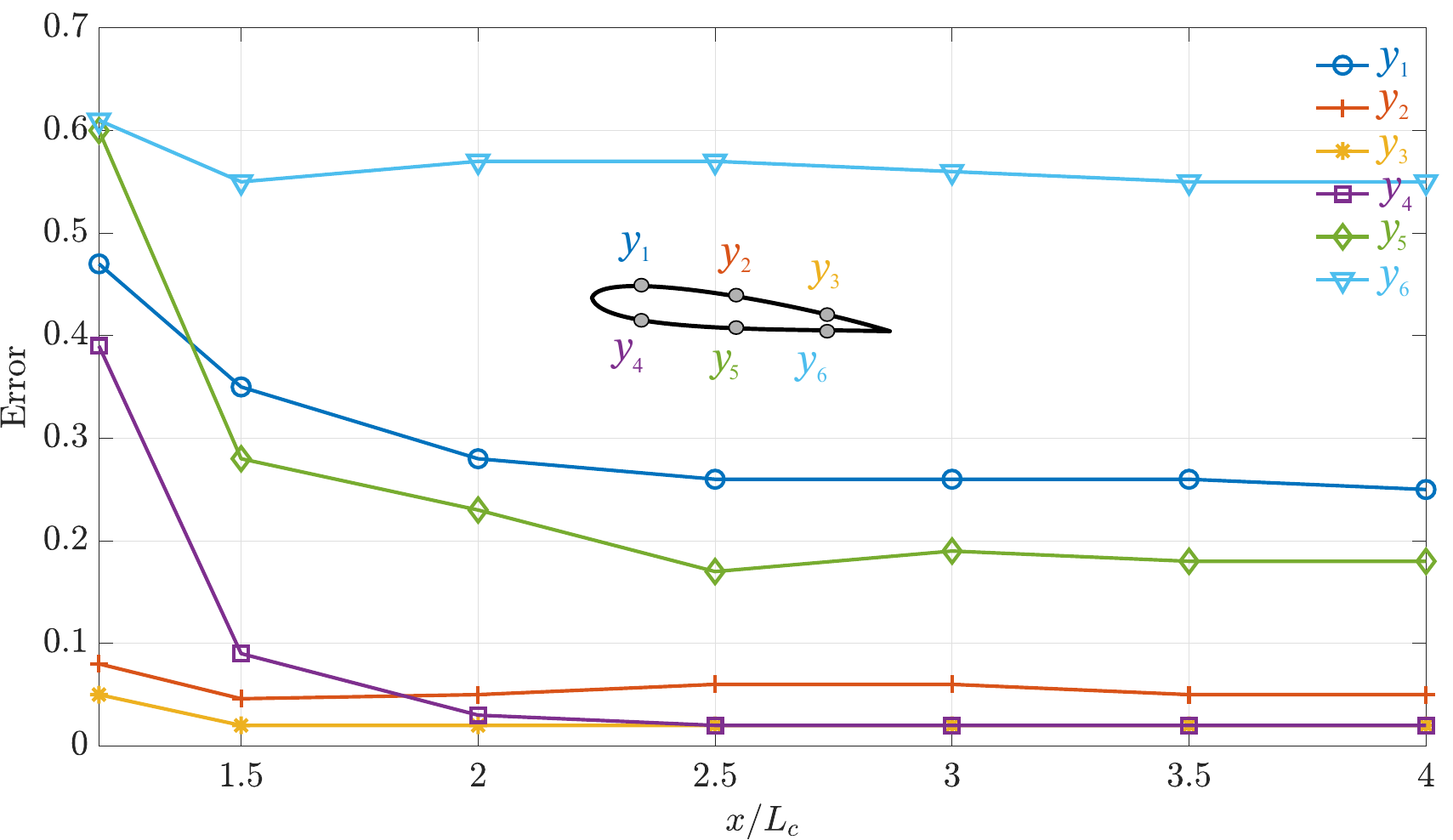}
        };
      \end{tikzpicture}
    \caption[]{\label{Fig_Est_Lin_eror}  Estimation error for the linear system as a function of the target location $x/L_{c}$ and $y/L_{c} = 0$.}
  \end{center}
\end{figure}
Since the estimator was designed for this linear system, the causal estimates are theoretically optimal. Figure \ref{Fig_estResult} shows examples of the true and estimated target readings as a function of time. The result indicates that the target near the trailing edge, positioned in a more complex flow region, is estimated less accurately. In contrast, the other three targets (\(z_2, z_3,\) and \(z_4\)) show better estimation with an error of \(0.07-0.08\). Overall, the frequency and amplitude of the fluctuations are well estimated. 

\begin{figure}[!t]
  \begin{center}
      \begin{tikzpicture}[baseline]
        \tikzstyle{every node}=[font=\small]
        \tikzset{>=latex}
        \node[anchor=south west,inner sep=0] (image) at (0,0) {
          \includegraphics[scale=0.9,width=0.9\textwidth]{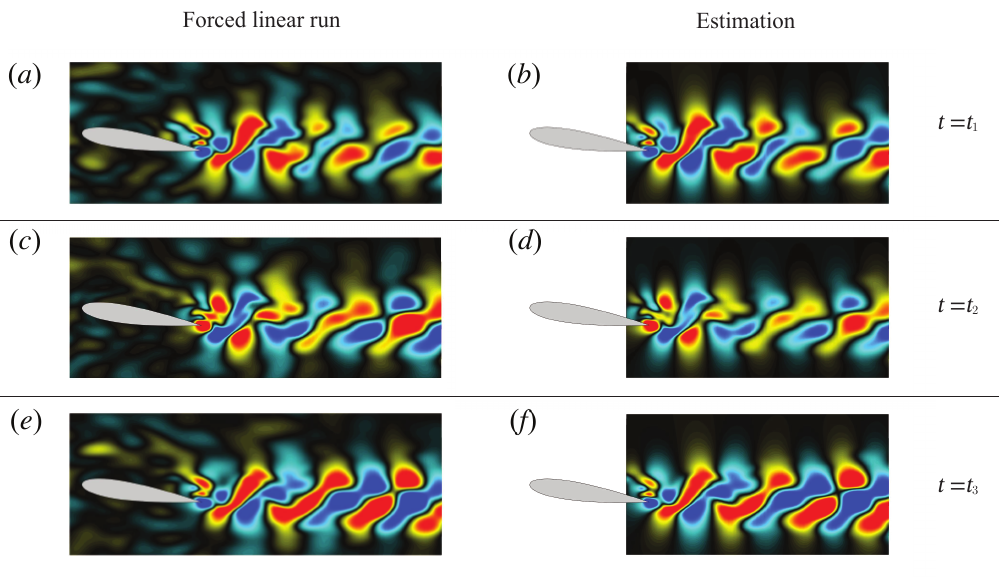}
        };
      \end{tikzpicture}
    \caption[]{\label{Fig_Est_linear_highrank}  Estimation of streamwise velocity fluctuation for the extended target region at three different time steps for the linear system using the sensor \( y_{3} \), as shown in figure \ref{Fig_Est_Lin_eror}.}
  \end{center}
\end{figure}
\begin{figure}[!t]
  \begin{center}
      \begin{tikzpicture}[baseline]
        \tikzstyle{every node}=[font=\small]
        \tikzset{>=latex}
        \node[anchor=south west,inner sep=0] (image) at (0,0) {
          \includegraphics[scale=1,width=0.7\textwidth]{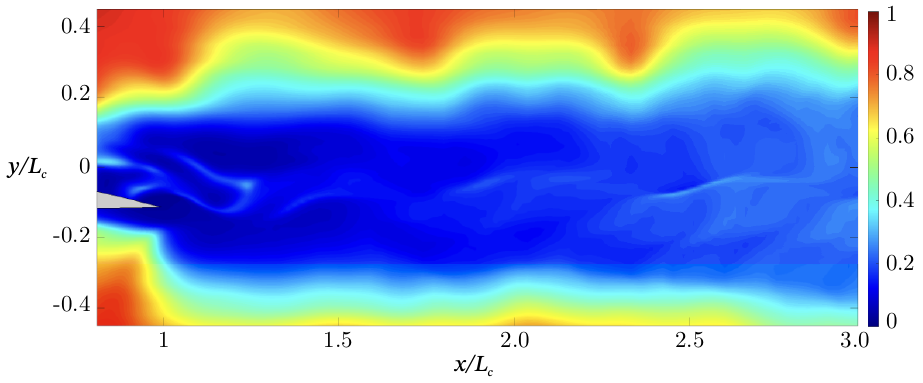}
        };
      \end{tikzpicture}
    \caption[]{\label{Fig_linear_est_highrank_error}Estimation error in the extended target regions for the linear system using the sensor \( y_{3} \).}
  \end{center}
\end{figure}

Next, we explore the impact of the sensor location on the estimation accuracy. Figure \ref{Fig_Est_Lin_eror} shows the estimation error as a function of the target location \( x/L_{c} \) (aligned with the trailing edge) for six different sensor locations on the airfoil surface. For targets near the trailing edge ($x/L_{c}<1.5$), the front sensors on the suction side (\( y_{1} \)) and pressure side (\( y_{6} \)) produce inaccurate estimates. In contrast, the rear sensors (\( y_{3} \) and \( y_{4} \)) and the middle sensor \( y_{2} \), which is the location used to demonstrate building the kernels in the previous section, result in better estimation accuracy. The suction-side sensors \( y_{2} \) and \( y_{3} \) accurately capture the flow dynamics near the trailing edge generated by the separation bubble over the airfoil. While \(y_{4}\) faces challenges in capturing flow information from the bottom of the airfoil, it can still estimate the downstream targets ($x/L_{c}>2$) as effectively as the sensors \(y_{2}\) and \(y_{3}\).

Finally, we estimate the state in an extended region of the flow rather than at individual, discrete targets, allowing us to evaluate the estimation accuracy in different regions. The estimation kernel in this context takes the form
\begin{equation}\label{eq6.4}
\begin{+bmatrix} 
\hat{\boldsymbol{z}}_{1}\\
\hat{\boldsymbol{z}}_{2}\\
\vdots\\
\hat{\boldsymbol{z}}_{n_{z}}\\
\end{+bmatrix}
=
\begin{+bmatrix} 
\hat{\boldsymbol{T}}_{z_{1}y_{1}}&\hat{\boldsymbol{T}}_{z_{1}y_{2}}&\hdots&\hat{\boldsymbol{T}}_{z_{1}n_{y}}\\
\hat{\boldsymbol{T}}_{z_{2}y_{1}}&\hat{\boldsymbol{T}}_{z_{2}y_{2}}&\hdots&\hat{\boldsymbol{T}}_{z_{2}n_{y}}\\
 \vdots&\vdots&\ddots&\vdots&\\
\hat{\boldsymbol{T}}_{z_{n_{z}}y_{1}}&\hat{\boldsymbol{T}}_{z_{n_{z}}y_{2}}&\hdots&\hat{\boldsymbol{T}}_{z_{n_{z}}y_{n_{y}}}\\
\end{+bmatrix}
\begin{+bmatrix} 
\hat{\boldsymbol{y}}_{1}\\
\hat{\boldsymbol{y}}_{2}\\
\vdots\\
\hat{\boldsymbol{y}}_{n_{y}}\\
\end{+bmatrix}
\end{equation}
where \( n_{y} \) and \( n_{z} \) represent the number of sensors and targets, respectively. Based on the optimal results from figure \ref{Fig_Est_Lin_eror}, we use just one sensor located on the suction side, \( y_{3} \). Figure \ref{Fig_Est_linear_highrank} shows snapshots of the estimates in the extended target regions. The three time steps \( t_{1}, t_{2}, t_{3} \) are selected to represent different phases of the vortex shedding. Figure \ref{Fig_linear_est_highrank_error} shows the estimation error as a function of position within the extended target region. As expected, the error increases as the target moves downstream or laterally away from the wake.  

\subsection{Nonlinear system}\label{subsec_Nonlinear system}

\begin{figure}[t]
  \begin{center}
      \begin{tikzpicture}[baseline]
        \tikzstyle{every node}=[font=\small]
        \tikzset{>=latex}
        \node[anchor=south west,inner sep=0] (image) at (0,0) {
          \includegraphics[scale=0.8,width=0.9\textwidth]{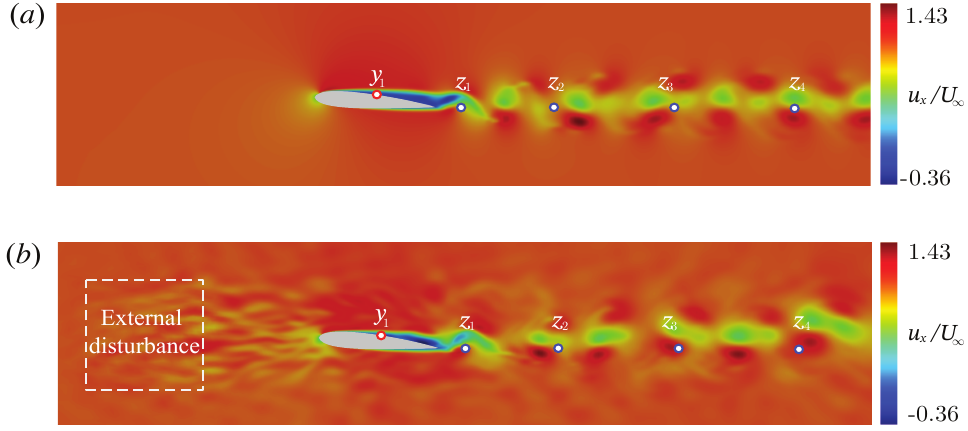}
        };
      \end{tikzpicture}
    \caption[]{\label{Fig_est_setup_NL} Instantaneous snapshot of the streamwise velocity $u_{x}$ for (\textit{a}) the clean and (\textit{b}) the noisy DNS cases. The symbols show the sensor and target locations. The noisy freestream is generated by a random forcing within the region $x/L_{c} \in [-2,-1]$ and $y/L_{c} \in [-0.5,0.5]$.}
  \end{center}
\end{figure}

In the previous section, we confirmed that the resolvent-based kernels, derived from the resolvent operator, provide accurate estimates for the linear system. We now shift our focus to the actual (nonlinear) system, where it is crucial to statistically account for the nonlinear terms using colored forcing.  We do so by using the data-driven approach described in $\S$\ref{subsec_Data-driven approach}, which is equivalent to the operator-based method with the appropriate forcing CSD $\hat{\mathsfbi{F}}$. As discussed earlier, we consider both clean and noisy freestream conditions for the nonlinear system, as illustrated in figures \ref{Fig_est_setup_NL}(\textit{a}) and (\textit{b}). The flow is simulated using DNS with the same numerical setup described in \S\ref{sec_Problem}. Since the estimator is defined in terms of perturbations to the mean, the mean is removed from the sensor readings before convolution with the estimation kernels.

\subsubsection{Nonlinear response to the external forcing}\label{subsub_NL_response_to_ExtForcing}

\begin{figure}[t]
  \begin{center}
      \begin{tikzpicture}[]
        \tikzstyle{every node}=[font=\small]
        \tikzset{>=latex}
        \node[anchor=south west,inner sep=0] (image) at (0,0) {
          \includegraphics[scale=1,width=0.9\textwidth]{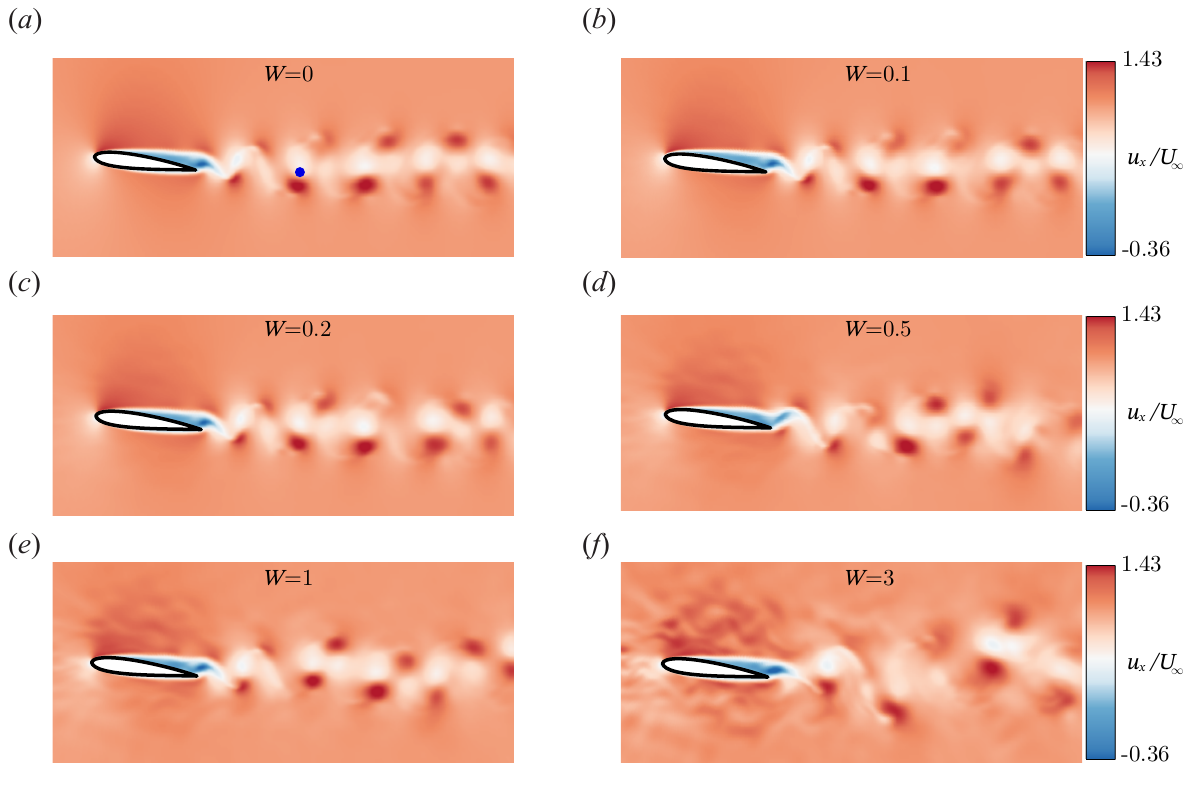}
        };
      \end{tikzpicture}
    \caption[]{\label{Fig_NL_externalF} Instantaneous snapshots of the streamwise velocity $u_x$ for varying freestream noise intensities, as determined by the forcing amplitude $W$: (\textit{a}) $W=0$ (clean), (\textit{b}) $W = 0.1$, (\textit{c}) $W = 0.2$, (\textit{d}) $W = 0.5$, (\textit{e}) $W = 1$, and (\textit{f}) $W = 3$. The blue dot in (\textit{a}) indicates the location for which the PSD is analyzed in figure~\ref{Fig_PSD_externalforcing}.}
  \end{center}
\end{figure}
\begin{figure}[!t]
  \begin{center}
      \begin{tikzpicture}[]
        \tikzstyle{every node}=[font=\small]
        \tikzset{>=latex}
        \node[anchor=south west,inner sep=0] (image) at (0,0) {
          \includegraphics[scale=1,width=0.6\textwidth]{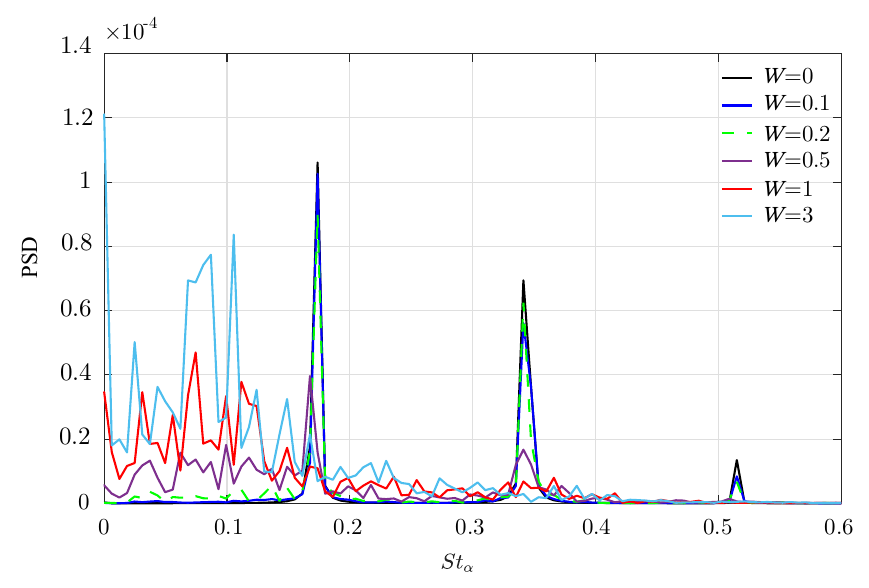}
        };
      \end{tikzpicture}
    \caption[]{\label{Fig_PSD_externalforcing} Power spectral density of the streamwise velocity $u_{x}$ for the nonlinear system in terms of the noise level $W$ at the point $[x/L_{c},y/L_{c}]=[2.11,-0.11]$ in figure \ref{Fig_NL_externalF}.}
  \end{center}
\end{figure}

The nonlinear system subject to external forcing can be expressed as
\begin{equation}\label{eq6.5}
\frac{\partial \boldsymbol{q}}{\partial t}= \mathcal{F} (\boldsymbol{q}) + \mathsfbi{B}_{f,ext} \boldsymbol{f}_{ext}.
\end{equation}
We choose the external forcing $\boldsymbol{f}_{ext} (x,t)$ to be white noise in space and time, generated using random vectors with each entry uniformly distributed over $[-1, 1]\times W$, where $W$ controls the variance of the random vector, adjusting the noise level of the freestream. The CSD matrix for this forcing is $\hat{\mathsfbi{F}}_{ext} = W^{2} \mathsfbi{I}$. In contrast, the linear system models the nonlinear terms as white noise (an identity matrix) \citep{mckeon_critical-layer_2010}.

To help us select an appropriate forcing amplitude $W$, we analyze the PSD of the state for varying noise levels, focusing on the impact on the vortex shedding. The snapshots in figure \ref{Fig_NL_externalF} demonstrate that while vortex shedding persists at all noise levels, the spatial periodic pattern in the streamwise direction is disrupted. Figure \ref{Fig_PSD_externalforcing} makes this point quantitative by showing the PSD of the state at the location indicated in figure~\ref{Fig_NL_externalF}(\textit{a}).  The results confirm that the vortex shedding frequencies ($St_{\alpha}\approx 0.17 \times n$ with $n=1,2,3$) are suppressed for $W \geq 1$, indicating a strong nonlinear modification to the linear dynamics.  We select $W = 1$ as the noise level for our estimation and control studies.


\subsubsection{Building estimation kernels with colored-forcing statistics}\label{subsub_BuildingESTkernel_w_coloredforcing}

\begin{figure}[t]
    \centering
    \includegraphics[scale=1,width=0.9\textwidth]{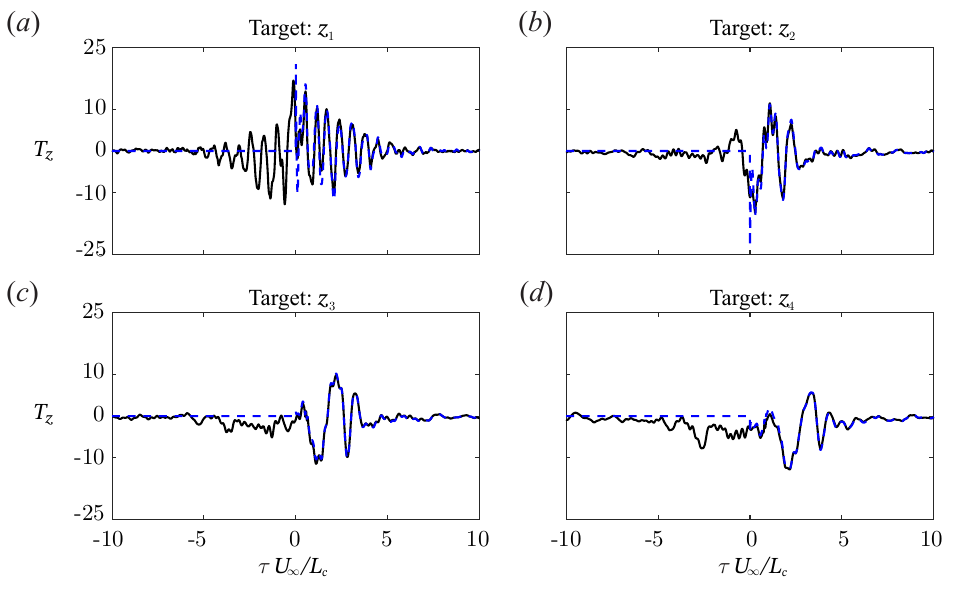}
    \caption[Estimation kernels with a colored forcing statistics]{\label{Fig_ESTKernel_DD} Estimation kernels with a colored forcing statistics: non-causal (black solid line) and causal (blue dashed line) kernels between (\textit{a}) $y_{1}$ [$x/L_{c} = 0.5$] and $z_{1}$ [$x/L_{c} = 1.2$], (\textit{b}) $z_{2}$ [$x/L_{c} = 2.0$], (\textit{c}) $z_{3}$ [$x/L_{c} = 3.0$], and (\textit{d}) $z_{4}$ [$x/L_{c} = 4.0$].}
\end{figure}

In $\S$\ref{subsub_Est_kernels_whitenoise}, we assumed the forcing CSD matrix $\hat{\mathsfbi{F}}_{nl}$ to be white noise (an identity matrix), resulting in kernels equivalent to a Kalman filter. To enhance the accuracy of the estimation kernels, the resolvent-based framework enables us to incorporate colored forcing statistics via a non-identity $\hat{\mathsfbi{F}}_{nl}$, which can be obtained in $\S$\ref{subsubsec_Extract_nlterms}. Once the nonlinear terms $\hat{\boldsymbol{f}}_{nl}$, such as those shown in figure \ref{Fig_NLterm_PSD} in appendix \ref{appx_nonlinear_terms}, are available, we compute $\mathsfbi{R}_{yf}\hat{\mathsfbi{F}}_{nl}\hat{\mathsfbi{R}}_{yf}^{\dagger}$ and $\mathsfbi{R}_{zf}\hat{\mathsfbi{F}}_{nl}\hat{\mathsfbi{R}}_{yf}^{\dagger}$ during the two-stage run outlined in \ref{subsubsec_Two-stage run}. 

Alternately, we can implement the data-driven approach \citep{martini_resolvent-based_2022} to obtain estimation kernels that automatically include the influence of the colored-forcing statistics \citep{zare_colour_2017,towne_resolvent-based_2020, martini_resolvent-based_2022}. The necessary sensor and target data are directly collected from DNS, and Welch's method \citep{welch_use_1967} is employed to obtain the CSD tensors required to construct the estimation kernels. The time series are divided into 64 blocks of length $t U_{\infty} /L_{c}=5$, with $50\%$ overlap. All other parameters for building the kernels and Wiener-Hopf factorization are consistent with those used in the operator-based approach.

The estimation kernels that account for colored nonlinear forcing are shown in figure \ref{Fig_ESTKernel_DD} for the same sensor and target configuration as was used for the white noise case, i.e., the sensor $y_{1}$ [$x/L_{c}=0.5$] and the targets $z_{1}$, $z_{2}$, $z_{3}$, and $z_{4}$ [$x/L_{c}=1.2, 2.0, 3.0, 4.0$], as shown in figure \ref{Fig_est_setup_NL}. The figure shows results for the noisy freestream case. As discussed in $\S$\ref{subsub_NL_response_to_ExtForcing} and shown in figures \ref{Fig_NLterm_PSD}(\textit{d}) in appendix \ref{appx_nonlinear_terms}, there is a higher degree of nonlinear interaction near the trailing edge. In this region, the estimation kernel with colored forcing peaks at $\tau U_{\infty} / L_{c}<0$, i.e., the kernel is highly non-causal. This can be alleviated by using the Wiener-Hopf method to optimally enforce causality. Further downstream in the wake in figure \ref{Fig_ESTKernel_DD}(\textit{b}), (\textit{c}), and (\textit{d}), the kernels exhibit simpler behavior and distinct peaks, which indicate the travel time of the fluctuations. The difference between the causal and non-causal kernels is small when the target is set further downstream in the wake, where the convective nature of the flow makes the kernels naturally (almost) causal. This trend is similar to the backward-facing step flow reported by \cite{martini_resolvent-based_2022}.

\subsubsection{Single-sensor estimation results for the nonlinear system}\label{subsubsec_SensorPlacement}
\begin{figure}[!t]
    \centering
    \includegraphics[scale=1,width=1\textwidth]{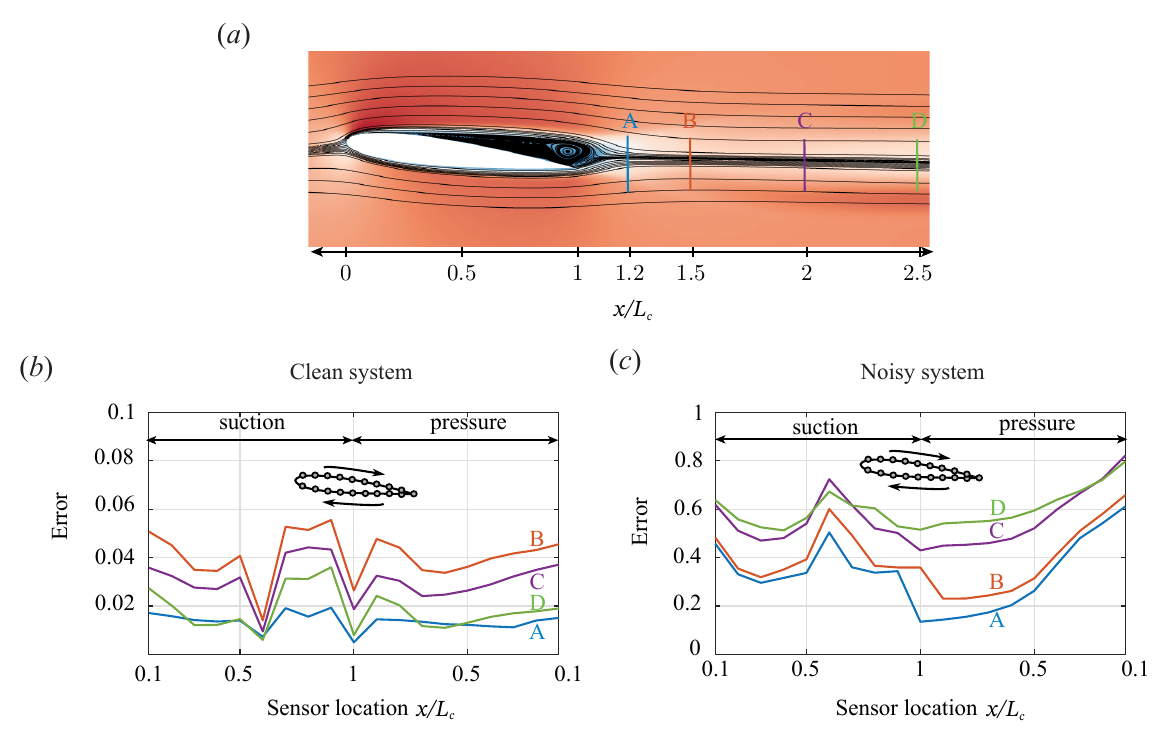}
    \caption[Sensors and targets placement using the streamlines]{\label{Fig_streamlines2} Streamline of the flow is shown along with four sets of targets: A ($x/L_c = 1.2$), B ($x/L_c = 1.5$), C ($x/L_c = 2.0$), and D ($x/L_c = 2.5$). Averaged estimation errors are reported for these four lines (A, B, C, and D), which are based on sensor locations on the airfoil surfaces. Panel (\textit{b}) illustrates the clean system, while panel (\textit{c}) depicts the noisy system.}
\end{figure}

We assess the estimation accuracy for the nonlinear system for a single sensor as a function of its location on the surface of the airfoil. Figures \ref{Fig_streamlines2}(\textit{b}) (clean freestream) and (\textit{c}) (noisy freestream) show the averaged estimation errors for four sets of targets (A, B, C, and D) shown in figure \ref{Fig_streamlines2}(\textit{a}). The errors are about an order of magnitude lower for the clear freestream than for the noisy freestream.  Imposing external forcing leads to higher-energy fluctuations in the wake, as illustrated in figure \ref{Fig_extF}(\textit{c}), which increases the impact of nonlinearity and deteriorates the estimation accuracy as the target is moved downstream (estimation error: $D > C > B > A$ in figure \ref{Fig_streamlines2}(\textit{b})). In contrast, the error for the clean freestream is non-monotonic with downstream distance: it is lowest for the nearest set of targets, increases, and then decreases again. This latter decrease in error is likely due to the increasingly low-rank behavior of the wake with increasing downstream distance for the clean inflow case.  

The recirculation bubble impacts the sensors differently for the clean and noisy freestream cases. For the clean freestream results in figure \ref{Fig_streamlines2}(\textit{b}), sensors positioned within the recirculation region $0.6 < x/L_{c} < 1$ \citep{marquet_hysteresis_2022}, show reduced accuracy. However, this effect is not as evident for the noisy freestream case in figure \ref{Fig_streamlines2}(\textit{c}). 
This suggests that, whereas the recirculation bubble shields the sensors in the clean freestream case, incoming fluctuations from the noisy freestream instantaneously penetrate the bubble and later contribute to the wake dynamics, providing useful information to the sensors. Among the sensors on the suction surface for the noisy freestream in figure \ref{Fig_streamlines2}(\textit{c}), those positioned before the separation bubble ($0.2<x/L_{c}<0.5$) and near the trailing edge ($0.8<x/L_{c}<1.0$) most effectively predict the wake dynamics. Notably, rear sensors on the pressure surface also show high estimation accuracy. 

\begin{figure}[!t]
    \centering
    \includegraphics[scale=1,width=1\textwidth]{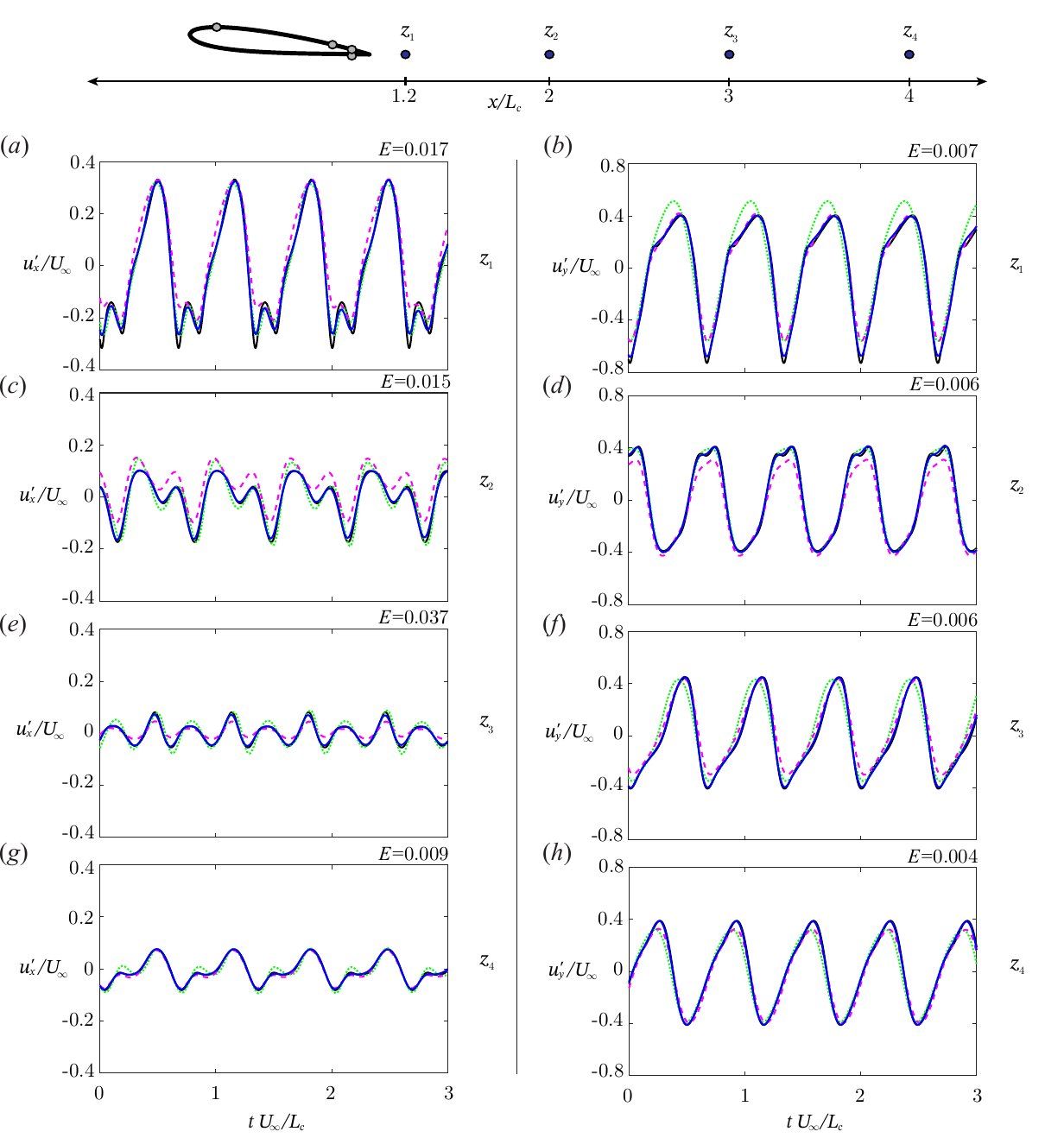}
    \caption[Estimation for the nonlinear system with clean freestreams]{\label{Fig_NLEST_pert_clean}  Estimation of $u_{x}'$ (left column) and $u_{y}'$ (right column) for the nonlinear system with a clean freestream for four targets (columns).  Lines: (black solid) true target from the DNS; (green dashed) Kalman filter estimates;  (magenta dashed) truncated non-causal estimates; (blue solid) resolvent-based causal estimates. The target locations are: (\textit{a}), (\textit{b}) [$z_{1}=x/L_{c}, y/L_{c}$] = [1.2, -0.11]; (\textit{c}), (\textit{d}) [2.0, -0.11]; (\textit{e}), (\textit{f}) [3.0, -0.11]; and (\textit{g}), (\textit{h}) [4.0, -0.11], as shown in the top figure and figure \ref{Fig_est_setup_NL}. The estimation errors for the resolvent-based method are noted in the top-right corner of each panel.}
\end{figure}

\begin{figure}[!t]
    \centering
    \includegraphics[scale=1,width=1\textwidth]{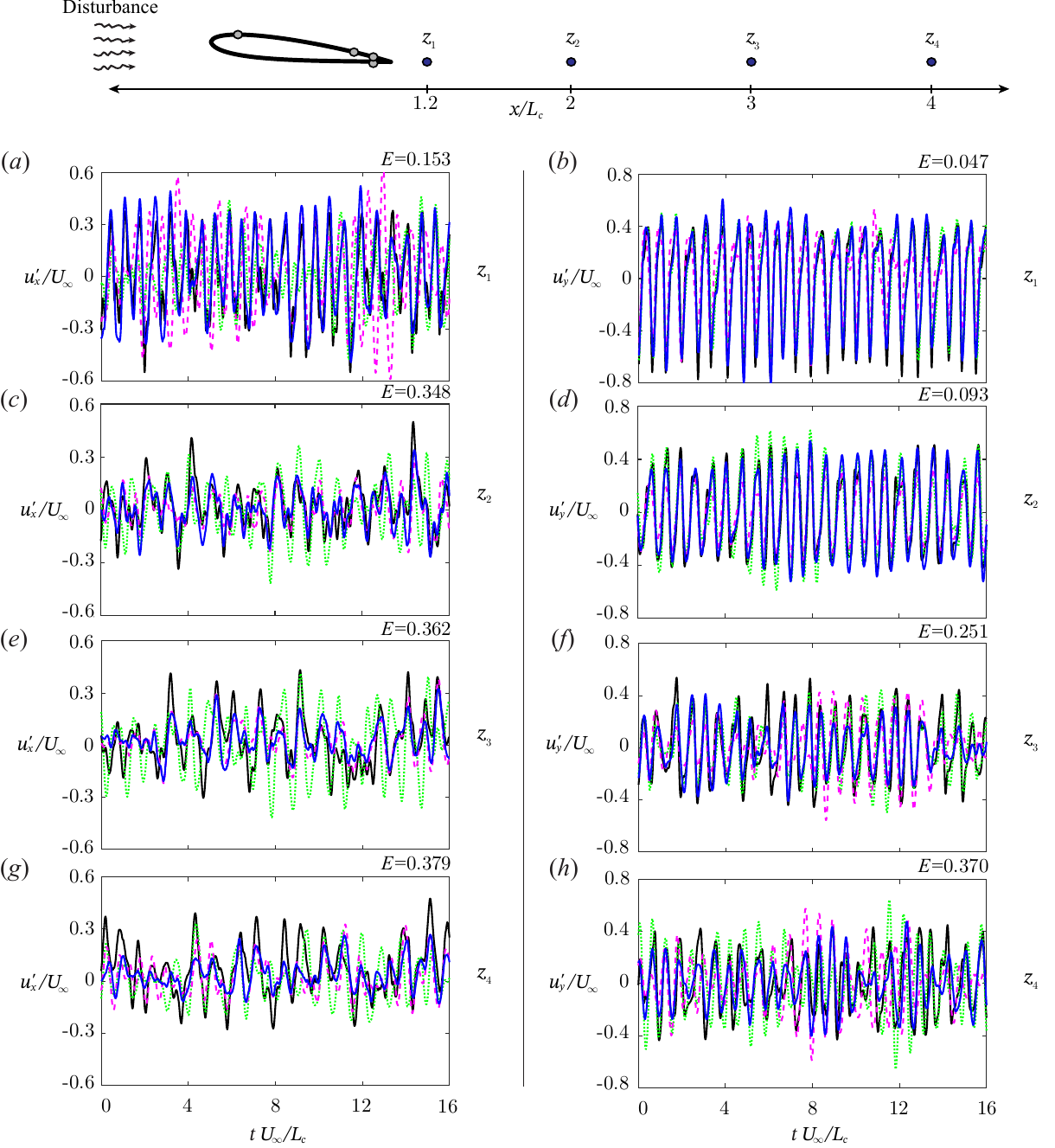}
    \caption[Estimation of $u_{x}'$ and $u_{y}'$ for the nonlinear system with the noisy freestream inflows]{\label{Fig_NLEST_pert_noisy}  Estimation of $u_{x}'$ (left column) and $u_{y}'$ (right column) for the nonlinear system with the noisy freestream. The black (solid) line shows the true DNS, while the other lines represent different methods: green (dashed) for Kalman filter, magenta (dashed) for truncated non-causal estimation, and blue (solid) for causal estimation (our method). The target locations are: (\textit{a}), (\textit{b}) at [$z_{1}=x/L_{c}, y/L_{c}$] = [1.2, -0.11]; (\textit{c}), (\textit{d}) at [2.0, -0.11]; (\textit{e}), (\textit{f}) at [3.0, -0.11]; and (\textit{g}), (\textit{h}) at [4.0, -0.11], as shown in the top figure and figure \ref{Fig_est_setup_NL}. The estimation errors for the causal method are noted in the top right corner of each panel.}
\end{figure}
\subsubsection{Multi-sensor estimation results for the nonlinear system}\label{subsub_Est_results}


Next, we present estimation results for the nonlinear system with clean and noisy freestream conditions using multiple sensors. In appendix \ref{appx_sensorplacement}, we empirically evaluate six candidate sensor configurations guided by the single-sensor results from the previous section. Ultimately, we selected candidate 6, as defined in Table \ref{table_optimalsensorplacement}. Since we are interested in vortex shedding, we report estimation results for both components of velocity that would be needed to compute vorticity.  

Figures \ref{Fig_NLEST_pert_clean} and \ref{Fig_NLEST_pert_noisy} present the causal resolvent estimation of $u_{x}'$ (left column) and $u_{y}'$ (right column) comparing with other methods and the true target reading (from DNS) for the clean and noisy freestream inflows at the target points ($x/L_{c} = 1.2, 2.0, 3.0, 4.0$) aligned with the trailing edge, shown in the top of each figure. The clean DNS system is well estimated using the causal resolvent-based approaches. The Kalman filter captures the dominant frequency's high-energy parts effectively but lacks spatial and temporal detail due to treating the nonlinear terms as white noise. The target near trailing edge, where nonlinearity is strongest in figure \ref{Fig_extF} in appendix \ref{appx_nonlinear_terms}, is poorly estimated using the Kalman filter, shown in \ref{Fig_NLEST_pert_noisy}(\textit{a}), while the poor estimates obtained using the TNC approach are due to the presence of substantial amplitude of the noncausal kernels in the non-causal part $\tau U_{\infty} / L_{c}<0$). However, the truncated non-causal kernels include the impact of the colored forcing statistics, making them effective when the the kernels are mostly naturally causal for the further downstream target locations, e.g., in the case of \ref{Fig_NLEST_pert_noisy}(\textit{e}) and (\textit{g}). Comparing left and right panels in figures \ref{Fig_NLEST_pert_clean} and \ref{Fig_NLEST_pert_noisy}, we observe that the cross-stream velocity is estimated more accurately than the streamwise velocity. 

\begin{figure}[!t]
    \centering
    \includegraphics[scale=1,width=1\textwidth]{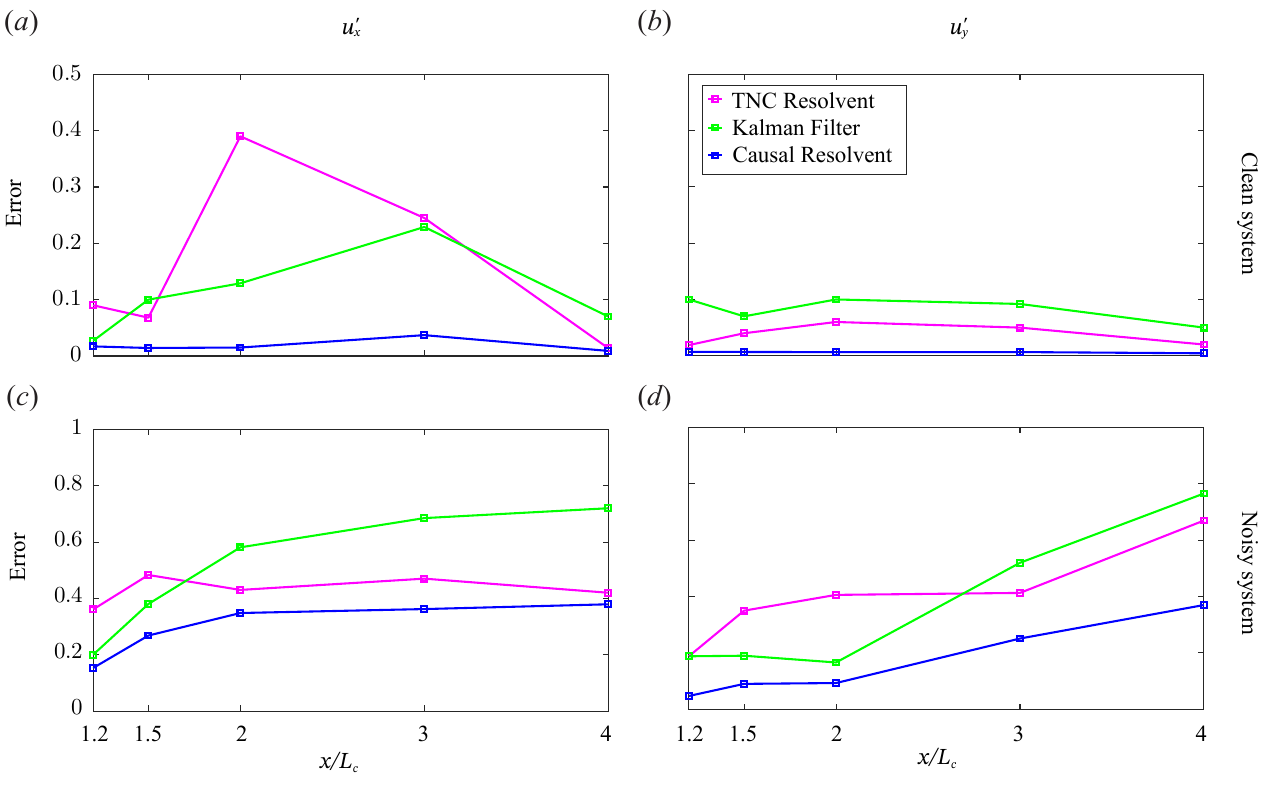}
    \caption[Estimation errors for the nonlinear systems]{\label{Fig_NLEST_DownstreamErrors} Estimation errors for nonlinear systems: (\textit{a}) $u_{x}'$ and (\textit{b}) $u_{y}'$ for clean freestream; (\textit{c}) $u_{x}'$ and (\textit{d}) $u_{y}'$ for noisy freestream. Blue lines represent causal resolvent-based estimation, while magenta and green lines denote truncated non-causal estimation using colored forcing and a Kalman filter (white noise), respectively.}
\end{figure}
\begin{figure}[!t]
    \centering
    \includegraphics[scale=0.8,width=0.9\textwidth]{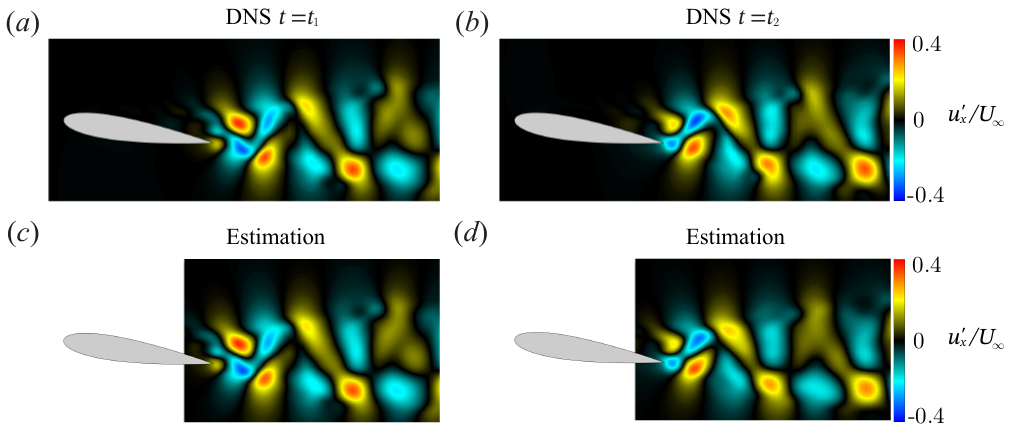}
    \caption[Estimation of the wake region for $u_{x}'$ for the clean system]{\label{Fig_NL_NoF_est_highrank_result} Estimation of the streamwise velocity fluctuation $u_{x}'$ in an extended wake region for the nonlinear system with a clean freestream at two times: top row shows DNS results, and the bottom row shows results from causal resolvent-based estimation.}
\end{figure}

\begin{figure}[!t]
    \centering
    \includegraphics[scale=1,width=0.9\textwidth]{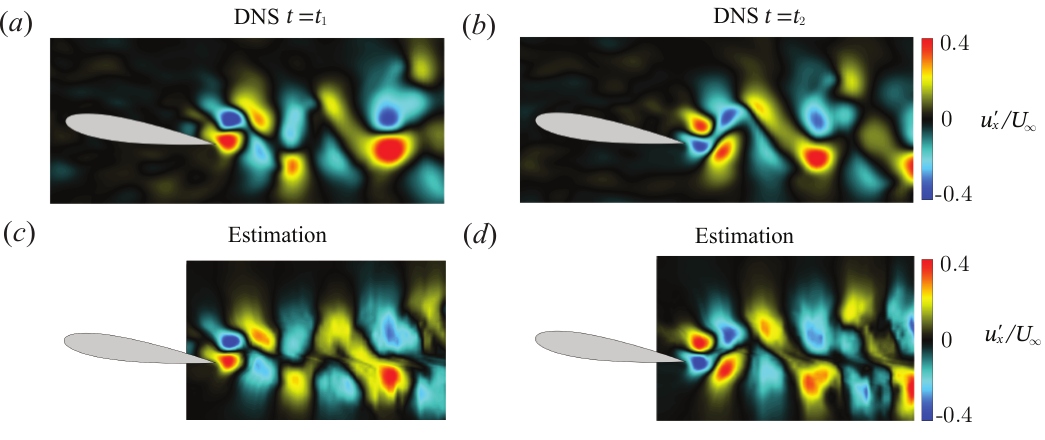}
    \caption[Estimation of the wake region for $u_{x}'$ for the clean system]{\label{Fig_NL_F_est_highrank_result_ux} Estimation of the streamwise velocity fluctuation $u_{x}'$ in an extended wake region for the nonlinear system with a noisy freestream at two times: top row shows DNS results, and the bottom row shows results from causal resolvent-based estimation.}
\end{figure}
Figure \ref{Fig_NLEST_DownstreamErrors} provides a more quantitative assessment of the estimation performance by showing the estimation error for each method as a function of the target position for both velocity components in the clean and noisy freestream cases.  Notably, the causal resolvent-based approach estimates both velocity components in the clean system with high accuracy. In the noisy system, the estimation accuracy decreases as the distance between the sensor and the target increases downstream. The causal resolvent-based approach enhances estimation accuracy near the trailing edge, where the lack of causality of the optimal noncausal estimator and the effects of nonlinearity are most significant.


Using the same four shear stress sensors, we estimate the velocity fluctuations $u_{x}'$ and $u_{y}'$ for an extended region of the wake in both clean and noisy nonlinear systems. The sensors are positioned near the trailing edge, allowing us to estimate the region behind $x/L_{c} > 0.8$. We present two snapshots of the estimation, selected to represent different phases of the vortex shedding. For the clean system, our estimation results for $u_{x}'$, as shown in figure \ref{Fig_NL_NoF_est_highrank_result}, are highly accurate. In the noisy nonlinear system, however, the estimation accuracy decreases due to perturbations that disrupt the vortex structure. Despite the challenges posed by chaotic fluctuations within the wake, the causal resolvent-based approach effectively estimates the wake flow, as demonstrated in figures \ref{Fig_NL_F_est_highrank_result_ux}.

\section{Resolvent-based control results}\label{sec_Resolvent-based control}

In this section, we use the resolvent-based controller described in $\S$\ref{subsec_Resolvent-based control} to suppress velocity fluctuations in the wake. 
Active flow control methods using blowing/suction, synthetic jets, and plasma actuators have been successfully used to suppress laminar vortex shedding behind bluff bodies in both experimental \citep{ffowcs_williams_active_1989, tao_flow_1996, strykowski_formation_1990, min_suboptimal_1999, fujisawa_feedback_2001} and numerical studies \citep{roussopoulos_nonlinear_1996, lin_feedback_2024}. Suppressing vortex shedding significantly reduces lift and drag fluctuations, attracting considerable engineering interest. We seek to mitigate velocity fluctuations in the wake, which naturally includes the influence of vortex shedding.

Resolvent analysis has been fruitfully used to guide open-loop control efforts by suggesting effective forcing frequencies and actuator locations to which the flow is responsive \citep{yeh_resolvent-analysis-based_2019,gross_laminar_2024} for open-loop control. Instead, we wish to design closed-loop controllers that react in real time to broadband frequency content within the flow. Using the estimated flow state from the resolvent-based estimator, implicitly included in the resolvent-based controller, the controller determines the actuation that most effectively cancels the target values, an approach also known as reactive control or the wave-canceling method \citep{sasaki_wave-cancelling_2018, sasaki_closed-loop_2018, morra_realizable_2020, martini_resolvent-based_2022, audiffred_experimental_2023}. As before, we first control the linear system before considering the actual nonlinear problem under noisy freestream conditions. We show that the optimal causal resolvent-based controller significantly outperforms a truncated non-causal controller, especially when using a pair of nested controllers to account for mean-flow modifications.

\subsection{Control set-up}

\begin{figure}[!h]
    \centering
    \includegraphics[scale=0.8,width=0.9\textwidth]{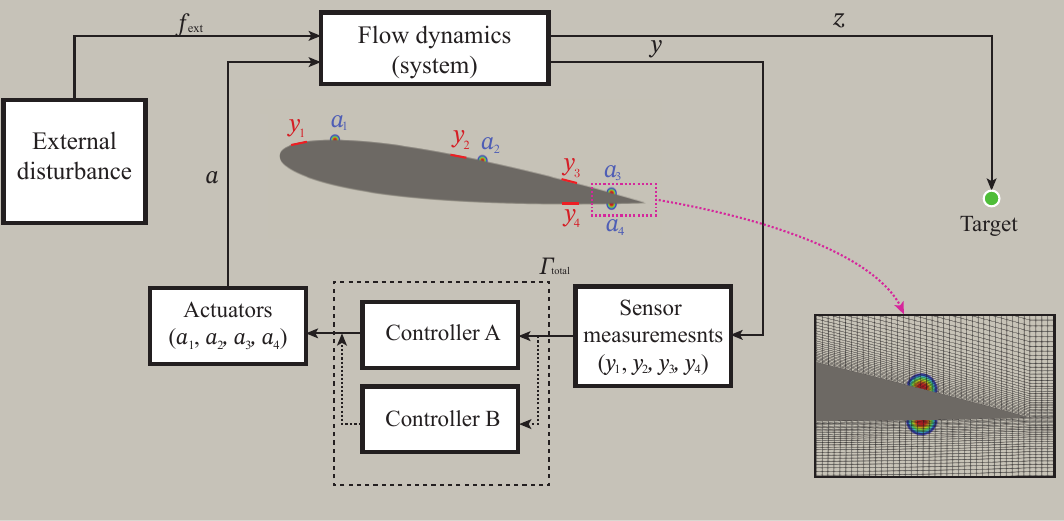}
    \caption[Control scheme for resolvent-based control of a NACA0012 airfoil]{\label{Fig_Block diagram} Control scheme for resolvent-based control of the flow around a NACA0012 airfoil. Red markers indicate shear-stress sensors, while contoured circles represent actuators in the form of momentum sources with Gaussian spatial support on the airfoil surface. The green circle marks the target location. For the control of the nonlinear system, we use a second nested controller (controller B) designed for the modified mean flow produced by the original controller (A). The insert highlights the grid resolution around the rear actuators.}
\end{figure}

Our overall control architecture is shown schematically in figure \ref{Fig_Block diagram}. Some closed-loop controllers used to suppress wake fluctuations rely on impractically located sensors, e.g., behind the body in the wake itself. Our resolvent-based controller avoids this issue by implicitly using the resolvent-based estimator to predict the evolution of disturbances using shear-stress sensors on the airfoil surface. Our actuators mimic unsteady blowing and suction at discrete positions on the surface of the airfoil, modeled as compact source terms (again with Gaussian spatial support) in the momentum equations. As shown earlier in figure~\pageref{Fig_FlowChart_RSVTOOLS}, the control signal is computed and applied online during the simulation.  

In choosing the placement of our sensors and actuators, we account for the combined amplifier and oscillator characteristics of this flow \citep{schmid_linear_2016}; the linear operator is globally stable, such that upstream disturbances are convectively amplified as they travel downstream, but it also contains a single lightly damped mode that generates amplifier-like vortex shedding. The sensor and actuator near the leading edge are designed to disrupt the recirculation bubble and suppress the global vortex-shedding behavior \citep{broglia_output_nodate,deda_neural_2023}.  The sensors and actuators near the trailing edge are designed to mitigate freestream fluctuations before they are amplified in the wake. We position the front sensors and actuators near the leading edge similarly to prior works\citep{colonius_control_2011, broglia_output_nodate, yeh_resolvent-analysis-based_2019, asztalos_modeling_2021}, while the positions of the rear sets are motivated by our estimation work.

For the nonlinear problem, the actuation modifies the mean flow via triadic interactions, even if the actuation itself is zero-mean. This presents a challenge since the controller is derived based on linearization about the original (uncontrolled) mean flow and is designed to minimize fluctuations to that mean. However, our primary objective remains to minimize the flow unsteadiness at the targets, i.e., the velocity fluctuations about the modified (controlled) mean flow. To address this challenge, we use an iterative approach similar to that of \cite{leclercq_linear_2019}. After applying the controller based on the original (uncontrolled) mean flow (labeled as controller A in figure \ref{Fig_Block diagram}), we design a second controller based on the new mean flow (controller B) and wrap it around the system that is still under the influence of controller A. The combined influence of the two controllers can be thought of as a single new nested controller that takes in the sensor measurements and drives the actuators. This iterative process can be repeated any number of times, but we achieve satisfactory results with just two nested controllers.

Following our previous work \citep{martini_resolvent-based_2022}, we quantify the controller performance via the metric
\begin{equation}\label{eq_con_perf}
\varepsilon_{\text{con}} = 1- \frac{\sum_{i} \int (\boldsymbol{z}_{i,\text{con}}(t))^{2}  dt}{\sum_{i} \int (\boldsymbol{z}_{i,\text{uncon}}(t))^{2}  dt},
\end{equation}
which measures the reduction in fluctuation energy at the targets compared to the uncontrolled flow.

\subsection{Control of the linear system}\label{subsec_Linear system in control}

We first consider the linear system obtained by linearizing the Navier-Stokes equations about the uncontrolled mean flow subject to external forcing, as in $\S$\ref{subsec_Linear system}. For the linear system, we use only controller A and just two actuators ($a_{3}$ and $a_{4}$). The resolvent-based control kernels are obtained using the operator-based approach described in \ref{subsec_Operator-based approach}. The direct and adjoint runs are conducted over the interval $t U_{\infty}/L_{c}\in [-48, 48]$.

\begin{figure}[!t]
    \centering
    \includegraphics[scale=0.9,width=1\textwidth]{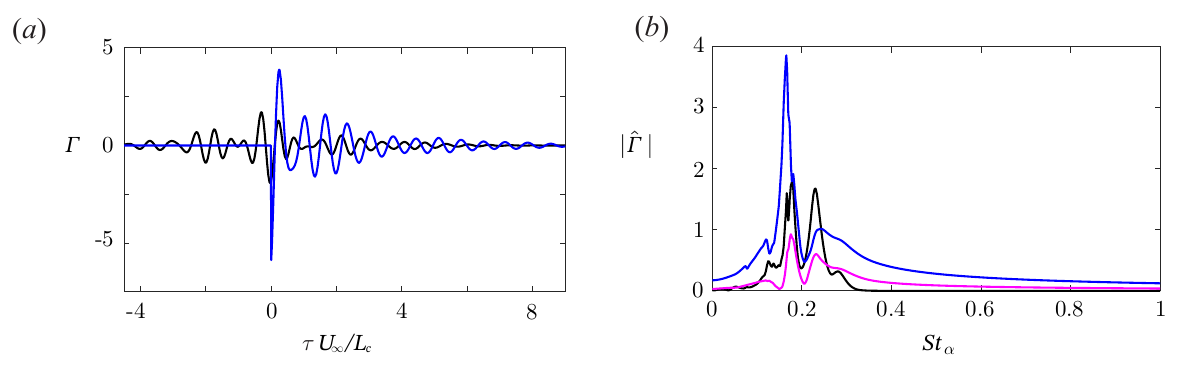}
    \caption[Control kernels in the time and frequency domains]{\label{Fig_controlkernels_linear} Control kernels with the sensor positioned near the trailing edge ($y_{3}$) and the target $z$ located at $x/L_{c}=1.5$ in the time (\textit{a}) and frequency (\textit{b}) domains. The black line represents the non-causal control kernel, the magenta line shows the truncated non-causal kernel, and the blue line depicts the causal control kernel computed using the Wiener-Hopf method.} 
\end{figure}

Figure \ref{Fig_controlkernels_linear} shows the non-causal and causal control kernels kernel between the $y_3$ sensor and $a_3$ actuator for a target $z$ at $x/L_{c}=1.5$ in the time and frequency domains. In figure \ref{Fig_controlkernels_linear}(\textit{a}), due to the close proximity of the sensor and actuator, the non-causal control kernel contains relatively large values in the non-causal part $\tau U_{\infty} / L_{c}<0$.  This issue is moderated using the Wiener-Hopf method, similar to the estimation kernels. For the causal kernel, the current measurement significantly impacts the actuation signal \citep{morra_realizable_2020, martini_resolvent-based_2022}. Figure \ref{Fig_controlkernels_linear}(\textit{b}) presents the control kernels in the frequency domain. The truncated non-causal control kernel (magenta line) shows a considerable loss in magnitude, while the causal control kernel significantly amplifies the sensor measurement at the vortex shedding frequency $St_{\alpha} = 0.17$. As expected, increasing the actuation cost $\mathsfbi{P}$ reduces the relative magnitude of the control kernel, leading to a smaller actuation force. We set $\mathsfbi{P}=\epsilon I$ with $\epsilon = 10^{-1}$ of the maximum value of $\mathsfbi{R}_{za}$. 

\begin{figure}[!t]
    \centering
    \includegraphics[scale=0.8,width=0.9\textwidth]{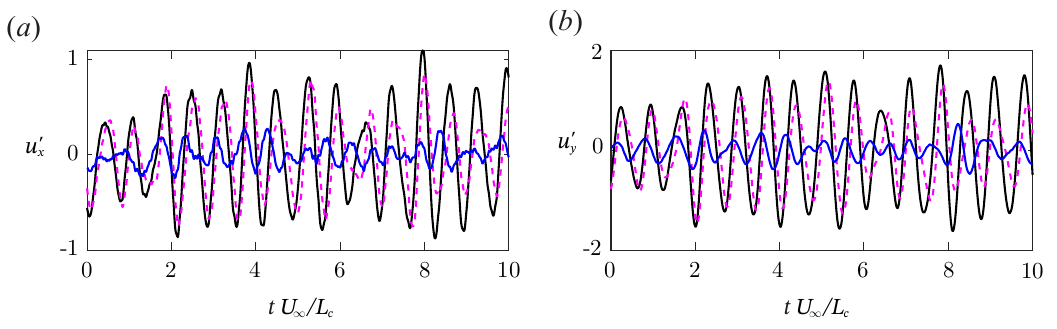}
    \caption[Time-series of velocity fluctuations for the uncontrolled and controlled linear systems]{\label{Fig_control_xy_linearsystem} Time-series of velocity fluctuations for the uncontrolled and controlled linear systems: (a) $u_{x}'$; (b) $u_{y}'$. Lines: uncontrolled (black line), truncated non-causal control (magenta dashed line), and causal control (blue line).}
\end{figure}

\begin{figure}[!t]
    \centering
    \includegraphics[scale=0.9,width=1\textwidth]{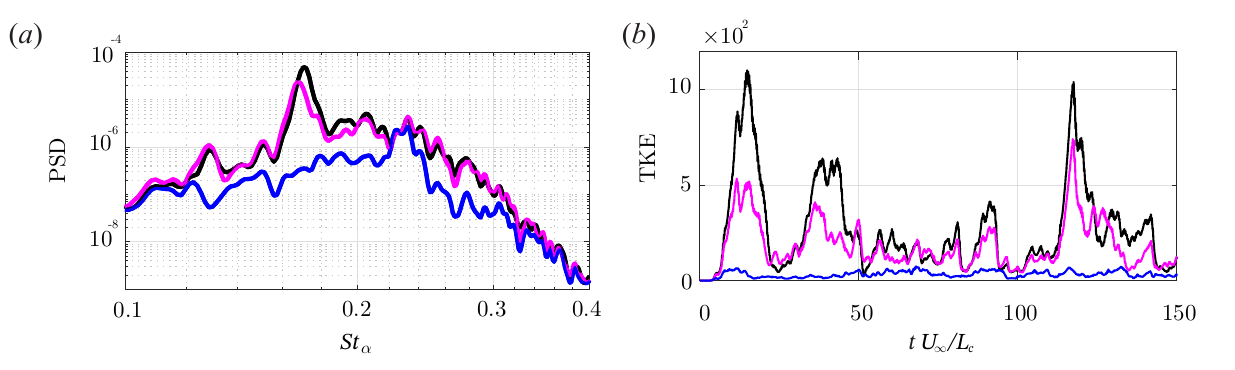}
    \caption[Control performance of the linear system]{\label{Fig_ControlPerformance_linear}Control performance for the linear system: (\textit{a}) PSD of the streamwise velocity fluctuations $u_{x}'$ for the controlled (blue) and uncontrolled (black) cases, with the magenta line representing the truncated non-causal control approach; (\textit{b}) Turbulent kinetic energy (TKE) integrated over the wake region.}
\end{figure}

Figure \ref{Fig_control_xy_linearsystem} presents the time-series data for the controlled and uncontrolled streamwise and cross-streamwise velocity fluctuations at the target. 
While the truncated non-causal controller achieves modest improvements, the causal resolvent-based control significantly reduces both velocity components. The power spectral density (PSD) of the uncontrolled and controlled streamwise velocity is shown in figure \ref{Fig_ControlPerformance_linear}(\textit{a}). The causal resolvent-based controller effectively suppresses the dominant vortex shedding frequency, while the truncated non-causal controller fails to accomplish this. Using the causal approach, the two actuators at the trailing edge reduce the turbulent kinetic energy of the velocity fluctuations at the target, as measured by (\ref{eq_con_perf}), by 85\%. In contrast, the truncated non-causal approach achieves only a 27\% reduction. In terms of RMS velocities, the causal controller achieves a 62\% decrease, compared to around 14\% for the truncated non-causal controller.

\begin{figure}[!t]
    \centering
    \includegraphics[scale=1,width=1\textwidth]{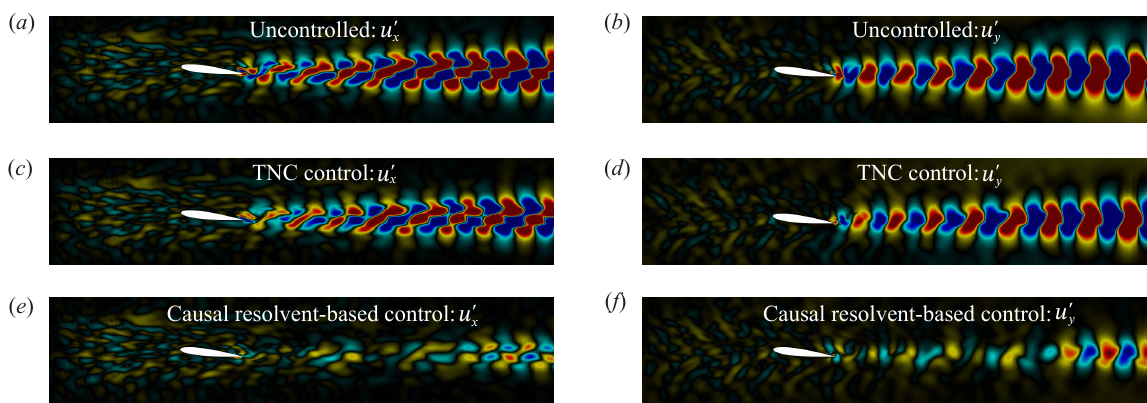}
    \caption[ The streamwise and cross streamwise velocity fluctuations field of the controlled and uncontrolled linear system]{\label{Fig_snapshot_lincon}  Snapshots of the streamwise and cross-streamwise velocity fluctuation fields for the linear system: (a,b) uncontrolled; (c,d) truncated noncausal (TNC) control; (e,f) causal resolvent-based control.}
\end{figure}

While the controller is designed to minimize the velocity fluctuations at the target, it also does so over an extended region of the wake. Figure \ref{Fig_snapshot_lincon} presents instantaneous snapshots of $u_{x}'$ and $u_{y}'$ for the controlled and uncontrolled systems. The wake modes excited by the upstream-generated external disturbance are effectively suppressed by the actuation at the trailing edge, as illustrated in figure \ref{Fig_snapshot_lincon}(\textit{e}) and (\textit{f}). To make this reduction quantitative, the turbulent kinetic energy(TKE) integrated over the wake region ($x/L_{c}=[1.1,5], y/L_{c}=[-1,1]$) is shown as a function of time in figure \ref{Fig_ControlPerformance_linear}(\textit{b}).

\subsection{Control of the nonlinear system}\label{subsec_Nonlinear system in control}

The ultimate goal of this work is to reduce the wake fluctuations in the nonlinear system (DNS) using the optimal linear resolvent-based controller. As discussed earlier, we use a second controller (controller B) to account for the impact of the first controller on the mean flow. After turning on the first controller, we wait until the flow achieves a new statistical steady state before turning on the second controller. The actuator outputs are determined by resolvent-based control kernels, with an additional constant forcing applied to the two front actuators to destroy the separation bubble. The data-driven approach is used to build the resolvent-based kernels to account for the colored nonlinear forcing. The operator approach is used to compute $\mathsfbi{R}_{ya}$ and $\mathsfbi{R}_{za}$, required in (\ref{eq_Gamma_nc_c_D}). Alternatively, these terms can be obtained using the data-driven approach by running a series of impulse response simulations, as discussed in $\S$\ref{subsec_Data-driven approach}. We show results only for the noisy freestream, as our controller entirely stabilizes the flow for the clean freestream case, resulting in a steady flow maintained by a steady actuation signal.

\begin{figure}[!t]
    \centering
    \includegraphics[scale=0.8,width=1\textwidth]{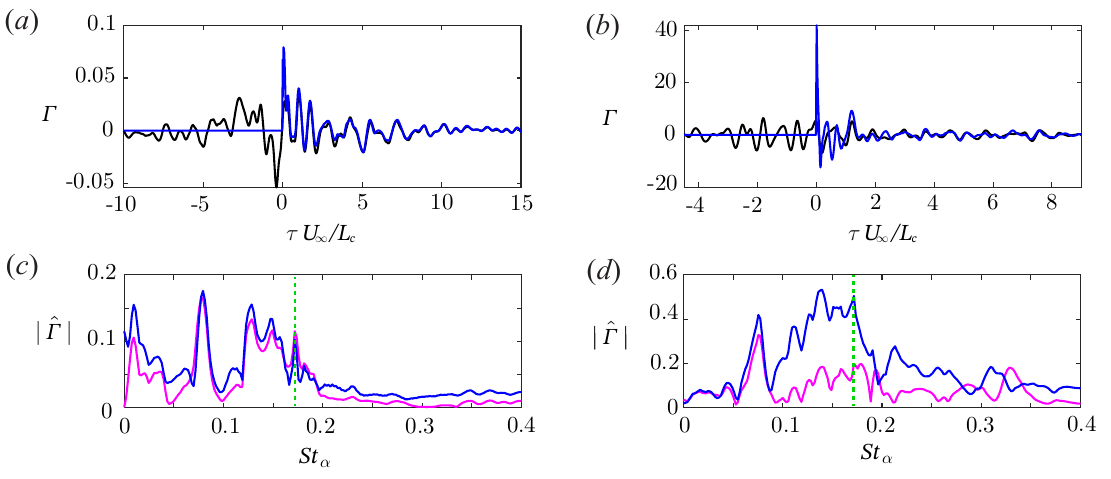}
    \caption[Control kernels for the nonlinear system]{\label{Fig_NLCON_kernels}Control kernels for the nonlinear system: (\textit{a}) and (\textit{b}) display the kernels in the time domain, while (\textit{c}) and (\textit{d}) present them in the frequency domain. Specifically, (\textit{a}) and (\textit{c}) correspond to $y_{3}$ and $a_{3}$, and (\textit{b}) and (\textit{d}) correspond to $y_{4}$ and $a_{4}$, as shown in figure \ref{Fig_Block diagram}. The green dashed line in $\textit{c}$ and $\textit{d}$ indicates the vortex shedding frequency.}
\end{figure}
\begin{figure}[!t]
    \centering
    \includegraphics[scale=0.8,width=0.7\textwidth]{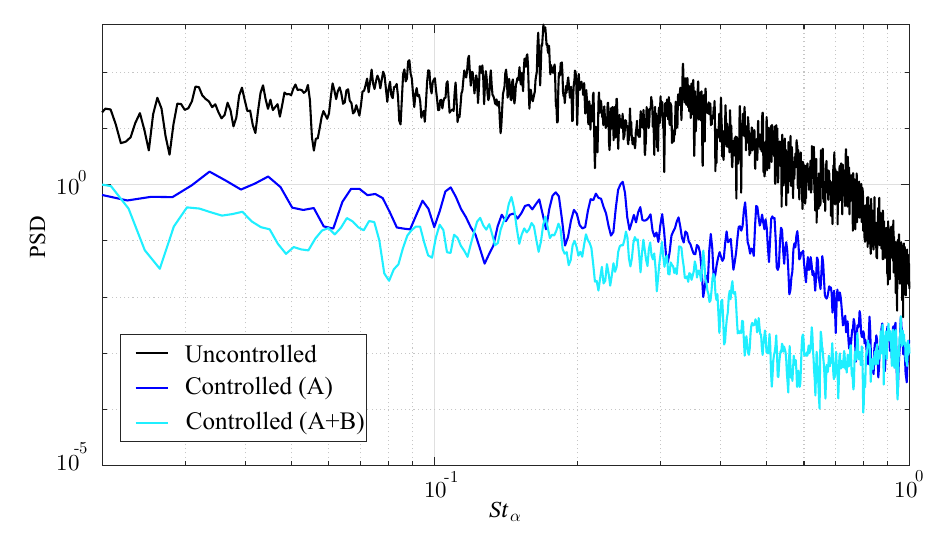}
    \caption[Power spectral density of the streamwise velocity fluctuations $u_{x}'$ at the target location]{\label{Fig_NLCON_PSD} PSD of the streamwise velocity fluctuation $u_{x}'$ at the target $z$ located at $[x/L_c, y/L_c] = [1.5, -0.11]$. The black solid line represents the uncontrolled flow; the blue line shows the controlled flow using Controller A; the cyan line depicts the controlled flow using both controllers (Controller A + Controller B).}
\end{figure}

Figure~\ref{Fig_NLCON_kernels} shows the control kernels for two sensor-actuator pairs in the time and frequency domains. The sensor, actuator, and target combination of figure \ref{Fig_NLCON_kernels}(\textit{a}) and (\textit{c}) are equivalent to those considered for the linear system in figure \ref{Fig_controlkernels_linear}. Since our sensor and actuator locations are positions close to each other, the peaks of the non-causal kernels are near $\tau U_{\infty} / L_{c}=0$. A notable difference between the kernels for the linear and nonlinear systems (which are different because of the colored nonlinear forcing) is that the vortex-shedding frequency is considerably less prevalent in the nonlinear case, presumably due to the disruption of the vortex shedding by the noisy freestream. Accordingly, the control kernels will amplify a wide range of frequencies in the sensor readings, producing broadband actuation signals.  

\begin{figure}[!t]
    \centering
    \includegraphics[scale=1,width=1\textwidth]{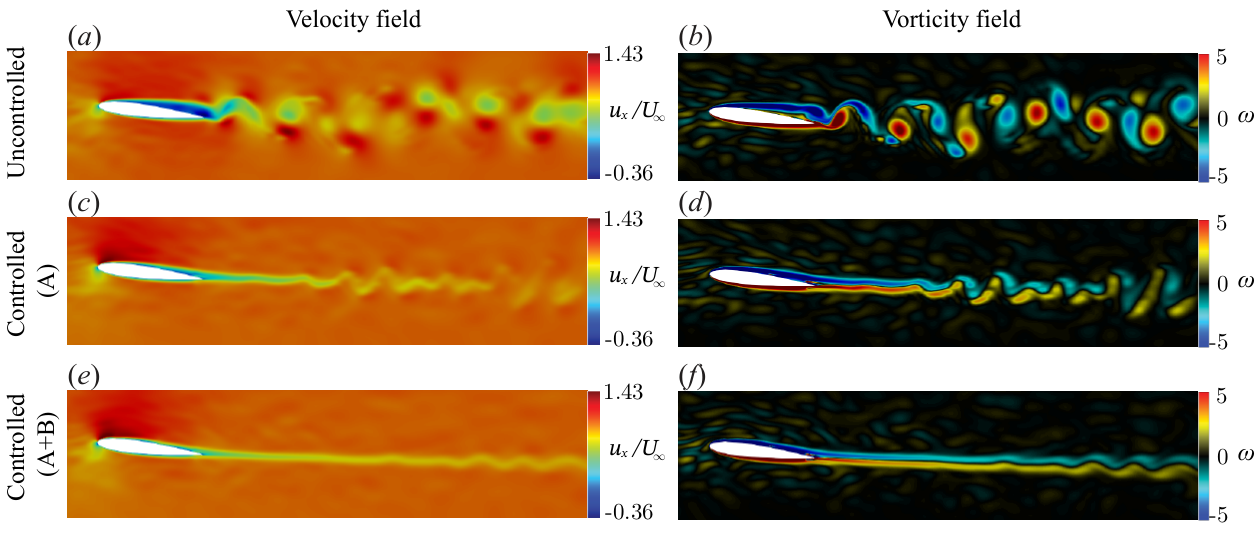}
    \caption[Velocity ($u_{x}$) and vorticity ($\omega$) fields for the system with noisy freestream inflow]{\label{Fig_NLCON_NL_F1}Velocity ($u_{x}$) and vorticity ($\omega$) fields for the system with noisy freestream inflow. (\textit{a}) and (\textit{b}) illustrate the uncontrolled flows; (\textit{c}) and (\textit{d}) show the controlled flows using Controller A; (\textit{e}) and (\textit{f}) present the controlled flows using both controllers (Controller A + Controller B).}
\end{figure}
\begin{figure}[!t]
    \centering
    \includegraphics[scale=0.9,width=0.8\textwidth]{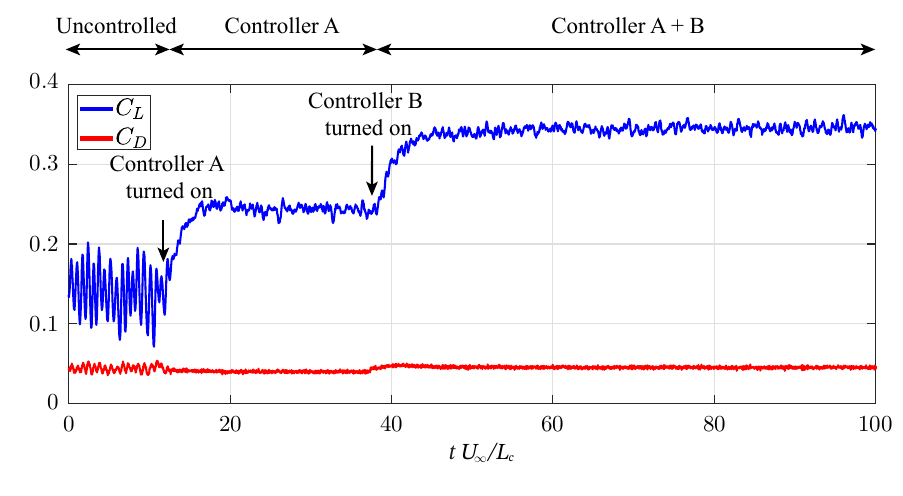}
    \caption[Controlled aerodynamic coefficients]{\label{Fig_NLCON_CDCL} Lift and drag coefficients for the uncontrolled and controlled flow over time.}
\end{figure}

Figure \ref{Fig_NLCON_PSD} shows the PSD of the streamwise velocity at the target for the uncontrolled and controlled flows. Both the single and nested controllers significantly reduce the energy of velocity fluctuations for $St_{\alpha}<1$. The control performance, measured by the metric in (\ref{eq_con_perf}) representing the reduction in velocity fluctuation energy, reaches approximately 94\% with Controller A. By incorporating a second controller (Controller A + B), the performance improves further, achieving 98\% reduction in fluctuation energy at the target. The reduction in RMS velocity is 78\% using Controller A and 90\% using both Controllers A and B. As shown in figure \ref{Fig_NLCON_PSD}, this improvement further mitigates target fluctuations at higher frequencies ($0.4 < St_{\alpha} < 0.7$), which originate further downstream, as we can verify in figure \ref{Fig_NLCON_NL_F1}.

Figure \ref{Fig_NLCON_NL_F1} presents both uncontrolled and controlled snapshots of the streamwise velocity and vorticity. As shown in figure \ref{Fig_NLCON_NL_F1}(\textit{a}) and (\textit{b}), chaotic vortex shedding is prominent in the wake of the uncontrolled flow. However, this can be significantly mitigated by controller A. Introducing controller B further suppresses the oscillating flow downstream ($x/L_{c} > 2$). In the controlled flow, the mean flow changed, which is essential as the wake fluctuations from vortex shedding originate from the separated flow. 

Finally, figure \ref{Fig_NLCON_CDCL} demonstrates the impact of the controllers on the aerodynamic coefficients. The time-averaged lift coefficient ($\bar{C}_{L}$) is improved by 143\% with the use of Controllers A and B, while the time-averaged drag coefficient ($\bar{C}_{D}$) remains largely unchanged. The fluctuations in both coefficients are effectively suppressed by the controllers. Although improving the aerodynamic coefficients was not the primary objective, mitigating wake fluctuations improved the lift coefficient as a welcome byproduct.

\section{Conclusions}\label{sec_Conclusion}
This work demonstrates the successful application of resolvent-based estimation and control to a two-dimensional NACA 0012 airfoil at $Ma_{\infty} = 0.3$, $Re_{L_{c}} = 5000$, and $\alpha = 6.5^{\circ}$. Under similar assumptions, our estimator and controller converge to the Kalman filter and LQG controller. However, our approach can incorporate the nonlinear terms of the Navier-Stokes equations using colored-in-time statistics, leading to significantly higher estimator accuracy and improved controller performance.\\
\indent To construct the estimator and controller, we employed two computational approaches: an operator-based approach and a data-driven approach. The operator-based approach is computationally efficient, does not require a priori model reduction, and accounts for the colored statistics of nonlinear terms from the Navier-Stokes equations that act as a forcing on the linear dynamics. The data-driven approach, which circumvents the need to construct linearized Navier-Stokes operators, naturally incorporates these colored statistics of the nonlinear terms. We utilized the Wiener-Hopf formalism to enforce causality, enabling the evaluation of only available measurements, which is an optimal strategy for real-time estimation and control.\\
\indent This work makes significant contributions to the implementation of resolvent-based frameworks. Specifically, it demonstrates the integration and execution of resolvent-based estimation and control approaches within a compressible flow solver designed for high-performance computing for large-scale problems. First, the linearized Navier-Stokes operator developed in this work is accurate and applicable for parallel computing, making it a powerful tool for large-scale applications. Second, using a parallel time-stepping approach significantly accelerates computations when handling large-scale linearized Navier-Stokes operators, including using a parallel adjoint solver efficiently. Additionally, the streaming Fourier transform within a solver is valuable for saving memory for constructing cross-spectral densities, which are necessary for building kernels. Another key feature of the tool is its ability to extract the nonlinear terms of the Navier-Stokes equations, offering valuable insights for other methodologies. Lastly, we solve the Wiener-Hopf problems directly within the solver to minimize reliance on post-processing tools such as MATLAB and to enhance computational efficiency by enabling faster routines with reduced memory usage.\\
\indent Before applying estimation and control to the laminar airfoil, we obtained the mean flow through direct numerical simulation and performed global stability and resolvent analysis around this mean flow. Random upstream perturbations were introduced to disrupt the periodic limit cycle caused by vortex shedding, inducing chaotic fluctuations. We then conducted resolvent-based estimation and control on both linear and nonlinear systems under these conditions.\\
\indent Our results demonstrate that resolvent-based kernels are effective in estimating and controlling chaotic fluctuations in the wake of an airfoil. The performance of both estimation and control is enhanced when sensors and actuators are strategically placed in effective locations. To determine these effective placements, we investigated estimation errors and employed a streamline strategy. While we addressed the estimation across the entire wake region, we found that controlling the entire region yields results similar to targeting a single point. In the linear system, the estimation error is approximately 8\% with two sensors, and the control performance reaches 85\% using two actuators. For the nonlinear system, the estimation error increases to 30\% with four sensors, while the control performance improves to 98\% using four actuators and two controllers. These accomplishments are significant, as they demonstrate the feasibility of applying a new closed-loop control method in a numerical setup and achieving reliable estimation. This work provides a solid foundation for extending the resolvent-based framework to other turbulent flow scenarios.\\ 
\indent Future work will focus on optimizing sensor and actuator placements, as well as applying control strategies to turbulent wakes behind an airfoil. In this study, we evaluated effective sensor placement based on estimation error, leading to satisfactory estimation and control results. However, the placement and the number of sensors and actuators have not yet been optimized. Addressing this will require formulating and solving a mathematical optimization problem, which will be considered in future research. Additionally, applying estimation and closed-loop control strategies to turbulent wakes is another area for future exploration. The positive effects of reducing turbulent wakes in terms of aerodynamic performance should be further investigated.






\begin{appendices}

\section{Resolution study of the baseflow in terms of eigenvalues}\label{appA}

Here, we study the grid convergence of the frequency of the vortex-shedding frequency in the DNS and the eigenvalue of the linear operator $\mathsfbi{A}$.  $N_{s}$ is the number of grid points along the suction surface, and $N_{w}$ is the number of grid points along the streamwise axis in the wake. Due to the need for a fine grid at the downstream target locations, we use a finer grid in the wake region compared to the typical grids with fewer points used in previous studies \citep{kojima_resolvent_2020,marquet_hysteresis_2022}. The Strouhal number (based on the angle of attack) $St_{\alpha,DNS}$ is the vortex-shedding frequency observed in the DNS, and $St_{\alpha,\mathsfbi{A}}$ is the frequency of the least-stable eigenvalue of the operator $\mathsfbi{A}$ linearized about the corresponding DNS mean flow. Based on our convergence study, we selected Mesh 5 for this work.

\begin{table}[h]
\centering
\begin{tabular}{ c c c c c c c}
\hline

Mesh & $N_{s}$ & $N_{w}$ & Total cells &  $St_{\alpha,\text{DNS}}$& $St_{\alpha,\mathsfbi{A}}$\\
\hline
1 & 100 &90 &  29,704 & 0.169650 & 0.168319   	\\
2 & 100 &180 &  43,924 & 0.168541 & 0.167249   \\
3 & 150 &135 &  67,354 & 0.169124 & 0.168933 	\\
4 & 200 &180 &  120,204 & 0.169525 & 0.169140  	\\
5 & 200 &268 &  148,188 & 0.169288 & 0.169241  	\\
6 & 200 &300 &  158,364 & 0.169211 & 0.169222  	\\
7 & 200 &360 &  177,444 & 0.169246 & 0.169207  	\\
  
 \hline
\end{tabular}
\caption[Grid convergence for resolution study]{Grid convergence of the DNS vortex shedding frequency and the least-stable eigenvalue of the linear operator $\mathsfbi{A}$.}
\label{tabA1}
\end{table}

\section{Wiener-Hopf Method}\label{appB}
\subsection{Theoretical Wiener-Hopf decomposition}
The Wiener-Hopf method \citep{noble1958} is a mathematical technique extensively used in applied mathematics. It enables the decomposition of arbitrary functions into components corresponding to the upper and lower halves of the complex plane. This paper leverages the Wiener-Hopf method to impose causality on estimation and control kernels, following methodologies outlined by \cite{daniele_fredholm_2007, martinelli_feedback_nodate,martini_resolvent-based_2022}.

First, we define the Fourier transform as
\begin{equation}\label{eq_FourierTransform_inWH}
 \boldsymbol{\hat{f}}(\omega) = \int_{-\infty}^{+\infty} \boldsymbol{f}(t) e^{-i\omega t} \, dt,
\end{equation}
where \(\boldsymbol{\hat{f}}(\omega)\) represents an arbitrary function in the frequency domain. This function can be decomposed into \(\boldsymbol{\hat{f}}_{+}(\omega)\) and \(\boldsymbol{\hat{f}}_{-}(\omega)\), which are analytic functions in the lower and upper complex half-planes, respectively. These can also be analyzed in the time domain as:
\begin{equation}\label{eq_causality}
 \boldsymbol{\hat{f}}_{+}(\omega) = \int_{0}^{+\infty} \boldsymbol{f}(t) e^{-i\omega t} \, dt,
\end{equation}
\begin{equation}\label{eq_noncausality}
 \boldsymbol{\hat{f}}_{-}(\omega) = \int_{-\infty}^{0} \boldsymbol{f}(t) e^{-i\omega t} \, dt,
\end{equation}
where the (\(+\)) subscript indicates that the function contains values only in the positive time domain, while the (\(-\)) subscript denotes that the function contains values only in the negative time domain. Thus, the subscripts (\(+\)) and (\(-\)) impose causality and non-causality on the function, respectively.

Now, we consider the two Wiener-Hopf problems \citep{martini_resolvent-based_2022} related to this paper's work,
\begin{equation}\label{eq_WHproblems1}
 \mathsfbi{\hat{H}}(\omega) \mathsfbi{\hat{\Gamma}}_{+}(\omega)  = \mathsfbi{\hat{\Lambda}}_{-}(\omega) + \mathsfbi{\hat{G}}(\omega),
\end{equation}
\begin{equation}\label{eq_WHproblems2}
 \mathsfbi{\hat{K}}(\omega) \mathsfbi{\hat{\Gamma}}_{+}(\omega)\mathsfbi{\hat{H}}(\omega) = \mathsfbi{\hat{\Gamma}}_{-}(\omega) + \mathsfbi{\hat{L}}(\omega)\mathsfbi{\hat{G}}(\omega),
\end{equation}
where $\omega$ = $i 2 \pi f$ with frequency $f$, and \( \mathsfbi{\hat{H}} \), \( \mathsfbi{\hat{G}} \), \( \mathsfbi{\hat{K}} \), \( \mathsfbi{\hat{L}} \) are known matrices (or functions), while \( \mathsfbi{\hat{\Gamma}}_{+}(\omega) \) and \( \mathsfbi{\hat{\Lambda}}_{-}(\omega) \) are the unknown matrices (or functions). The objective of the Wiener-Hopf problem is to determine \( \mathsfbi{\hat{\Gamma}}_{+}(\omega) \) and \( \mathsfbi{\hat{\Lambda}}_{-}(\omega) \).


To solve the two Wiener-Hopf problems, we employ two types of factorizations. The additive factorization decomposes the matrix into two \( \pm \) components, separated only through the addition process
\begin{equation}\label{eq1.22}
\mathsfbi{\hat{\Gamma}}(\omega) = (\mathsfbi{\hat{\Gamma}}(\omega))_{-} + (\mathsfbi{\hat{\Gamma}}(\omega))_{+},
\end{equation}
using the parenthesis $(\cdot)_{\pm}$.
The multiplicative factorization, which is not commutative, is performed as
\begin{equation}\label{eq1.23}
\mathsfbi{\hat{\Gamma}}(\omega) = \mathsfbi{\hat{\Gamma}}(\omega)_{-} \mathsfbi{\hat{\Gamma}}(\omega)_{+}.
\end{equation}
where the subscripts (\( \pm \)) are used without parentheses.
The solutions of the first Wiener-Hopf problems in (\ref{eq_WHproblems1}) is given by
\begin{equation}\label{eq_causal1}
\mathsfbi{\hat{\Gamma}}_{+}(\omega) = \left( \mathsfbi{\hat{G}}(\omega) \mathsfbi{\hat{H}}_{-}^{-1}(\omega) \right)_{+} \mathsfbi{\hat{H}}_{+}^{-1}(\omega),
\end{equation}
where \( \mathsfbi{\hat{H}}_{-} \) is obtained from the reverse multiplicative factorization. For (\ref{eq_WHproblems2}), the solutions is 
\begin{equation}\label{eq_causal2}
\mathsfbi{\hat{\Gamma}}_{+}(\omega) = \mathsfbi{\hat{K}}_{+}^{-1}(\omega) \left( \mathsfbi{\hat{K}}_{-}^{-1}(\omega) \mathsfbi{L}(\omega) \mathsfbi{G}(\omega) \mathsfbi{G}_{-}^{-1}(\omega) \right)_{+} \mathsfbi{\hat{H}}_{+}^{-1}(\omega).
\end{equation}

\subsection{Numerical Wiener-Hopf decomposition}

To ensure a seemly, fully parallelized workflow, we solve the Wiener-Hopf problems \citep{noble1958} used to enforce causality within the CFD solver. While the addictive factoration is straightforward to solve numerically, the multiplicative factorization is more complicated. A numerical solution for the multiplicative factorization was provided in \cite{martini_resolvent-based_2022}. The solution of the multiplicative factorization can be independently formed as 
\begin{equation}\label{eq3.14}
\hat{\mathsfbi{G}}(\omega) \hat{w}_{i,+} (\omega) = \hat{w}_{i,-} (\omega), 
\end{equation}
where
\begin{subequations}\label{eq3.15}
\begin{alignat}{2}
\hat{\mathsfbi{G}}_{-}(\omega) &=[\hat{w}_{1,-} (\omega),\hat{w}_{2,-} (\omega), \cdots, \hat{w}_{n_{freq},-} (\omega)],\\
\hat{\mathsfbi{G}}_{+}(\omega) &=[\hat{w}_{1,+} (\omega),\hat{w}_{2,+} (\omega), \cdots, \hat{w}_{n_{freq},+} (\omega)]^{-1},
\end{alignat}
\end{subequations}
where $\hat{\mathsfbi{G}}\in \mathbb{C}^{n_{y}\times n_{y}\times n_{\text{freq}}}$ or $\hat{\mathsfbi{G}}\in \mathbb{C}^{n_{a}\times n_{a}\times n_{\text{freq}}}$. To solve (\ref{eq3.14}), a Fredholm integral equation of the second kind \citep{daniele_fredholm_2007} is derived as   
\begin{equation}\label{eq3.16}
\hat{\boldsymbol{x}}_i(\omega)+\frac{1}{2 \pi \mathrm{i}} \int_{-\infty}^{\infty} \frac{\hat{\boldsymbol{G}}^{-1}(\omega) \hat{\boldsymbol{G}}(u)-1}{u-\omega} \hat{\boldsymbol{x}}_i(u) \mathrm{d} u=\hat{\boldsymbol{G}}^{-1}(\omega) \frac{\hat{\boldsymbol{w}}_{i,-}\left(\omega_0\right)}{\omega-\omega_0}.
\end{equation}

Due to the difficulty of the formation of the integration path, \cite{martini_resolvent-based_2022} constructed a linear problem with the size $\mathsfbi{G}\in \mathbb{C}^{n_{y}\times n_{y} \times n_{\text{freq}}}$, given by
\begin{equation}\label{eq3.17}
\hat{\boldsymbol{x}}_i(\omega)+\frac{1}{2 \mathrm{i}} \mathcal{H}\left(\hat{\boldsymbol{x}}_i\right)(\omega)-\frac{1}{2 \mathrm{i}} \hat{\boldsymbol{G}}^{-1}(\omega) \mathcal{H}\left(\hat{\boldsymbol{G}} \hat{x}_i\right)(\omega)=\hat{\boldsymbol{G}}^{-1}(\omega) \frac{\hat{\boldsymbol{w}}_{i,-}\left(\omega_0\right)}{\omega-\omega_0},
\end{equation}
where
\begin{equation}\label{eq_numericalWH_eqn}
\mathcal{H}(\hat{\boldsymbol{x}}) = P.V. \frac{1}{\pi} \int_{-\infty}^{\infty} \frac{1}{\omega - u} \hat{\boldsymbol{x}}(u) du,
\end{equation}
represents the Hilbert transform of $\hat{\boldsymbol{x}}(\omega)$. Equation (\ref{eq_numericalWH_eqn}) can be solved using the Generalized Minimal Residual (GMRES) iterative method \citep{saad_gmres_1986}. We solve (\ref{eq_numericalWH_eqn}) directly within the solver to minimize reliance on post-processing tools such as MATLAB and to enhance computational efficiency by enabling faster routines with reduced memory usage.
\section{Statistics of the nonlinear terms}\label{appx_nonlinear_terms}

\begin{figure}[!t]
  \begin{center}
      \begin{tikzpicture}[]
        \tikzstyle{every node}=[font=\small]
        \tikzset{>=latex}
        \node[anchor=south west,inner sep=0] (image) at (0,0) {
          \includegraphics[scale=1,width=1\textwidth]{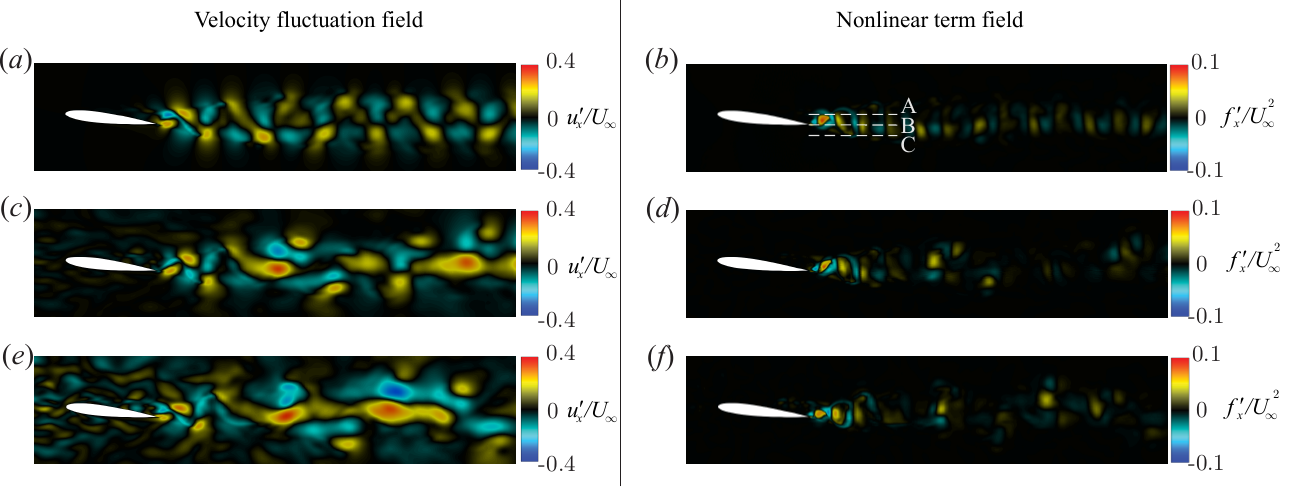}
        };
      \end{tikzpicture}
    \caption[]{\label{Fig_extF} Instantaneous velocity fluctuation field \( u_{x}^{'}\) ((\textit{a})(\textit{c})(\textit{e})) and the corresponding nonlinear terms $f_{x}'$ ((\textit{b})(\textit{d})(\textit{f})) computed from \ref{subsubsec_Extract_nlterms}. (\textit{a}), (\textit{b}): no forcing ($W =0$); (\textit{c}), (\textit{d}): \(W = 1\); and (\textit{e}), (\textit{f}): \(W = 3\).}
  \end{center}
\end{figure}

We explore how much the nonlinearity interaction is developed from the external forcing $\mathsfbi{F}_{ext}$ for the laminar airfoil flow with the intensity of the external disturbance. If the intensity of the external forcing is sufficiently small, (\ref{eq_Syy1_Szy1}) is approximately equivalent to (\ref{eq_Syy2_Szy2}). Figure \ref{Fig_extF} illustrates the instantaneous streamwise velocity fluctuations $u_{x}'$ field with $W=[0,1,3]$ in (\textit{a}), (\textit{c}), and (\textit{e}), and the corresponding the same time step's nonlinear terms in (\textit{b}), (\textit{d}), and (\textit{f}). The nonlinear terms are extracted using our application $\S$\ref{subsubsec_Extract_nlterms}. The clean system ($W=0$) shown in figure \ref{Fig_extF}(\textit{a}) and (\textit{b}) and the noisy systems ($W=[1,3]$) depicted in figures \ref{Fig_extF}(\textit{c}) - (\textit{f}) demonstrate that nonlinear terms predominantly exist in the wake near the trailing edge and in regions where the gradients of the velocity component are significant.

\begin{figure}[!t]
  \begin{center}
      \begin{tikzpicture}[]
        \tikzstyle{every node}=[font=\small]
        \tikzset{>=latex}
        \node[anchor=south west,inner sep=0] (image) at (0,0) {
          \includegraphics[scale=0.8,width=0.8\textwidth]{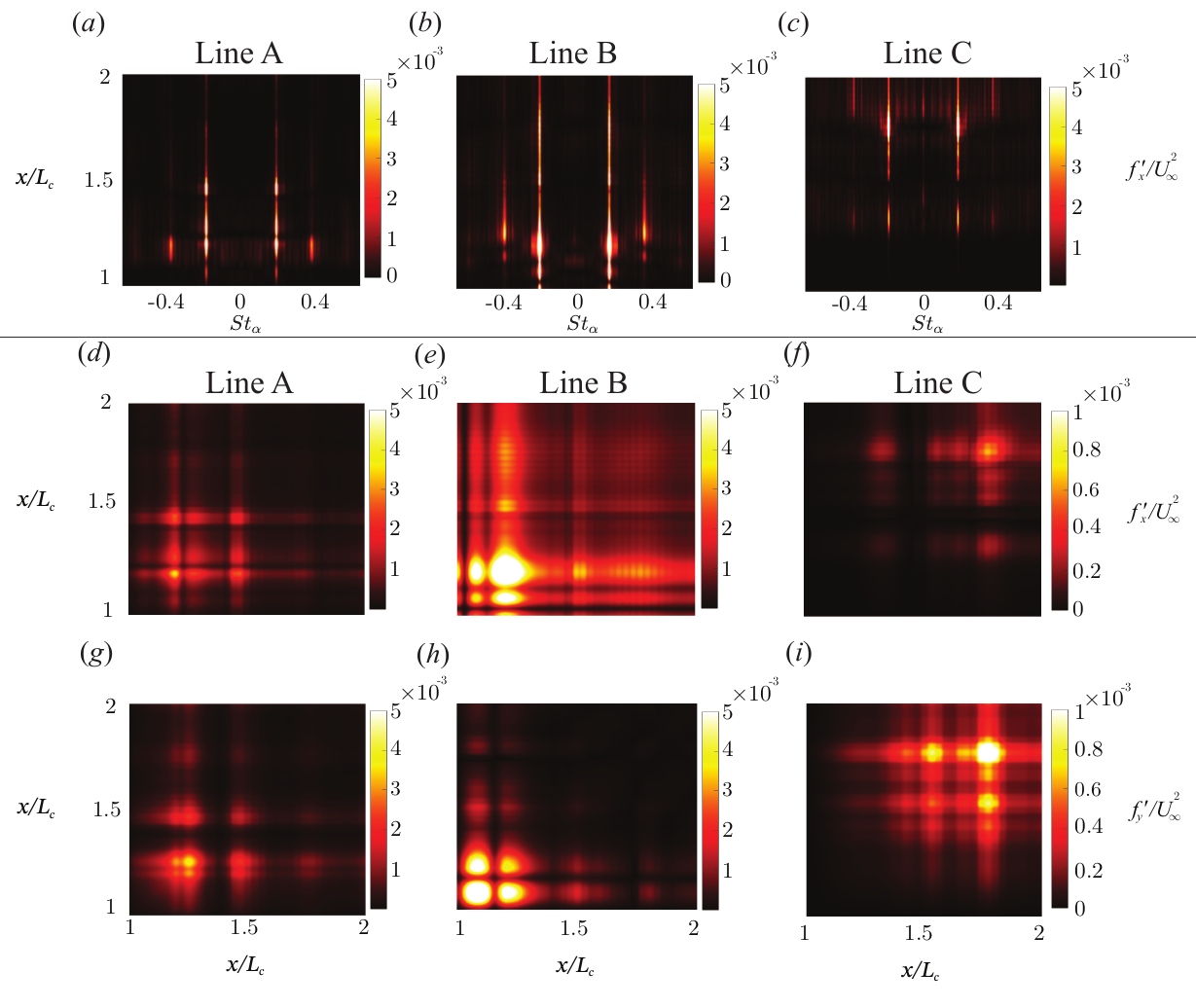}
        };
      \end{tikzpicture}
    \caption[]{\label{Fig_NLterm_PSD} Power Spectral Densities (PSDs) and Cross-Spectral Densities (CSDs) of the nonlinear terms. The top three panels (\textit{a})–(\textit{c}) display PSDs along Lines A, B, and C in figure \ref{Fig_extF}(\textit{b}). Panels (\textit{d})–(\textit{f}) show the CSDs of $f_{x}'$, and panels (\textit{g})–(\textit{i}) present the CSDs of $f_{y}'$ for each corresponding line.}
  \end{center}
\end{figure}

Figure \ref{Fig_NLterm_PSD} presents the PSDs and the CSDs at vortex shedding frequency ($St_{\alpha}=0.17$) of the nonlinear terms along lines A ($y/L_{c}=0.01$), B ($y/L_{c}=-0.11$), and C ($y/L_{c}=-0.21$) behind the airfoil, shown in figure \ref{Fig_extF}(\textit{b}). In figure \ref{Fig_NLterm_PSD}(\textit{a})–(\textit{c}), the nonlinear terms exhibit significant energy at the vortex-shedding frequencies. Strong energy is observed near the trailing edge ($x/L_{c}<1.5$) along the top (A) and middle (B) lines, as shown in figures \ref{Fig_NLterm_PSD}(\textit{a}) and (\textit{b}), while the bottom line (C) shows energetic regions further downstream, as seen in figure \ref{Fig_NLterm_PSD}(\textit{c}). This pattern arises because vortex shedding primarily originates from the separated flow on the suction surface, located slightly above the trailing edge. In figures \ref{Fig_NLterm_PSD}(\textit{d})–(\textit{i}), the nonlinear terms for both velocity components exhibit statistically significant strength in the region $1<x/L_{c}<1.2$ along line B, consistent with the observed deterioration in the estimation performance of the linear system. Examining the CSD tensors $\hat{\mathsfbi{F}}_{nl}$ is meaningful, as they are closely associated with coherent structures \citep{towne_spectral_2018}. By constructing the forcing CSD matrix $\hat{\mathsfbi{F}}_{nl}$, we can account for the nonlinear effects of the flow in the estimation kernels, potentially improving estimation accuracy in turbulent flows.

\section{Sensor placement for estimation}\label{appx_sensorplacement}

\begin{table}[!t]
  \centering
  \begin{tabular}{ c  c  c  c  }
    \multicolumn{2}{c}{Placement candidates}   &  \multicolumn{2}{c}{Errors} \\
    &   &  Clean system &  Noisy system \\
      \hline
    1 
    &
    \begin{minipage}{.3\textwidth}
      \includegraphics[scale=0.4]{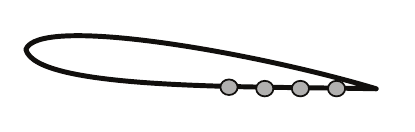}
    \end{minipage}
    &
    0.04
    &
    $0.54$\\
        2
    &
    \begin{minipage}{.3\textwidth}
      \includegraphics[scale=0.4]{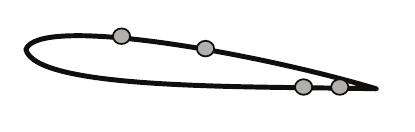}
    \end{minipage}
    &
    0.03
    &
    $0.40$\\
            3
    &
    \begin{minipage}{.3\textwidth}
      \includegraphics[scale=0.4]{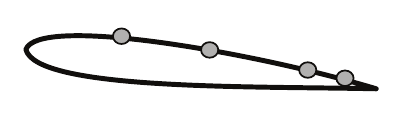}
    \end{minipage}
    &
    0.02
    &
    $0.36$\\
            4
    &
    \begin{minipage}{.3\textwidth}
      \includegraphics[scale=0.4]{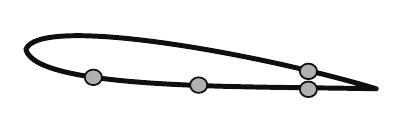}
    \end{minipage}
    &
    0.01
    &
    $0.35$\\
            5
    &
    \begin{minipage}{.3\textwidth}
      \includegraphics[scale=0.4]{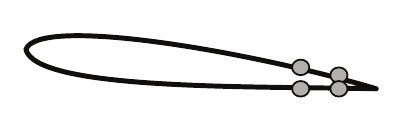}
    \end{minipage}
    &
    0.001
    &
    $0.33$\\
            6
    &
    \begin{minipage}{.3\textwidth}
      \includegraphics[scale=0.4]{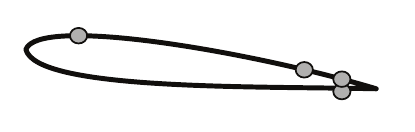}
    \end{minipage}
    &
    0.001
    &
    $0.33$\\

    \\ 
  \end{tabular}
  \caption[Sensor placement candidates for causal resolvent-based estimation]{Sensor placement candidates for causal resolvent-based estimation.}\label{table_optimalsensorplacement}
\end{table}

Based on the single-sensor results reported in $\S$\ref{subsubsec_SensorPlacement}, we propose several candidates for the most effective sensor placement or multiple sensors, shown in table \ref{table_optimalsensorplacement}. The targets are averaged points of interest for both clean and noisy nonlinear systems. We verified that the estimator can improve accuracy when the sensor is at the trailing edge. However, this configuration is impractical for the control problem, which will be discussed in the upcoming section. Ultimately, We adopted candidate 6 in table \ref{table_optimalsensorplacement}, where sensors are located upstream of the separation bubble on the suction surface and near the trailing edge on both the suction and pressure surfaces for estimation. The mathematical formulations for determining the optimal sensor placement are beyond the scope of this study and will be addressed in future research.

\end{appendices}

\bibliographystyle{plainnat}
\bibliography{arXiv_template}

\end{document}